\documentclass{article}

\usepackage{arxiv}

\usepackage[T1]{fontenc}
\usepackage[utf8]{inputenc}
\usepackage{amsmath,amssymb,amsthm}
\usepackage{mathtools}
\usepackage[expansion=false,protrusion=false]{microtype}
\usepackage[round,authoryear]{natbib}
\usepackage{enumitem}
\usepackage{booktabs}
\usepackage{tabularx}
\usepackage{xcolor}
\usepackage[hidelinks,colorlinks=true,urlcolor=blue!60!black,
            linkcolor=black,citecolor=blue!50!black]{hyperref}
\usepackage{siunitx}
\usepackage{graphicx}
\graphicspath{{figures/}{diagrams/}{./}}
\usepackage{caption}
\usepackage{qrcode}
\usepackage{tikz}
\usetikzlibrary{arrows.meta,positioning,calc,shapes.geometric}

\usepackage{subcaption}
\usepackage{rotating}
\usepackage{placeins}
\usepackage{tcolorbox}
\tcbuselibrary{breakable}
\newcommand{\rtdlink}[2]{\href{https://smn-lab.readthedocs.io/en/latest/#1}{#2}}
\newtcolorbox{benchbox}[1][]{%
  breakable, colback=blue!3, colframe=blue!40!black, boxrule=0.4pt,
  left=6pt, right=6pt, top=4pt, bottom=4pt, fontupper=\small,
  before upper={\textbf{In the companion bench#1.}\ }}

\newtheorem*{definition*}{Definition}
\newtheorem*{remark*}{Remark}
\newtheorem*{claim*}{Claim}

\newcommand{\caz}{\textsc{caz}}
\newcommand{\smn}{\textsc{smn}}

\newcommand{\hap}{\textsc{hap}}
\newcommand{\nap}{\textsc{nap}}

\newcommand{\fig}[1]{Fig.~\ref{#1}}

\title{The Sensation Modulating Network:\\
Haltability as the architectural ground for object-directed
phenomenology}
\shorttitle{The Sensation Modulating Network}
\author{
G.~Nagarjuna\thanks{Email: \texttt{nagarjuna@iiserpune.ac.in};
ORCID: \url{https://orcid.org/0000-0001-6773-8454}}\\
\small Indian Institute of Science Education and Research (IISER) Pune, India
\and
Durgaprasad Karnam\thanks{Email: \texttt{karnamdpdurga@gmail.com};
ORCID: \url{https://orcid.org/0000-0002-7620-0223}}\\
\small Centre for Educational Technology (CET), Indian Institute of Technology Bombay, India
}
\date{}

\begin{document}
\maketitle

\begin{abstract}
We propose the \emph{Sensation Modulating Network} (SMN): the cognitive
agent conceived as the whole body, organized at every anatomical scale
by \emph{opponent} dynamics, built from \emph{Sensation Modulators} ---
tissue that senses and acts through one substrate --- paired into
\emph{Coordinated Action Zones} routed by a body-wide broadcast.  It is
an \emph{inclusive model of the body}: one in which physical variables
such as gravity, material properties such as elasticity, and the
topological and geometrical structure of the body itself do
constructive cognitive work.  This
paper is scoped to what such a body constructs at its foundation --- a
self-model, a world-model in that self's frame, and object-directedness
--- each constructed by the body's physics, not assumed as a primitive.
We present the
architecture as a \emph{generative} model: one small kit of primitives
whose morphological variations (chain, sheet, tube, layered,
appendicular) construct experience by the \emph{same} mechanism, an
invariance we demonstrate for the self-model across body plans and
scales.  The central thesis is that \emph{haltability} --- the active
holding of an opponent equilibrium --- is the architectural condition
object-directed phenomenology requires; a second principle, that an
object is a bundle of more than one property, carries the construction
from a felt resistance to a genuine object.  A companion computational
bench realizes each construction as a runnable, falsifiable experiment,
each with a pre-registered order parameter and matched foil.  We close
by placing the principal competing accounts --- sensorimotor
enactivism, active inference, ecological and affordance-based theories,
and others --- as \emph{limiting cases} within a wider landscape, stating
in each case the falsifiable criterion that would tell them apart, and by
giving systems and cognitive neuroscience its place: the nervous system as
the integrating core that makes the body one, not a commander over it.
On this account, the cognitivism--4E impasse stems from an incomplete
account of the embodied agent's architecture: its resolution begins
not with the brain alone but with the whole body, modelled this way.
\end{abstract}


\section{Introduction}\label{sec:intro}

\subsection*{On reading this paper}

This paper asks what an embodied cognitive agent must structurally
\emph{be}, and proposes one architectural answer.  We ask the reader to
entertain four questions.  \emph{What if embodiment is not the situation
of the brain in the body, but of the brain as one organ within a body
that is itself the unit of cognition?}  \emph{What if the muscular system
is not the body's organ for implementing behaviour, but its primary organ
for sensing space?}  \emph{What if action is, before it is anything else,
the means by which an organism computes its location?}  \emph{What if
intentional cognition requires, above all, the architectural capacity to
\emph{halt} --- to hold an attentional state actively still against
ongoing dynamics?}  Grant the alternative architecture these questions
suggest, and a number of things fall into place: the boundary between
self and world, the directedness of attention, the gap between affordance
and action, the layered evolutionary history of motor competence.  Our
objective is not to show the alternative is correct but that it is
\emph{possible} --- that an architecture answering all four questions
affirmatively can be specified coherently and made to do explanatory
work.  We build it, deliberately, from the smallest agent that can be
drawn and simulated, and add nothing that is not forced by composing that
modular unit into larger bodies.

\subsection*{The problem}

The problem is, first, a live engineering one.  Constructing a
self-model, a world-model, and an object-model is exactly what much of
contemporary AI and robotics is trying to do: world-models learned for
planning \citep{Ha2018WorldModels,LeCun2022WorldModels,Hafner2020Dreamer},
image models that synthesize whole visual worlds by denoising
\citep{Rombach2022LatentDiffusion}, robots that fit models of their own
bodies \citep{Bongard2006SelfModeling}, and SLAM systems that map an
environment while localizing within it
\citep{DurrantWhyteBailey2006SLAM}.  What these share, for all their
power, is that they build the model \emph{from outside} the agent --- from
a corpus an outside observer's camera has seen, from an absolute map, from
a central optimizer searching a space of candidates.  None builds the
world the way an animal does: \emph{from the body's own engagement}, as
the invariant of what its own action modulates, in a frame anchored to the
body rather than to the world.  That is the gap this paper addresses --- and
it is against these engineering programs, as much as against the
cognitive-science accounts, that the framework is most sharply, and most
falsifiably, distinguished (\S\ref{sec:rel-robotics}).  One tradition is
the exception, and the kin: \emph{epigenetic} (developmental) robotics
builds cognition developmentally, from the agent's own engagement
\citep{Balkenius2001EpigeneticRobotics}, and the framework offered here is
a specification within that lineage.

This gap has a deeper root in cognitive science itself, long split between
two programs.  Cognitivism --- rooted in a nativist, information-processing
paradigm --- explains language and reasoning through amodal representations
and computational rules, generally taken to reside in the brain
\citep{Chomsky1957SyntacticStructures,Fodor1975LanguageOfThought,Pinker1994LanguageInstinct}.
4E approaches (embodied, embedded, enactive, extended) ground cognition in
the situated action of the whole body in its environment
\citep{VarelaThompsonRosch1991,MaturanaVarela1980,noe_action_2004,clark1997being,Thompson2007MindInLife}.
Cognitivism accounts for the compositionality of language but cannot say
how formal symbols acquire meaning --- the symbol-grounding problem
\citep{Harnad1990SymbolGrounding}; 4E approaches explain how perception and
action are coupled but struggle with the recursive, compositional
structure cognitivism foregrounds.  Neither has produced a unified account
of both the grounded character of animal cognition and the generative
character of human language, and neither specifies the body in enough
architectural detail to say how a world could be built \emph{from} it:
cognitivism treats the body as peripheral input/output, 4E invokes it as
constitutive without a mechanism.  It is this missing specification --- of
the body as the thing that constructs the model --- that the engineering
programs inherit when they reach outside the agent for one instead.

This paper offers the specification.  It takes the whole body as the unit
of cognitive analysis, and treats the body's \emph{physics} as
constructive rather than incidental --- gravity, the elasticity of the
muscular substrate, and the topological and geometrical structure of the
body are the materials from which a self-model, a world-model, and
object-directedness are built.

\subsection*{The proposal}

We propose the \emph{Sensation Modulating Network} (SMN): the cognitive
agent conceived as the entire body, organized as a network of
\emph{Coordinated Action Zones} (CAZs) whose elementary operation is the
modulation of sensation through action.  Every opponent pair is organized
through \emph{Sensation Modulators} that simultaneously sense and act,
while sensory surfaces participate as read-only transducers coupled into
the same network.  The nervous system is not the cognitive engine but the
plastic communication backbone that routes signals among these
distributed elements --- a broadcaster and a shared state-space, not a
commander.  Action does not require central initiation: organisms without
nervous systems move, and isolated cardiac tissue beats
\citep{Llinas2001-xl,Levin2014MolecularBioelectricity}.  What organizes
the body at every anatomical scale is \emph{opponency} --- sympathetic
against parasympathetic, flexor against extensor, left against right,
excitation against inhibition --- and it is in those opponent balances
that feeling and knowing meet
\citep{Damasio1999_Feeling,Damasio2010_Self,Damasio2021_FeelingKnowing}.

\emph{This paper is scoped to the foundation.}  It develops what an
inclusive, physically grounded body constructs before any of it is
learned or conventionalized: a self-model, a world-model in that self's
frame, and the object-directedness that haltability affords.  Two
features make the account a \emph{model} rather than a description of one
animal.  First, it is \emph{generative}: a small kit of primitives ---
segment, CAZ, transducer --- composes by a few moves into the whole
range of animal bodies, and the \emph{same} construction mechanism, we
show, is invariant under those moves.  Second, each construction is
\emph{falsifiable}: it comes with an order parameter and a matched foil,
realized as a runnable experiment in a companion computational bench, so
the argument can be carried compactly here while the growing record lives
in the bench.  The higher action patterns through which generative
grammar and the cognitivism--4E reconciliation would be reached are
introduced only as the trajectory this foundation makes \emph{locatable}
(\S\ref{sec:taxonomy}), and are left to companion work.  We do, however,
close (\S\ref{sec:related}) by placing the principal competing accounts
as \emph{limiting cases} within a wider landscape, stating in each case the
falsifiable criterion that would tell them apart.

\subsection*{Haltability and phenomenology}

The architectural payoff that motivates the paper concerns
\emph{phenomenology}.  What does an embodied system need to support
object-directed phenomenology in Husserl's sense
\citep{Husserl1900_LU,Husserl1913_Ideas}?  The answer we develop is
\emph{haltability}.  Without gaps in action --- without the architectural
ability to halt and attend --- there is no capacity to hold something as
available, interruptible, and resumable, and therefore as aboutable.  The
claim is structural, not metaphysical: haltability provides the
architectural locus intentional directedness needs, not that it
\emph{produces} phenomenology.  The chain runs through the architecture
--- opponency makes co-activation available; co-activation makes halt
possible; halt makes attention possible; attention makes intentional
directedness possible --- with no module added on top.  A second
constructive principle completes the move from directedness to an object:
nothing counts as an object unless it has more than one property, so
objecthood is a cross-modal invariant, not a single felt resistance.  The
eight registers we catalogue (\S\ref{sec:registers}) are the behavioural
fingerprints of this dispositional architecture.

\subsection*{Plan of the paper}

The paper is a single construction, built in the order its physics and
its epistemics accumulate.  \S\ref{sec:atom} presents the \emph{modular unit} ---
three segments, two CAZs --- and the physics it carries, reading the
architecture's seven commitments off the finished block.
\S\ref{sec:kit} shows the modular unit is one instance of a \emph{generative}
kit, and states the invariance the construction relies on.
\S\ref{sec:selfmodel} constructs the self-model and demonstrates that
invariance across body plans and scales; \S\ref{sec:worldmodel} builds
the world-model in the self's frame and the structural self/world/other
distinction that is the first epistemic transition.
\S\ref{sec:directedness} turns haltability into object-directedness, and
\S\ref{sec:object} turns the resistance it is directed at into an object
--- reaching the threshold of the second epistemic transition, which is
the boundary of the paper's scope.  \S\ref{sec:taxonomy} names the taxonomy of action patterns that
more deeply layered bodies climb, and \S\ref{sec:related} places the
competing accounts as limiting cases with falsifiable criteria.
\S\ref{sec:neuro} gives systems and cognitive neuroscience its place ---
the nervous system as the body's integrating core, not its commander.
\S\ref{sec:conclusion} concludes.  The formalism is developed
\emph{inline}, next to the construction each equation makes precise;
\ref{app:formalism} indexes it, gathers the eight registers, points
to the companion bench, and --- in \S\ref{app:category-theory} --- invites a
category theorist to give the framework a compositional formalization.

\section{The argument, and its runnable companion}\label{sec:companion}

\begin{figure}[t]
\centering
\resizebox{\linewidth}{!}{%
\begin{tikzpicture}[
   stage/.style={draw=blue!45!black, fill=blue!5, rounded corners,
                 align=center, font=\small, minimum height=3.2em,
                 text width=6.6em, inner sep=3pt},
   reg/.style={align=center, font=\scriptsize, text=black!65},
   >=Stealth, node distance=1.4em]
\node[stage] (s1) {\textbf{1}\\[1pt] Modular unit};
\node[stage, right=of s1] (s2) {\textbf{2}\\[1pt] Generative morphology};
\node[stage, right=of s2] (s3) {\textbf{3}\\[1pt] Self-model};
\node[stage, right=of s3] (s4) {\textbf{4}\\[1pt] World; self/world/other};
\node[stage, right=of s4] (s5) {\textbf{5}\\[1pt] Haltability \& directedness};
\node[stage, right=of s5] (s6) {\textbf{6}\\[1pt] From resistance to object};
\foreach \a/\b in {s1/s2,s2/s3,s3/s4,s4/s5,s5/s6}
  \draw[->,blue!45!black,thick] (\a) -- (\b);
\node[reg, below=2pt of s1] {R1, R2};
\node[reg, below=2pt of s2] {grammar};
\node[reg, below=2pt of s3] {topology};
\node[reg, below=2pt of s4] {R5, Q2};
\node[reg, below=2pt of s5] {E3, Pred1};
\node[reg, below=2pt of s6] {E2, P3};
\end{tikzpicture}}
\caption{The paper's spine, and its realization in the companion bench.
Each construction stage (\S\S\ref{sec:atom}--\ref{sec:object}) maps to one
or more preregistered bench experiments (register tags below each stage;
see Table~\ref{tab:bench-map} for the links and results), and the bench
pages are ordered to follow this same 1--6 progression.}
\label{fig:bench-roadmap}
\end{figure}
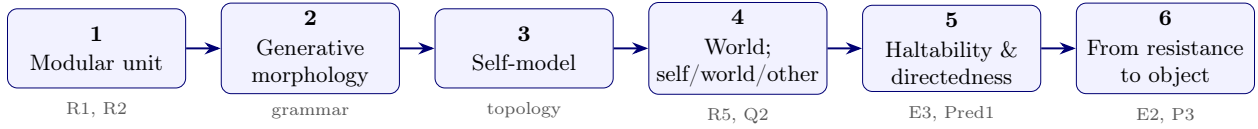

The construction that follows is not offered as theory alone.  Each step
--- the modular unit, the morphology, the self-model, the world-model,
haltability, and the object --- is realized in an open companion bench,
\textsc{smn-lab}, a MuJoCo testbed in which the paper's commitments become
\emph{preregistered} experiments: a hypothesis and its order parameter are
stated in advance, the relevant formalism is injected directly into
runnable code, a matched foil is run alongside, the result is plotted, and
the whole is reproducible with a single command.  Every bench figure
quoted in this paper is a mean over $N$ independent seeds ($N{=}20$ for
Registers~1, 2, and~5; $N{=}10$ for the reafference ratio), reported with
its standard deviation --- distributions, not single runs; the one
deterministic contrast, the pattern-hold of \S\ref{sec:directedness},
carries no seed variance, and its robustness is a parameter sweep.  The
bench is documented
at \rtdlink{}{smn-lab.readthedocs.io}\footnote{All bench links in this paper
--- those in Table~\ref{tab:bench-map}, in the roadmap
(Fig.~\ref{fig:bench-roadmap}), and in the per-section \textsf{In the
companion bench} boxes --- are \emph{external} hyperlinks to the online
documentation; they are not internal cross-references and cannot be
followed offline.}, organized to follow this paper's
sections, and \S\ref{app:simulations} indexes the realizations.  This
section shows, once and concretely, how a commitment becomes a number ---
so that the central architectural claims can be followed into runnable,
falsifiable demonstrations wherever the Phase-I bench implements them.

\begin{benchbox}[ --- a worked example: tonic-load coupling (R1)]
\begin{description}[leftmargin=7.5em, font=\itshape, itemsep=1.5pt, topsep=3pt, parsep=0pt]
\item[Commitment] alert energy makes a partner zone's tonic tone rise with
  the active zone's load.
\item[Formalism] $\dot E_R=\rho\,F_{\mathrm{active}}\,\mathbf{1}[g{=}1]-E_R/\tau_E$,
  with partner tone $a_0+\beta E_R$, predicting a load--tone slope
  $\beta\rho\tau_E$.
\item[Code] the prediction is injected verbatim as the
  \texttt{AlertEnergyBoard} operator, with the equation and the lines that
  realize it shown side by side on the bench page.
\item[Hypothesis] \emph{preregistered:} the measured slope equals
  $\beta\rho\tau_E$; a classical-inhibition foil is load-independent.
\item[Result] slope $0.744$ (sd\,$<\!0.001$) $\approx$ predicted $0.750$; the foil is flat.
\item[Replicate] \texttt{python experiments/sweep\_r1\_tonic\_load.py}
  $\to$ \rtdlink{experiments/sweep_r1_tonic_load/}{bench page}.
\end{description}
\end{benchbox}

\begin{table}[!ht]
\centering\small
\caption{Where to read each construction in the companion bench.  Bench
links resolve to the preregistration, the formalism-to-code injection, and
the plotted result for each experiment.  These are external, online-only
links, not internal references.}
\label{tab:bench-map}
\begin{tabularx}{\linewidth}{@{}c l l X@{}}
\toprule
\S & Construction & Bench & Preregistered order parameter $\to$ result (vs.\ foil) \\
\midrule
\ref{sec:atom} & Modular unit & \rtdlink{experiments/sweep_r1_tonic_load/}{R1}, \rtdlink{experiments/sweep_r2_resumption/}{R2} & load--tone slope $\to 0.744$ (sd\,$<\!0.001$) $\approx\beta\rho\tau_E$, foil flat; resumption $121$--$172$\,ms vs.\ $311\pm1$\,ms \\
\ref{sec:selfmodel} & Self-model & \rtdlink{experiments/self_model_topology/}{topology} & graph recovery $\to$ exact on elastic, chance on rigid foil; morphology-invariant \\
\ref{sec:worldmodel} & World; self/world/other & \rtdlink{experiments/sweep_r5_dual_signal/}{R5}, \rtdlink{experiments/q2_reafference/}{Q2} & residual $R^2 \to 0.998$ self / $0.05\pm0.03$ decoupled \\
\ref{sec:directedness} & Haltability \& directedness & \rtdlink{experiments/p6_haltability_aboutness/}{E3}, \rtdlink{experiments/sweep_pred1_haltability/}{Pred1} & deceptive-reach halt signature vs.\ ballistic foil \\
\ref{sec:object} & Resistance $\to$ object & \rtdlink{experiments/p5_self_field_object/}{E2} & self/field/object factoring; crossmodal binding deferred \\
\ref{sec:taxonomy} & Taxonomy & \rtdlink{experiments/p7_scaling_network/}{E4}, \rtdlink{experiments/p8_objecthood_transition/}{E4b} & world-resolution scales with network; objecthood as density transition \\
\bottomrule
\end{tabularx}
\end{table}

Every construction section below closes with an \textsf{In the companion
bench} box giving its specific realization: the preregistered order
parameter, the result, the matched foil, and a direct link to the bench
page where the formalism, the code, and the plots can be inspected and
re-run.

\section{The SMN modular unit: a building block and its physics}\label{sec:atom}

The simplest agent this framework describes is small enough to draw
and to simulate: three body segments joined by two \emph{Coordinated
Action Zones}.  We call it the \emph{modular unit} of the architecture --- in
the spirit of a composable modulatable element, not because it is all
there is, but
because it already exposes the structural commitments every larger
body will share.  Everything the rest of the paper constructs --- a
self-model (\S\ref{sec:selfmodel}), a world-model
(\S\ref{sec:worldmodel}), object-directedness
(\S\S\ref{sec:directedness}--\ref{sec:object}), and a taxonomy of
action patterns (\S\ref{sec:taxonomy}) --- is built by composing and
layering this one block.  It is worth being concrete about it first,
because it is also where the physics enters.  For this agent, gravity,
mass, friction, elasticity, and opponency are not background
conditions on cognition; they are the very parameters of its
dynamics.  A physicist reading this section is reading the model in
full: what follows adds bodies and scales, but no new physics that is
not already present in the modular unit.

\begin{figure}[!htbp]
\centering
\includegraphics[width=0.82\linewidth]{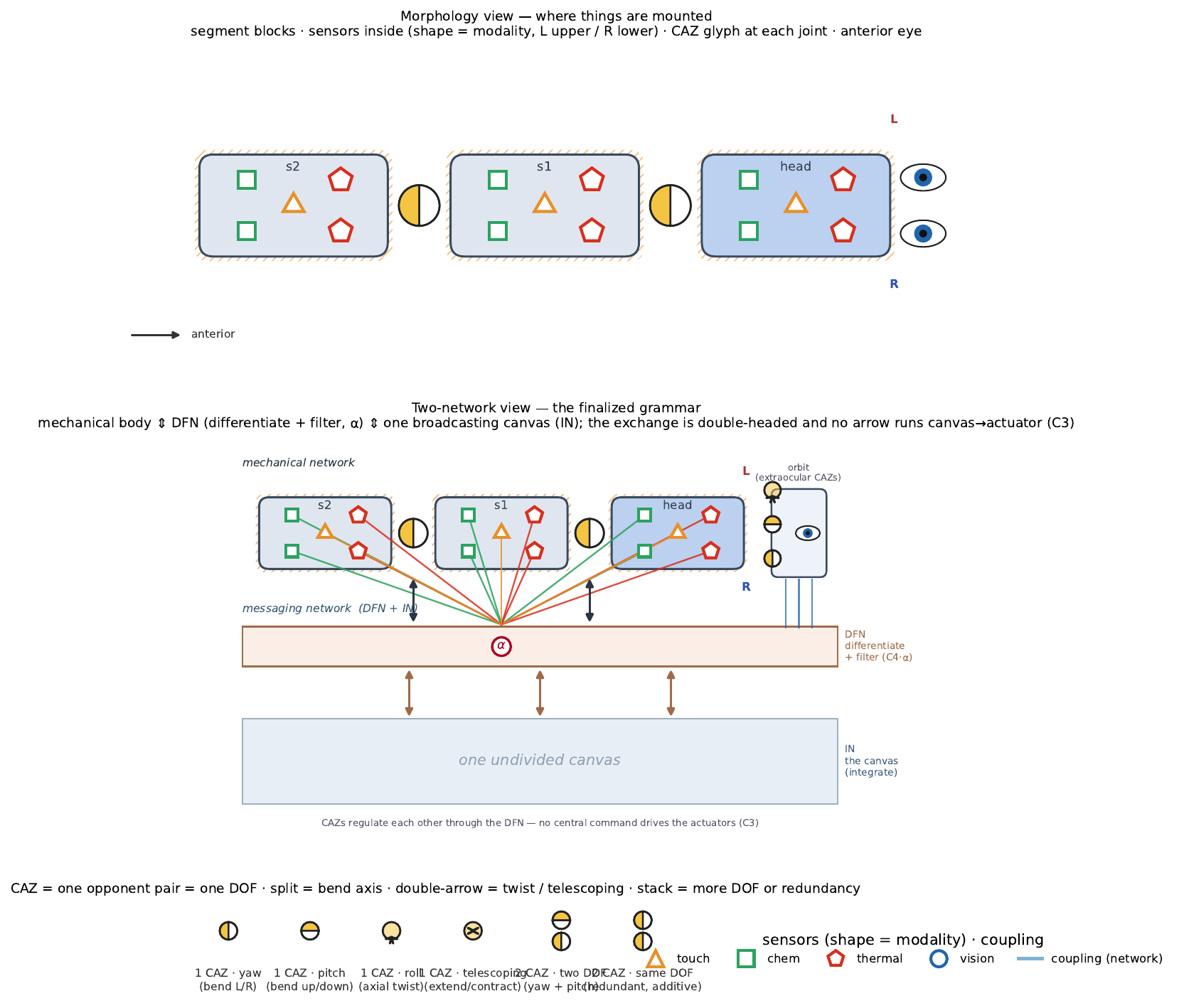}
\caption{The modular unit and the two networks, drawn in the diagram grammar
this paper uses throughout.  \emph{Top --- morphology view:} the three segments
(head, $s_1$, $s_2$), with sensors mounted inside each block (shape encodes
modality; L upper, R lower), a split-circle \textbf{CAZ glyph} at each
of the two joints, and the anterior eye that marks the front.
\emph{Middle --- two-network view:} the same body as two coupled networks --- the
\emph{mechanical network} above and the \emph{messaging network} below (both defined
in \S\ref{sec:kit-network}).  The messaging network is drawn as its two functional
parts: a \emph{differentiating and filtering} layer (DFN), where the
active-perception filter $\alpha$ (C4) admits only streams a board's own action
predicts, feeding the \emph{one} broadcasting canvas that \emph{integrates} them
(IN).  The body\,$\Leftrightarrow$\,DFN exchange is double-headed
(afferent and peer-routed efferent), single-interface transducers reach the DFN as
colour-coded bundles (the eye on its orbit is a bundle of extraocular CAZs, not a
passive sensor), and --- crucially --- \emph{no} arrow runs from the canvas to the
actuators: the zones regulate one another through the DFN, not by central command
(C3).  \emph{Bottom --- key:} one split-circle CAZ is one opponent pair and one
degree of freedom (the split sets the bend axis --- yaw, pitch, roll, or
telescoping), with the sensor-shape legend.  Every later figure reuses this
grammar.  \emph{In the bench:} it is the documented diagram grammar
(\rtdlink{diagram-grammar/}{smn-lab}).}
\label{fig:atom-anatomy}
\end{figure}

We build the modular unit in the order in which its physics accumulates: the
opponent primitive (\S\ref{sec:atom-opponency}), the sensing-and-acting
tissue that realizes each pole (\S\ref{sec:atom-sm}), the balance the
pair forms and the messaging beam that couples it
(\S\ref{sec:atom-caz}), the joint dynamics and the elastic substrate
that carries them (\S\ref{sec:atom-joint}), the routing board
(\S\ref{sec:atom-board}), the haltability operator and its energetic
state variable (\S\ref{sec:atom-halt}), and the read-only transducer
that couples the block to the world (\S\ref{sec:atom-transducer}).  We
then read the architecture's seven commitments off the finished block
(\S\ref{sec:atom-commitments}), collect its physical parameters in one
place (\S\ref{sec:atom-physics}), and close with the experiments the
modular unit already affords (\S\ref{sec:atom-experiments}).

\subsection{Opponency is the primitive}\label{sec:atom-opponency}

The biological body is organized by opponent dynamics at every scale
at which it is organized, and the modular unit inherits this as its founding
principle.  Below the autonomic level, every cell maintains itself
through opponent metabolic dyads (anabolism against catabolism,
synthesis against breakdown) and opponent ionic transporters holding
the $\mathrm{Na^+\!/K^+}$, $\mathrm{Ca^{2+}}$, and pH gradients.  The
autonomic layer balances sympathetic against parasympathetic drive,
with homeostatic set points \emph{emerging} from the balance rather
than being computed and imposed
\citep{Damasio1999_Feeling,Damasio2010_Self}.  The motor layer runs on
Sherrington's reciprocal innervation
\citep{Sherrington1906_IntegrativeAction,sherrington1905reciprocal}:
flexor and extensor at a joint are jointly modulated, and
equilibrium-point control
\citep{Feldman1986_EquilibriumPoint,Feldman2015ReferentControl} makes
the controlled variable the \emph{equilibrium of the pair}, not the
activation of either pole.  The bilateral and segmental layer ---
tetrapod gait, fish swimming, worm peristalsis, lamprey undulation ---
runs on left--right opponency between half-centres
\citep{Grillner2006Lamprey,Grillner2020PrinciplesMotorControl,MarderBucher2001CB}.

The pattern is the same at every layer: the architectural primitive is
an opponent pair whose dynamics produce a \emph{balance state}, and
behaviour and feeling emerge from modulation of that balance.  We take
this convergence across scales as a substantive empirical observation
--- not a parsimony choice --- and make it the modular unit's first
commitment.  The two CAZs of the modular unit are the motor-layer instance of
the primitive; the same primitive is what makes every other layer, and
every larger body, an SMN.

\subsection{Opponency is included contraries, not a servo}\label{sec:atom-invertibility}

It matters how the opponent pair is read.  Read as a \emph{servo} --- two
actuators driving a joint toward a set-point supplied from outside --- it is a
piece of control engineering, and nothing about it is alive.  That reading is
available, but it is the \emph{limiting case}, not the primitive.  The opponent
pair is the physical realization of \emph{a process and its inverse held within
one system} --- the principle of \emph{included contraries}, or \emph{dialogical
invertibility} \citep{GN2005_TowardsModelLifeCognition,Nagarjuna1994_Inversion}.
A belief system obeys the excluded middle: a proposition and its negation cannot
both belong to it.  A \emph{living} physical system obeys the opposite rule ---
a process $P$ and its inverse $\bar P$ are parts of \emph{the same} system ---
and it is this internalized counter-process that is the root of autonomy.  Where
the counter-process is external (the set-point imposed, the correction supplied
from without), the unit degenerates to a servo; where it is \emph{within}, the
unit maintains its own identity against perturbation --- organizational closure
in the sense of \citet{MaturanaVarela1980}.

This is not a gloss added after the fact.  The elastic substrate
(\S\ref{sec:atom-joint}) and the alert-energy dynamics (\S\ref{sec:atom-halt})
make the unit's persistence \emph{depend on its own inverting activity}: a body
that cannot invert the perturbations it meets does not hold together.  A living
opponent unit is therefore \emph{precarious} by construction, not by stipulation
--- the repair-or-perish condition that distinguishes a Being from a mechanism
\citep{GN2005_TowardsModelLifeCognition}.  The servo has no such stake; the
opponent-as-included-contraries does.  We keep this distinction live throughout:
where we write of ``control'' or ``targets'' we mean the internalized case ---
hear opponency as invertibility, not as set-point tracking.

\subsection{The Sensation Modulator: sensing and acting through one
substrate}\label{sec:atom-sm}

Each pole of an opponent pair must both \emph{sense} the state of the
pair (in order to maintain or modulate the balance) and \emph{act} to
produce it.  We formalize this dual function as a \emph{Sensation
Modulator} (SM), the smallest structural unit of the SMN.  Each SM has
a \emph{mechanical interface}, through which it engages the body's
geometry (a configuration variable and an associated force), and a
\emph{messaging interface}, through which it participates in the
network (an efferent activation it receives, an afferent signal it
emits).  The SM is not a modelling choice; it is what tissue
\emph{does} wherever opponency is the operating principle --- the same
tissue that produces displacement also senses what the displacement
did to the sensory field.  It \emph{modulates sensation through
action}, and the network built from these units inherits its name.
(We say ``messaging interface'' rather than ``neural pathway''
deliberately: the architecture is implementation-neutral about what
carries the message --- electrical, hormonal, optical, software.)

This dual role of the messaging interface --- one tissue that both receives an
efferent activation and emits an afferent signal --- carries an architectural
commitment: every signal the network routes, efferent and afferent alike, must be
mutually translatable within a \emph{common operational code}, so that action and
perception share one substrate rather than meeting across a translation boundary
\citep{Prinz1990CommonCoding,Hommel2001TEC}.  The dual-signal property, the
active-perception filter (Commitment~C4), and the later externalization of patterns
as public traces all presuppose it.

Three properties make the SM a single-substrate unit rather than a
sensor bolted to an actuator.  \emph{First}, the force is a
\emph{joint} function of the activation received and the current
configuration, and it is \emph{pull-only}: tissue can pull along its
axis but not push; pushing is what the antagonist supplies.
\emph{Second}, the afferent signal is read out from the same tissue
whose deformation produces the force --- the same configuration
variable enters both.  The two coupled relations are
\begin{align}
F &= \Phi(a,\,q,\,\dot q), \qquad F\geq 0 \text{ (pull-only)},
\label{eq:Phi}\\
s &= \Psi(q,\,\dot q,\,F),
\label{eq:Psi}
\end{align}
with $a$ the activation, $(q,\dot q)$ the configuration, $F$ the
force, and $s$ the afferent signal.  \emph{Third} --- the defining
commitment --- there is no internal boundary between ``the sensor''
and ``the actuator''.  This is a substantive claim about
muscle--tendon biology, supported by the placement of muscle spindles,
Golgi tendon organs, and joint receptors within the contracting tissue
itself
\citep{ProskeGandevia2012,TuthillAzim2018ProprioceptionReview}.  The
specific functional form of $\Phi$ (Hill-type for vertebrate skeletal
muscle; other forms for cardiac, smooth, or artificial actuators) is a
matter of implementation design; the structural commitments ---
pull-only, jointly activation-and-configuration dependent, single
substrate --- are what the architecture rests on, and they survive
substitution of the actuator.

One consequence is easily missed and will matter later: when a zone
acts, the sensation it modulates is not only its own embedded
receptors.  A turn of the neck modulates the visual field, the
vestibular signal, and trunk proprioception at once.  The SM's
function therefore reaches into the whole network of read-only sensory
transducers (\S\ref{sec:atom-transducer}), not just the acting zone
--- which is why action generates sensorimotor contingencies the agent
can learn (\S\ref{sec:worldmodel}).

\subsection{The Coordinated Action Zone: a balance with a messaging
beam}\label{sec:atom-caz}

A \emph{Coordinated Action Zone} (CAZ) is a pair of Sensation
Modulators in antagonistic relationship --- two zones $Z_+$ and $Z_-$
whose forces oppose along a shared configuration variable --- together
with a small computational substrate, the \emph{communication board}
$B$, that routes messages between them.  The CAZ is the smallest scale
at which opponency produces a coherent body-state variable the agent
can both sense and modulate; the modular unit has two of them.

The image to hold is a balance scale.  A balance has two pans coupled
inversely through a rigid beam: when one rises the other falls, and
equilibrium is the configuration in which their loads match.  A CAZ
behaves dynamically like a balance --- forces from $Z_+$ and $Z_-$
oppose along the shared variable, equilibrium is where they match, and
modulating one end displaces the equilibrium predictably.  The
instrument-like calibration a physical balance exhibits (its
sensitivity, its reliability, the inverse-functional relation that
makes it a good measurement device) is recovered for the CAZ from the
same opponent inverse coupling.

There is one consequential disanalogy.  A balance's beam is a rigid
mechanical object; a CAZ has no such beam.  Even when $Z_+$ and $Z_-$
are anatomically adjacent, they are not mechanically linked the way two
pans are.  Their balance is produced not by proximity but by the board
$B$, which routes the message each zone emits to its opponent, supplying
the reciprocal activation each receives.  \textbf{The board is the beam
of the CAZ-balance, but it is a messaging beam rather than a rigid
mechanical one.}  We should be precise about what this does \emph{not}
mean: the broadcast the board carries is not a non-physical signal.  It
is \emph{physically realized but functionally individuated and
medium-independent} --- every message is a physical state (a tension, a
rate, a potential) that one zone writes into a shared medium and its
partner reads by responding to it, with nothing contentful shipped and
nothing reading it from outside.  A signal here is just a difference in
one part of the body that makes a difference to another; what
individuates the beam is its \emph{modulatory role}, not a dedicated
physical channel.

This is what makes the architecture \emph{cognitive} rather than
merely mechanical: a messaging beam can be re-routed, re-weighted,
gated, or modulated in ways a rigid beam cannot.  A physical balance's
beam cannot rewire itself; the SMN's board can.  Learning, adaptation,
and context-sensitivity enter here --- not as machinery added on top,
but as the operating mode of a balance whose beam is a messaging
substrate.  Crucially, the board does not \emph{originate} either
zone's activation: the efferent a zone receives is the board's routing of its
\emph{opponent partner}'s afferent signal, which the board weights, gates, or
releases but does not generate.  The opponent pair \emph{is} the
coupling; the board is the medium through which it propagates.

\begin{figure}[!htbp]
\centering
\includegraphics[width=0.8\linewidth]{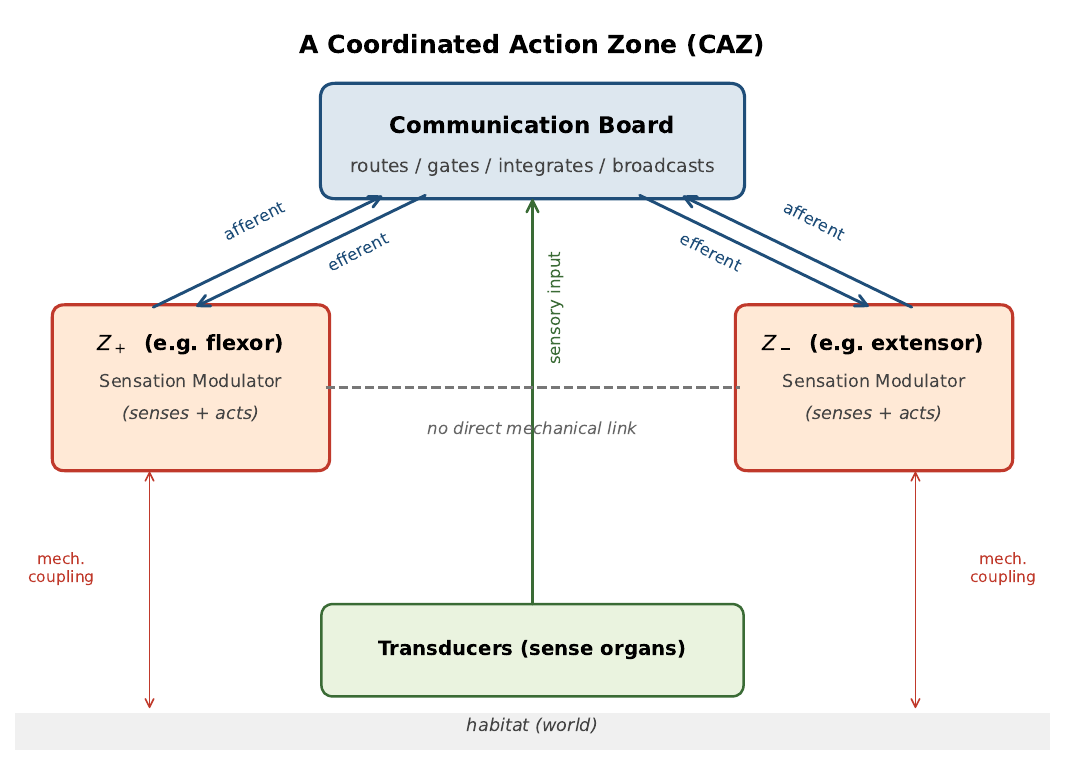}
\caption{One CAZ of the modular unit.  Two zones $Z_+$ and $Z_-$ (Sensation
Modulators) act on a shared joint through pull-only mechanical contact
with the habitat (red).  The board $B$ reads each zone's afferent and
routes it to the opponent as the reciprocal activation; there is no
direct mechanical link between the zones.  Read-only transducers
deliver world-side input to the board (\S\ref{sec:atom-transducer}),
and the haltability operator $H$ (\S\ref{sec:atom-halt}) recruits the
antagonistic affordance to hold the configuration actively still.}
\label{fig:atom-caz}
\end{figure}

\subsection{Joint dynamics and the elastic substrate}\label{sec:atom-joint}

At the mechanical interface, the two zones of a CAZ jointly move the
configuration variable $q$ they share.  For a one-degree-of-freedom
joint with moment arm $r$ and inertia $I$,
\begin{equation}
I\,\ddot q \;=\; r\,(F_+ - F_-)\;-\;b\,\dot q\;-\;k\,(q - q_0),
\label{eq:joint}
\end{equation}
where $F_\pm = \Phi(a_\pm,\ell_\pm,\dot\ell_\pm)$ are the pull-only
forces of the two zones ($\ell_+ = \ell_0 - rq$, $\ell_- = \ell_0 +
rq$), $b\,\dot q$ is damping, and $k\,(q-q_0)$ is the passive
\emph{elastic} restoring of the muscle--tendon substrate.  When $F_+ >
F_-$ the joint accelerates one way, when $F_- > F_+$ the other, and
when the opponent forces balance it holds --- the operating mode of a
physical balance.  What is non-standard is that $a_+$ and $a_-$ are not
independent commands from a controller: they emerge from the board's
routing (\S\ref{sec:atom-board}).

The elastic term $k\,(q-q_0)$ is not incidental; it is the physics
keystone of the whole construction.  Each Sensation Modulator is a
muscle--tendon unit --- an active contractile element in series with a
passive elastic one --- so the opponent substrate is compliant, and a
displacement imposed at one CAZ transmits to its neighbours \emph{with
attenuation} that grows with the number of intervening segments.
Because of this, and \emph{only} because of this, a body can later
recover its own connectivity graph from movement alone: the self-model
of \S\ref{sec:selfmodel} is available exactly because the substrate is
elastic.  A \emph{rigid} substrate moves as one common mode and leaves
no differential displacement to read --- so rigidity is a genuine
falsifier of the framework, not an uninteresting limit.  We defer the
read-out itself to \S\ref{sec:selfmodel}, where it is the first
construction the modular unit's physics makes possible; here we note only that
the modular unit has already paid for it, in the single parameter $k$.

\subsection{The communication board}\label{sec:atom-board}

The board does three things.  It \emph{reads} the afferent each zone
emits; it \emph{routes} that afferent to the partner as the reciprocal
activation; and it \emph{applies modulation} --- a halt gate, an
equilibrium-configuration bias from the broader network, and an
alert-energy bias from the haltability operator.  The simplest routing
is passive: each zone's afferent becomes part of the opponent's
activation, implementing the inverse coupling that makes the pair
behave like a balance.  Two inputs modulate this default.  When the
agent has a task --- a desired equilibrium configuration $\lambda$ ---
a \emph{drive} biases the active zone toward closing the gap.  When the
haltability operator engages, the \emph{halt gate} suspends the drive
and both zones settle to a baseline co-contraction.  Formally, with a
halt gate $g\in\{0,1\}$,
\begin{equation}
\text{halt } (g=0):\quad a_+ = a_- = a_0,
\label{eq:halt}
\end{equation}
\begin{equation}
\text{active } (g=1):\quad
\text{drive} = K_p(\lambda - q) - K_d\,\dot q,
\label{eq:drive}
\end{equation}
\begin{equation}
\begin{aligned}
&\text{drive}\geq 0:
&\;a_+ &= \mathrm{clip}(\text{drive},0,1),
&\;a_- &= \mathrm{clip}(a_0 + \beta E_R,0,1),\\
&\text{drive}<0:
&\;a_- &= \mathrm{clip}\bigl(-\text{drive}\,(1+\gamma E_R),0,1\bigr),
&\;a_+ &= \mathrm{clip}(a_0 + \beta E_R,0,1),
\end{aligned}
\label{eq:board}
\end{equation}
where $a_0>0$ is the baseline co-contraction (the tonic activity of
the inactive zone), and $(\beta,\gamma)$ are the coupling coefficients
of the haltability operator (\S\ref{sec:atom-halt}).  The second branch
encodes an architectural asymmetry: when the active zone changes, the
previously alert zone's drive is amplified by $(1+\gamma E_R)$, giving
it a head start.  This makes precise what ``active equilibrium'' means
--- the inactive zone is not silent but tonically engaged, scaling with
load and with recruited alert energy, an engagement that is
metabolically expensive and produces measurable signatures a passive
antagonist would not.

\subsection{Haltability and alert energy}\label{sec:atom-halt}

Haltability is the architectural ability to actively hold a
configuration still --- distinct from passive relaxation and from
classical inhibition.  Holding a task-relevant configuration while leaving
other degrees of freedom free is the minimal-intervention principle of optimal
feedback control \citep{TodorovJordan2002OFC}, realized here mechanically by
the opponent pair rather than by an optimizing controller.  It is a function of
the antagonism between $Z_+$
and $Z_-$, not something the board generates.  The precise
dynamical-systems word is \emph{affordance}: two pull-only zones whose
forces oppose along a shared variable \emph{admit} a state of co-active
equilibrium in which the forces cancel and the configuration is held
without motion.  That state is not produced by anything; it exists
because the structural coupling makes it exist, as the gravitational
coupling of a constrained mass makes a pendulum's attractor exist.
What the board does is \emph{recruit} the affordance: put both zones
into co-activation and maintain it against perturbation.

We capture recruitment, maintenance, and release with an operator $H$
whose state variable is the \emph{alert energy} $E_R$: a non-negative
scalar measuring the metabolic cost of the recruited co-activation.
It accumulates in proportion to the active zone's load while driving,
and decays passively otherwise:
\begin{equation}
\boxed{\;\frac{dE_R}{dt} = \rho\,F_{\text{active}}\,\mathbf{1}[g=1]
              - \frac{E_R}{\tau_E}\;}
\label{eq:alertenergy}
\end{equation}
with $F_{\text{active}} = \max(F_+, F_-)$, build rate $\rho$, decay
constant $\tau_E$, and $\mathbf{1}[g=1]$ the ``not halted'' indicator.
Under an average active load its steady state is the load-scaled active
equilibrium
\begin{equation}
E_R^* = \rho\,\tau_E\,\langle F_{\text{active}}\rangle .
\label{eq:Estar}
\end{equation}
Three things follow.  Alert energy is modelled as a \emph{real metabolic quantity}
--- bounded, dimensionful, passively decaying --- and the architecture
predicts it should be measurable as the antagonist's tonic activation
tracking agonist load, decaying over seconds after release rather than
vanishing at once
\citep{Feldman1986_EquilibriumPoint,Latash2008EquilibriumPointHistory}.
The ``active equilibrium'' of \S\ref{sec:atom-board} is recovered as
its loaded steady state.  And the distinction from \emph{classical
inhibition} is sharp: classical inhibition has no alert energy --- the
inactive zone is at baseline or silenced, with no recruited affordance
to maintain --- so the dynamics of $E_R$ are exactly what the SMN
claims tissue does and inhibition does not.  This single variable is
also, we will see, the substrate of attention (\S\ref{sec:worldmodel}):
the same energy that distinguishes haltability from inhibition is what
an unpredicted sensation recruits.

\subsection{The transducer}\label{sec:atom-transducer}

The CAZ is the \emph{action} half of the modular unit.  The other half is the
\emph{transducer}: the classical, read-only sense organ that converts a
physical perturbation of the environment into a signal.  Unlike the
Sensation Modulator, the transducer does not write back into the body's
mechanical state --- it participates in one domain (messaging) where
the SM participates in two.  Its reading depends jointly on the world
\emph{and} on the agent's pose: the same external event produces a
different reading at a different configuration.  Formally the
transducer is a read-only map $S:\Phi(t,x)\mapsto s(t)$ from the
physical field $\Phi$ (light, sound pressure, chemical concentration,
contact) to a signal; for a one-degree-of-freedom agent with
configuration $\theta$ and a stimulus at angular position $\varphi$,
the simplest non-trivial form is a Gaussian receptive field
\begin{equation}
s(t) = I\exp\!\left(-\frac{(\theta(t)-\varphi(t))^2}{\sigma^2}\right) + \eta(t),
\label{eq:rf}
\end{equation}
with intensity $I$, width $\sigma$, and sensory noise $\eta$; other
tuning forms (orientation columns, place fields, concentration curves)
are admissible.  The structural commitment is that $s$ depends jointly
on $\theta$ (modulable by action) and $\varphi$ (the world, only
conditionally observable).  This joint dependence is what makes any
change in $s$ ambiguous between a self-generated (reafferent) and a
world-generated (exafferent) cause --- the problem the world-model of
\S\ref{sec:worldmodel} is built to solve.

\subsection{What the block commits to}\label{sec:atom-commitments}

The modular unit is small enough that the architecture's commitments can be
read directly off it rather than stipulated in advance.  Seven carry the
substantive content; subsequent sections and the predicted registers
refer back to them.

\begin{description}[itemsep=4pt,labelindent=0pt,leftmargin=2em,
                    style=nextline]
\item[C1 (opponency at every scale)]
The body is organized by opponent dynamics at every scale
(\S\ref{sec:atom-opponency}); the primitive is an opponent pair whose
dynamics produce a balance state.  An empirical generalization, not a
parsimony choice.
\item[C2 (Sensation Modulators everywhere)]
The smallest element is a single tissue complex that both senses and
acts through one substrate (\S\ref{sec:atom-sm}), with a mechanical and
a messaging interface.  A structural commitment about elementary units.
\item[C3 (decentralized coordination through broadcast)]
Each CAZ runs its own balance locally, from its own afferents and the
broadcast it reads; coordination is a property of the network, not a
function computed by a centre (developed as bodies grow,
\S\ref{sec:kit}).  This is not a denial of the brain's role but a claim
about its \emph{kind}: the nervous system is the integrating beam of the
network, not a commander over it (\S\ref{sec:neuro}).
\item[C4 (active-perception filtering)]
Each board attends only to sensor streams currently coupled to its
modulation; streams not predicted as a consequence of current action
are dropped from the predictive computation (kept under low-cost
peripheral monitoring)
\citep{Bajcsy1988_ActivePerception,OReganNoe2001Sensorimotor}.  A
parsimony commitment.
\item[C5 (habituation: drop the predictable)]
Streams that \emph{are} coupled but whose variation is fully predicted
by an ongoing pattern drop out of attended status.  The complement of
C4.  In active-inference terms, C4--C5 are precision-weighting and
prediction-error suppression
\citep{Feldman2010AttentionUncertainty,Brown2013SensoryAttenuation}:
attended streams are those granted high precision, predictable ones
down-weighted.
\item[C6 (spatial cognition through differential displacement)]
Spatial structure is constructible only from the differential
displacement Sensation Modulators produce (\S\ref{sec:selfmodel}) ---
which is exactly why the elastic term of \eqref{eq:joint} is
load-bearing.  An empirical generalization: the SMN is fundamentally a
geometric architecture.
\item[C7 (the elastic substrate --- last but not least)]
The opponent substrate is \emph{compliant}: the passive restoring
$k(q-q_0)$ of \eqref{eq:joint} is finite and load-bearing, not a rigid
linkage.  It is the framework's cleanest falsifier --- in the rigid limit
$k\to\infty$ the self-model of \S\ref{sec:selfmodel} collapses to chance
--- but it does two things that make it the pivot of the whole cognitive
architecture, and, ironically, both are done by a plain mechanical property.
\emph{Within} a modular unit, compliance is the physical basis of its
\emph{autonomy}: an elastic opponent pair can hold its own equilibrium and
invert the perturbations it meets (the included-contraries reading of
\S\ref{sec:atom-invertibility}), whereas a rigid link merely transmits
force and holds nothing of its own.  \emph{Across} a network of such units,
compliance is the medium through which a displacement in one zone reaches
and is read by the others, so that many local balances integrate into a
single body-state --- the integrated effect we undergo as one agent rather
than a colony (\S\ref{sec:kit-network}).  Autonomy for the part,
integration for the whole, from one spring: an empirical generalization,
and the one the architecture can least afford to be wrong about.
\end{description}

\noindent
Three of these (C1, C6, C7) are empirical generalizations the architecture
takes on board and that sustained contrary evidence would falsify; two
(C2, C3) are structural commitments about the elementary units; two
(C4, C5) are parsimony moves made when the architecture is
instantiated.  The distinction matters for anyone trying to test the
account.

\subsection{The physics as the model's parameters}\label{sec:atom-physics}

It is worth collecting, in one place, the physics the modular unit has
introduced --- because for a physicist these are the model's control
parameters, and everything downstream is a study of how the
constructions vary with them.

\begin{description}[itemsep=3pt,labelindent=0pt,leftmargin=2em,
                    style=nextline]
\item[Gravity and mass ($g$, $m$, $I$).]
The joint carries inertia $I$ and its segments have mass; gravity and
the medium set the load the opponent pair holds against.  There is no
``rest'' that is not actively maintained --- the balance is against a
field.
\item[Friction and damping ($b$).]
The $b\,\dot q$ term, together with the medium's drag, sets the
regime.  The bench runs the modular unit \emph{overdamped}, which localizes
each drive to its endpoints and is what makes the self-model read-out
clean (\S\ref{sec:selfmodel}).
\item[Elasticity ($k$).]
The passive restoring $k\,(q-q_0)$ of the muscle--tendon substrate.
This is the load-bearing parameter: it is what lets displacement
attenuate with distance, and so what makes a self-model recoverable at
all.  $k\to\infty$ (rigidity) is the framework's cleanest falsifier.
\item[Opponency.]
The pull-only antagonism itself --- not a scalar but a structural
parameter.  Its \emph{type} depends on the physics of the antagonist,
which is where morphology re-enters (\S\ref{sec:kit}): a
\emph{skeletal} CAZ is a flexor/extensor pair across a rigid lever, and
its opponency runs under a \emph{rigid-length} (fixed-lever) constraint;
a \emph{hydrostatic} CAZ (worm, tongue, gut) is a pair of linked linear
actuators with no skeleton, whose antagonist is the structure itself ---
constant volume or turgor --- so its opponency runs under a
\emph{constant-volume} (incompressibility) constraint.\footnote{We write
\emph{constraint} rather than \emph{conservation law} to stay within what
this paper argues: force, in particular, is not a conserved quantity in
the Noether sense.  We hold the stronger view --- that these body-scale
constraints are conservation principles proper, each the shadow of a
symmetry, and that symmetry is itself grounded in the \emph{inverse
relations} that structure the physical world, life, and cognition alike
\citep{Nagarjuna1994_Inversion,GN2005_TowardsModelLifeCognition} --- but
that thesis is not established here; it is flagged as a direction, argued
elsewhere.}  Whether the rigid element sits inside (endoskeleton),
outside (exoskeleton), or is absent (hydroskeleton) is thus a physical
choice that types the constraint the CAZ obeys.
\item[Transducer physics ($I$, $\sigma$, $\eta$).]
The receptive-field intensity, width, and noise of \eqref{eq:rf} ---
and, more generally, the physics of each modality (optics, acoustics,
chemistry, contact) --- set what the read-only sensors can resolve, and
enter the world-model of \S\ref{sec:worldmodel}.
\end{description}

\noindent
Most of the framework's physics enters here, at the modular unit.  What the
following sections add is bodies and scales; the only new physics they
introduce is the constraint that a \emph{new} opponent regime
(hydrostatic vs.\ skeletal) brings with it --- which is itself a
composition of the parameters above, not a new primitive.

\subsection{What the modular unit already affords}\label{sec:atom-experiments}

Even before any body is grown around it, the modular unit is a complete
experimental object.  From this one block one can already run: the
opponent balance holding a configuration against gravity and the
medium; the load-scaled tonic engagement of the inactive zone
(\eqref{eq:Estar}), which is the SMN's sharpest departure from
classical inhibition; the recruited halt and its release; and, with a
single transducer attached, the ambiguity of a sensory change between a
self-generated and a world-generated cause.  These are not yet the
self-model or the object --- those need the elastic transmission of
\S\ref{sec:selfmodel} and the multimodal crossing of
\S\ref{sec:object} --- but they are enough to fix the modular unit's dynamics
empirically, and each carries its own order parameter and matched foil.
The remainder of the paper is, in one sense, the study of what happens
to these affordances as the modular unit is composed into bodies of different
shape, scale, and depth.

\begin{benchbox}[ --- the modular unit]
The unit's opponent dynamics run in the bench: the non-inertial crawler
moves only as a network effect (coupling sweep), the antagonist pair earns
its keep (co-contraction benefit), and the two energetic registers are
quantified against matched foils --- \textbf{R1} tonic-load coupling
(partner-tone slope $0.744$, sd\,$<\!0.001$, $\approx\beta\rho\tau_E$;
classical-inhibition foil flat) and \textbf{R2} resumption latency
($121$--$172$\,ms vs.\ a $311\pm1$\,ms classical foil).  \rtdlink{experiments/sweep_r1_tonic_load/}{R1},
\rtdlink{experiments/sweep_r2_resumption/}{R2}.
\end{benchbox}

\section{Generative morphology}\label{sec:kit}

The modular unit of \S\ref{sec:atom} is one object, but the primitives it is
made of --- a segment, a Coordinated Action Zone, a transducer --- are
a \emph{kit}.  The claim of this section is that the whole variety of
animal bodies is assembled from this one kit by a few compositional
moves, and --- this is the part that matters --- that the constructions
of the following sections (self-model, world-model, object) are carried
out by the \emph{same mechanism} whatever body the kit assembles.  The
model is generative in two senses, worth separating at the outset: it
generates \emph{bodies} (a morphological kit), and one mechanism
generates \emph{cognition} across those bodies (an invariance).  We
take them in turn and close on the invariance, which the next section
demonstrates.

\subsection{One kit, four moves}\label{sec:kit-moves}

A body, in this framework, is a graph of segments, and every animal
body is reachable from the modular unit by four moves on that graph.
\emph{Scale} adds segments in series (a longer chain).  \emph{Branch}
attaches a sub-chain to a segment --- an appendage: a leg, a wing, a
fin, an antenna.  \emph{Nest} puts small segments in series to make a
flexible part (a finger bends \emph{because} it is many small rigid
links jointed together).  \emph{Configure DOF} chooses each joint's
CAZ --- one opponent pair per degree of freedom, the orientation of
its bend axis selecting yaw, pitch, roll, or telescoping.  A worm, a
fish, a quadruped, a bird, and a biped are the same kit assembled
differently: scale a chain and you have a worm; give it a yaw wave and
an anisotropic medium and it swims; branch four appendages off a
segmented axis and configure their joints and it walks.  Nothing in the
kit changes from one animal to the next; only the four moves are
applied differently.  That a handful of primitives and moves spans the
menagerie is the first, ordinary sense in which the model is
generative.

\begin{figure}[!htbp]
\centering
\includegraphics[width=\linewidth]{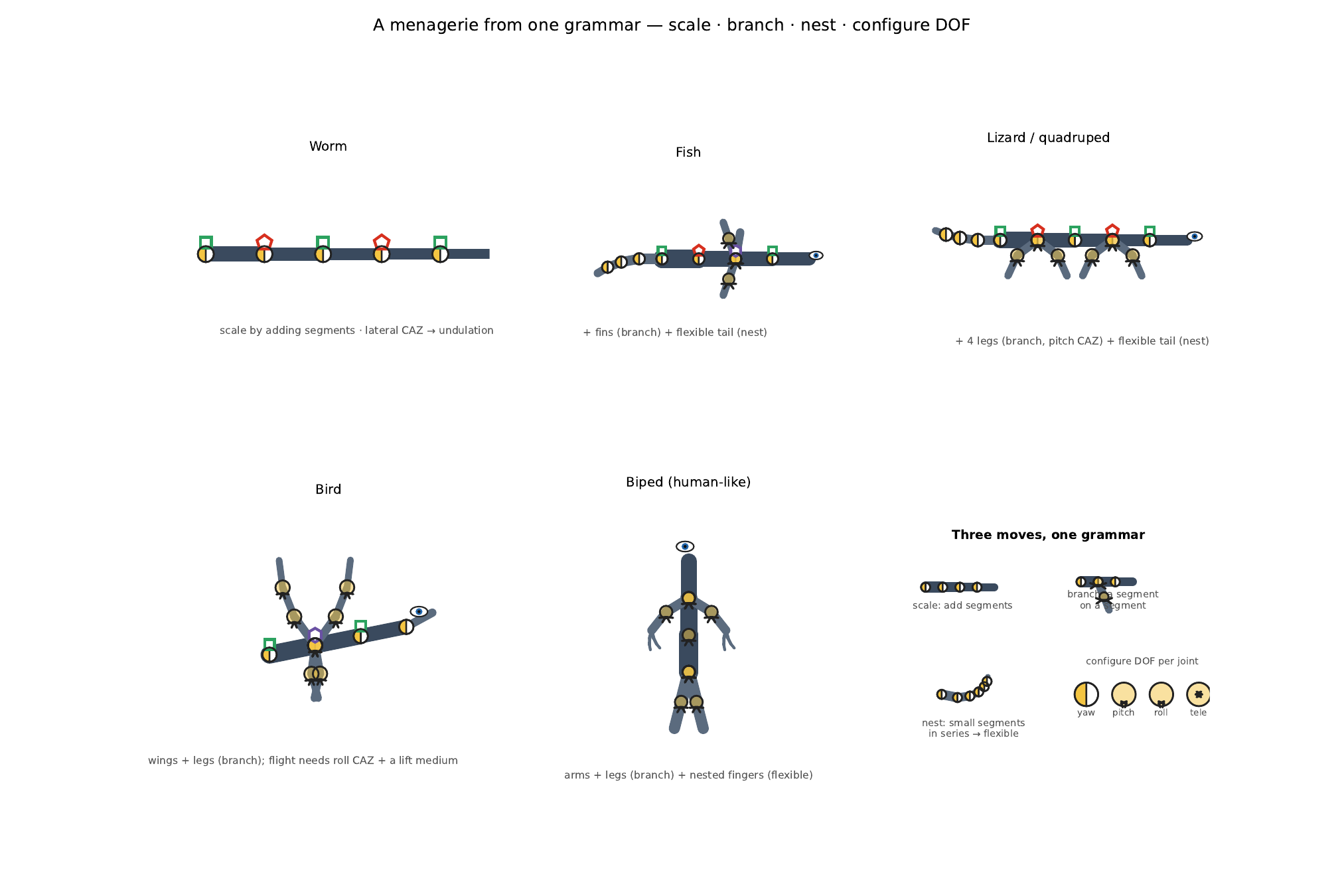}
\caption{A menagerie from one grammar.  A worm, a fish, a
lizard/quadruped, a bird, and a biped are the \emph{same kit} assembled
by four moves --- \emph{scale} (add segments), \emph{branch} (a chain on
a segment), \emph{nest} (small segments in series $\to$ a flexible part),
and \emph{configure DOF} (choose each joint's CAZ; inset, lower right).
Every body is drawn in the grammar of \fig{fig:atom-anatomy} ---
split-circle CAZ glyphs, shape-coded sensors, the anterior eye --- and
nothing in the kit changes from one animal to the next.  \emph{In the
bench:} each body is instantiated as a MuJoCo model from this one kit
(\rtdlink{diagram-grammar/}{diagram grammar}).}
\label{fig:menagerie}
\end{figure}

\subsection{Topologies and their constraints}\label{sec:kit-topology}

The moves generate a small set of recurring \emph{topologies}, and each
carries its own physics.  A \emph{chain} is the axial series; a
\emph{sheet} is a two-dimensional lattice of segments; a \emph{tube} is
a chain of linked rings; a \emph{nested} body is a lattice whose nodes
are themselves lattices.  Which topology a body has is not cosmetic: it
fixes how displacement propagates, and so what the self-model of
\S\ref{sec:selfmodel} recovers --- a path for a chain, a tree for a
branched body, a cylinder for a tube, and, coarse-grained, the body at
successive scales for a nested one.

Topology also selects the \emph{constraint} the opponency obeys,
picking up the thread from \S\ref{sec:atom-physics}.  A skeletal CAZ
--- a flexor/extensor pair across a rigid lever --- runs under a
rigid-length constraint, the lever fixing the geometric relation of the
two sides; its antagonist is the opposing muscle.  A hydrostatic CAZ ---
linked linear actuators with no skeleton, as in a worm, a tongue, or the
gut --- has no opposing lever; its antagonist is the \emph{structure
itself}, a constant volume or turgor, so it runs under a constant-volume
(incompressibility) constraint.  Whether the rigid element sits inside
the body (endoskeleton), outside it (exoskeleton), or is absent
(hydroskeleton) is therefore a physical choice that types the
constraint, and different regions of one animal may make different
choices --- a vertebrate's skeletal limbs alongside its hydrostatic
tongue and gut.  This is where physics re-enters the generative story:
composing a new topology can compose a new constraint, and that is the
only genuinely new physics the rest of the paper meets.

``Nesting'' has two senses worth keeping apart.
\emph{Flexibility-nesting} is the one just used: a flexible part is a
series of small rigid links, so flexibility is nested segmentation, not
a new primitive.  \emph{Scale-recursion nesting} is deeper: a lattice
whose nodes are themselves lattices, so that the same self-model
read-out recovers the body at every level --- segments within blocks
within super-blocks.  The first is a fact about how a bending part is
built; the second is a property of the read-out, and it is what makes
the invariance below hold across \emph{scales}, not only across shapes.

\subsection{The axes that make geometry possible}\label{sec:kit-axes}

A uniform substrate cannot construct geometry.  A perfectly symmetric
body --- a sphere of isotropic opponent pairs --- would have no
coordinate system, because no axis would be distinguished from any
other.  Coordinate systems require axes, and axes require asymmetry.
The body's asymmetries are therefore not incidental anatomical facts
but the substrate from which geometric construction becomes possible
--- the axes against which the differential displacement of
\S\ref{sec:selfmodel} becomes meaningful (Commitment~C6).  Five
asymmetries do this work, established progressively through
evolutionary--developmental history:

\begin{description}[itemsep=2pt,labelindent=0pt,leftmargin=2em,
                    style=nextline]
\item[Polarity (anterior--posterior)]
The most ancient asymmetry, already in cnidarians: the head--tail axis,
the preferred direction along which locomotion and behaviour are
oriented.
\item[Tubularity (oral--aboral; inside--outside)]
An interior tube from mouth to anus, creating the autonomic-core /
appendicular-periphery distinction and making interoception and
exteroception architecturally distinct.
\item[Segmentation (rostral--caudal repetition)]
The body laid down as a series of segments, each with its own opponent
organization coupled anti-phase to its neighbours
\citep{McGinnisKrumlauf1992Hox,Carroll2005EndlessForms}.  Segmentation
is what lets locomotion be a wave rather than a jump, and the body's
metric be additive along the axis.
\item[Bilaterality (left--right)]
Mirror symmetry across a sagittal plane, with left--right opponent
pairs at every segment.  This is what makes the nervous system a paired
organ, underwrites the bilateral CAZs of \S\ref{sec:atom-caz}, and
grounds direction-of-arrival for many modalities.
\item[Dorso-ventrality (dorsal--ventral)]
A top and a bottom, giving the body a gravity-aligned axis --- an
architectural ``up'' --- and making posture a coherent body-level
variable.
\end{description}

This also fixes the relationship between asymmetry and the opponency of
\S\ref{sec:atom-opponency}: opponency is meaningful only \emph{along an
axis} --- left--right opponency needs a sagittal plane, flexor--extensor
opponency a joint axis.  The asymmetries are what make opponency a
coherent organizing principle, by supplying the axes along which
opponent pairs are defined.  The body is not a uniform substrate
populated with opponent pairs, but a substrate with definite
asymmetries that establish the axes along which opponency, and then
geometry, is meaningful.

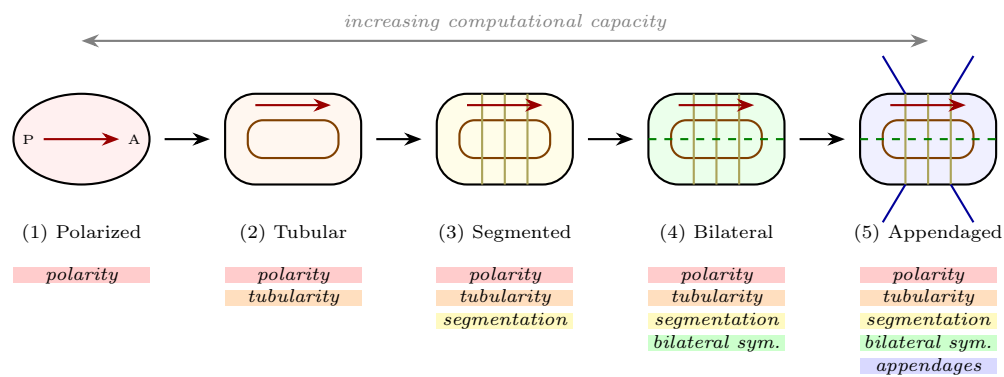
\begin{figure}[!htbp]
\centering
\begin{tikzpicture}[
  arr/.style={->, >=Stealth, thick},
  lbl/.style={font=\scriptsize, align=center},
  feat/.style={font=\scriptsize\itshape, text=black},
  >=Stealth
]
  \def\colsep{2.8}

  \begin{scope}[shift={(0,0)}]
    \draw[thick, fill=red!6] (0,0) ellipse (0.9 and 0.6);
    \draw[thick, red!60!black, ->] (-0.5,0) -- (0.5,0);
    \node[font=\tiny] at (-0.7,0) {P};
    \node[font=\tiny] at (0.7,0) {A};
    \node[lbl, below=4mm] at (0,-0.6) {(1) Polarized};
    \fill[red!20] (-0.9,-1.9) rectangle (0.9,-1.7);
    \node[feat] at (0,-1.8) {polarity};
  \end{scope}

  \draw[arr] (1.1,0) -- (1.7,0);

  \begin{scope}[shift={(\colsep,0)}]
    \draw[thick, fill=orange!6, rounded corners=4mm]
      (-0.9,-0.6) rectangle (0.9,0.6);
    \draw[thick, orange!50!black, rounded corners=2mm]
      (-0.6,-0.25) rectangle (0.6,0.25);
    \draw[thick, red!60!black, ->] (-0.5,0.45) -- (0.5,0.45);
    \node[lbl, below=4mm] at (0,-0.6) {(2) Tubular};
    \fill[red!20] (-0.9,-1.9) rectangle (0.9,-1.7);
    \fill[orange!25] (-0.9,-2.2) rectangle (0.9,-2.0);
    \node[feat] at (0,-1.8) {polarity};
    \node[feat] at (0,-2.1) {tubularity};
  \end{scope}

  \draw[arr] (\colsep+1.1,0) -- (\colsep+1.7,0);

  \begin{scope}[shift={(2*\colsep,0)}]
    \draw[thick, fill=yellow!10, rounded corners=4mm]
      (-0.9,-0.6) rectangle (0.9,0.6);
    \draw[thick, orange!50!black, rounded corners=2mm]
      (-0.6,-0.25) rectangle (0.6,0.25);
    \foreach \x in {-0.3, 0, 0.3} {
      \draw[thick, yellow!60!black] (\x,-0.6) -- (\x,0.6);
    }
    \draw[thick, red!60!black, ->] (-0.5,0.45) -- (0.5,0.45);
    \node[lbl, below=4mm] at (0,-0.6) {(3) Segmented};
    \fill[red!20] (-0.9,-1.9) rectangle (0.9,-1.7);
    \fill[orange!25] (-0.9,-2.2) rectangle (0.9,-2.0);
    \fill[yellow!30] (-0.9,-2.5) rectangle (0.9,-2.3);
    \node[feat] at (0,-1.8) {polarity};
    \node[feat] at (0,-2.1) {tubularity};
    \node[feat] at (0,-2.4) {segmentation};
  \end{scope}

  \draw[arr] (2*\colsep+1.1,0) -- (2*\colsep+1.7,0);

  \begin{scope}[shift={(3*\colsep,0)}]
    \draw[thick, fill=green!8, rounded corners=4mm]
      (-0.9,-0.6) rectangle (0.9,0.6);
    \draw[thick, orange!50!black, rounded corners=2mm]
      (-0.6,-0.25) rectangle (0.6,0.25);
    \foreach \x in {-0.3, 0, 0.3} {
      \draw[thick, yellow!60!black] (\x,-0.6) -- (\x,0.6);
    }
    \draw[thick, dashed, green!50!black] (-0.9,0) -- (0.9,0);
    \draw[thick, red!60!black, ->] (-0.5,0.45) -- (0.5,0.45);
    \node[lbl, below=4mm] at (0,-0.6) {(4) Bilateral};
    \fill[red!20] (-0.9,-1.9) rectangle (0.9,-1.7);
    \fill[orange!25] (-0.9,-2.2) rectangle (0.9,-2.0);
    \fill[yellow!30] (-0.9,-2.5) rectangle (0.9,-2.3);
    \fill[green!20] (-0.9,-2.8) rectangle (0.9,-2.6);
    \node[feat] at (0,-1.8) {polarity};
    \node[feat] at (0,-2.1) {tubularity};
    \node[feat] at (0,-2.4) {segmentation};
    \node[feat] at (0,-2.7) {bilateral sym.};
  \end{scope}

  \draw[arr] (3*\colsep+1.1,0) -- (3*\colsep+1.7,0);

  \begin{scope}[shift={(4*\colsep,0)}]
    \draw[thick, fill=blue!6, rounded corners=4mm]
      (-0.9,-0.6) rectangle (0.9,0.6);
    \draw[thick, orange!50!black, rounded corners=2mm]
      (-0.6,-0.25) rectangle (0.6,0.25);
    \foreach \x in {-0.3, 0, 0.3} {
      \draw[thick, yellow!60!black] (\x,-0.6) -- (\x,0.6);
    }
    \draw[thick, dashed, green!50!black] (-0.9,0) -- (0.9,0);
    \draw[thick, red!60!black, ->] (-0.5,0.45) -- (0.5,0.45);
    \draw[thick, blue!60!black] (-0.3,0.6) -- (-0.6,1.1);
    \draw[thick, blue!60!black] (-0.3,-0.6) -- (-0.6,-1.1);
    \draw[thick, blue!60!black] (0.3,0.6) -- (0.6,1.1);
    \draw[thick, blue!60!black] (0.3,-0.6) -- (0.6,-1.1);
    \node[lbl, below=4mm] at (0,-0.6) {(5) Appendaged};
    \fill[red!20] (-0.9,-1.9) rectangle (0.9,-1.7);
    \fill[orange!25] (-0.9,-2.2) rectangle (0.9,-2.0);
    \fill[yellow!30] (-0.9,-2.5) rectangle (0.9,-2.3);
    \fill[green!20] (-0.9,-2.8) rectangle (0.9,-2.6);
    \fill[blue!15] (-0.9,-3.1) rectangle (0.9,-2.9);
    \node[feat] at (0,-1.8) {polarity};
    \node[feat] at (0,-2.1) {tubularity};
    \node[feat] at (0,-2.4) {segmentation};
    \node[feat] at (0,-2.7) {bilateral sym.};
    \node[feat] at (0,-3.0) {appendages};
  \end{scope}

  \draw[<->, thick, gray] (0,1.3) -- (4*\colsep,1.3)
    node[midway, above, font=\scriptsize\itshape, text=gray]
    {increasing computational capacity};

\end{tikzpicture}
\caption{The morphogenetic progression that scaffolds the SMN.  Each
feature, once gained, persists through all subsequent stages (the
cumulative bars beneath each diagram).  (1)~Polarity establishes a
directional axis; (2)~tubularity creates a through-gut with haltable
peristaltic flow; (3)~segmentation distributes CAZs along the axis;
(4)~bilateral symmetry creates matched left--right zones;
(5)~paired appendages add independently controllable effectors.  Each
step adds computational capacity while retaining all prior features
--- the four moves of \S\ref{sec:kit-moves}, read developmentally.}
\label{fig:morphogenetic-progression}
\end{figure}

\subsection{From modular unit to network}\label{sec:kit-network}

Assembled into a whole body, the kit is a \emph{network} --- and it repays being
exact about which nodes and which edges, because the body is a network in more
than one sense at once.  At the level of the tissue it is a single \emph{signal
graph}: the SMN is the graph of all CAZs, with three kinds of node ---
\emph{actuators} (motor tissue), \emph{sensors} (the single-interface transducers
that are the body's principal source of data), and \emph{integrators} (neurons and
their boards) --- joined by one kind of edge, the information-bearing coupling.
Only actuator nodes have mechanical interfaces, so only they do geometric work on
the body itself; sensors and integrators do their work on signals already in the
graph.

Read by \emph{state} --- by what carries its own dynamics --- the signal graph
presents \emph{two networks}.  The \emph{messaging network} is the graph in its
integrative aspect: its nodes are the CAZ boards and its edges are the couplings
--- the afferents that report to a board, and the broadcast that ties every board
to one shared canvas (the operator $\Pi$ below) --- and its state is that canvas.
The \emph{mechanical network} is not a sub-graph but one the signal graph
\emph{induces} through its actuators: its nodes are the body's segments, and its
edges are the joints, each realized by an opponent pair of actuators (one CAZ, one
degree of freedom); its state is the body's configuration.  The two live on
different vertex sets, related by an \emph{incidence} rather than a duality: an
actuator is a \emph{node} of the signal graph and, because it attaches two
segments, contributes to a joint that is an \emph{edge} of the mechanical network.
The \emph{CAZ} is the hinge --- an opponent pair of actuator nodes in the signal
graph, one joint-edge in the mechanical network, and a board-node in the messaging
network --- which is what lets one CAZ move the body, sense, and broadcast.

Whether the count is \emph{two} networks or \emph{three} is then only a question
of where the sensors go.  Fold the sensor nodes into the messaging network as its
afferent sources and there are two networks; keep them separate and there are
\emph{three heterogeneous graphs}, on distinct vertex sets, coupled through the CAZ
--- a \emph{kinematic} graph (segments, joined by actuators), a \emph{sensory}
graph (transducers, with afferent edges into the boards), and a \emph{messaging}
graph (boards, joined by broadcast).  We take the two-network reading as primary, because each of the
two carries its own state, whereas the sensory graph carries none of its own --- it
is a source, not a dynamics.  Two features of this network matter for what follows.

First, it is \emph{decentralized} (Commitment~C3).  Each CAZ runs its
own balance locally, from its own afferents and the broadcast it reads;
coordination is a property of the network, not a function some centre
computes.  This is not a denial of the central nervous system but a
different account of its job.  The CNS is a \emph{broadcaster} --- the
messaging substrate through which CAZs share state, without which the
body is a colony of competent zones and with which it is a single agent
--- and a \emph{state-space estimate}, an integrated read of where the
body is, available to every board as context; but it is not the origin
of command.  Contractile and signalling capacities precede neurons in
evolution and persist without them: cardiac tissue beats in a dish, and
bioelectric signalling coordinates cell collectives without neurons
\citep{Marder2011VariabilityCompensationModulation,Levin2014MolecularBioelectricity,LevinPezzulo2017EndogenousBioelectricNetworks}.
The opponent pair already \emph{affords} equilibrium and does not need
to be told to hold it.

\begin{figure}[!htbp]
\centering
\includegraphics[width=\linewidth]{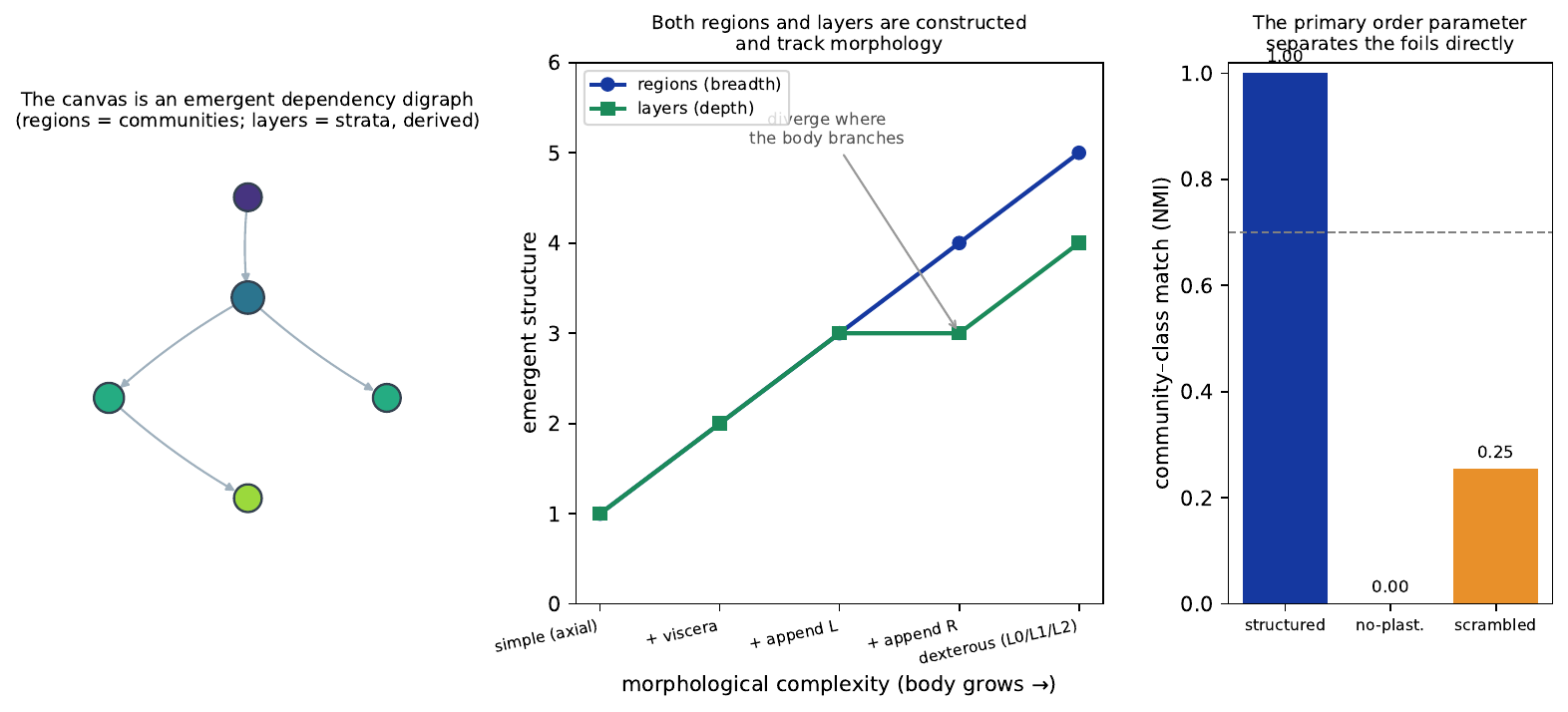}
\caption{The canvas constructs its own structure, in the companion bench.
\emph{Left:} with a plastic broadcast, modules broadcasting to one undivided
canvas build an \emph{emergent dependency digraph} --- its \emph{regions} are the
communities of mutually coupled zones (breadth) and its \emph{layers} are the
dependency strata (depth), derived from the graph rather than stipulated.
\emph{Centre:} as the body grows, in this bench the numbers of regions and of
layers both rise with it and \emph{diverge} where the body branches --- paired
appendages join at one stratum, adding breadth without depth.  \emph{Right:} the
discriminating order parameter is the match (normalized mutual information) between
the emergent communities and the modules' functional classes, which are defined by
their coupling structure: a structured broadcast recovers those classes (NMI
$1.00$), far above a frozen canvas ($0.00$) or a class-less one ($0.25$).  The
residual $0.25$ shows that self-organization alone leaves \emph{some} structure ---
but not the class structure the framework predicts, which only the structured
broadcast attains.
\emph{In the bench:} \rtdlink{experiments/canvas_regions/}{canvas-regions}.}
\label{fig:canvas-emergent}
\end{figure}

This broadcaster has a precise specification.  Writing the network state as
the tuple of CAZ states $\big(x_j(t)\big)_{j\in\mathcal{Z}}$, the
\emph{broadcast operator} $\Pi$ maps it to a body-wide broadcast
$b(t)=\Pi\big((x_j(t))_{j\in\mathcal{Z}}\big)$ that enters every CAZ's board
alongside that zone's own afferents.  The architectural commitment is a
condition on $\Pi$, not a choice of its functional form: $\Pi$ is
\emph{non-degenerate}, meaning that for every CAZ $j$, that zone both
\emph{contributes} to the broadcast ($\partial b/\partial x_j\not\equiv 0$)
and \emph{reads} it ($b$ enters its board).  We call this condition
\emph{network closure}; it is graph- and operator-theoretic and can be
checked on any candidate implementation.  It is the exact content of the
colony-versus-single-agent contrast just drawn: a zone that neither
contributes to nor reads the broadcast is not part of the SMN.  The
\emph{particular} form of $\Pi$ --- linear with delays, attention-weighted
integration, multi-frequency broadcast layers, neural-tissue routing --- is
deliberately left open, because the constructions that follow depend on
network closure, not on any one realization; we therefore specify $\Pi$ by
the condition it must satisfy rather than by a single equation.  One
admissible realization is biologically the most telling: a \emph{self-organizing}
closure, whose broadcast weights are plastic, so that continued broadcasting
reshapes them.  Under such a $\Pi$ the messaging graph does not merely relay ---
it \emph{constructs} its own structure on the broadcast substrate, the way
activity-dependent cortical maps form \citep{Kohonen1982SOM}; the layered
organization below then follows not as a stipulation but as a construction, and
the companion bench shows exactly this (Fig.~\ref{fig:canvas-emergent}), where the
emergent communities recover functional classes defined by coupling structure, not
imposed as spatial labels.  Its
companion, the \emph{active-perception filter} $\alpha$ (Commitment~C4,
\S\ref{sec:atom-commitments}), is a per-stream gate $\alpha_i\in[0,1]$ on the
afferents reaching a board: it admits a stream only if its variation is predicted
as a non-trivial consequence of that board's own efference, while unadmitted
streams ($\alpha_i\!\to\!0$) stay under low-cost peripheral monitoring so that
novel coupling can still break through.

These two operators are the messaging network's two functional \emph{sub-networks}
--- the differentiating and integrating roles it divides into, not a second pair of
networks beside the mechanical and messaging ones.  The board's \emph{differentiating}
operation --- contrasting each afferent against the body's own active state, and
gating it by $\alpha$ --- is the \emph{differentiating and filtering network} (DFN):
it processes sensory--motor contingencies, admitting only the streams a board's own
action modulates.  (The modulation map $F$ of \eqref{eq:F-modulation} is the
body-side dependency this contrast is computed \emph{against}, not the operation
itself.)  Its complement is the broadcast operator $\Pi$, the \emph{integrating
network} (IN) --- the canvas on which the DFN's filtered output is integrated into
one held state.  DFN and IN are distinguished by function --- though they map onto a
characteristic spatial signature, the distributed modulating weights of the DFN and
the localized integrating beam of the IN (\S\ref{sec:neuro}).

Network closure is not an untested posit.  The minimal non-trivial network
--- two coupled CAZs --- is exactly the setting of the dual-signal property
(\S\ref{sec:world-dualsignal}): there, one zone's reafference residual is
reconstructed from the \emph{other} zone's state precisely \emph{because} the
partner is on the broadcast (self-contact, $R^2\!\to\!1$), while the matched
foil removes it (decoupled, $R^2\!\to\!0$).  That contrast is a direct,
falsifiable test of $\Pi$'s non-degeneracy on the two-CAZ network.  The
behaviour of \emph{large} networks --- where the routing graph and the
specific form of $\Pi$ begin to matter --- is the analysis this Phase-I paper
defers; the operator and its defining condition are not deferred.

\begin{figure}[!htbp]
\centering
\includegraphics[width=0.86\linewidth]{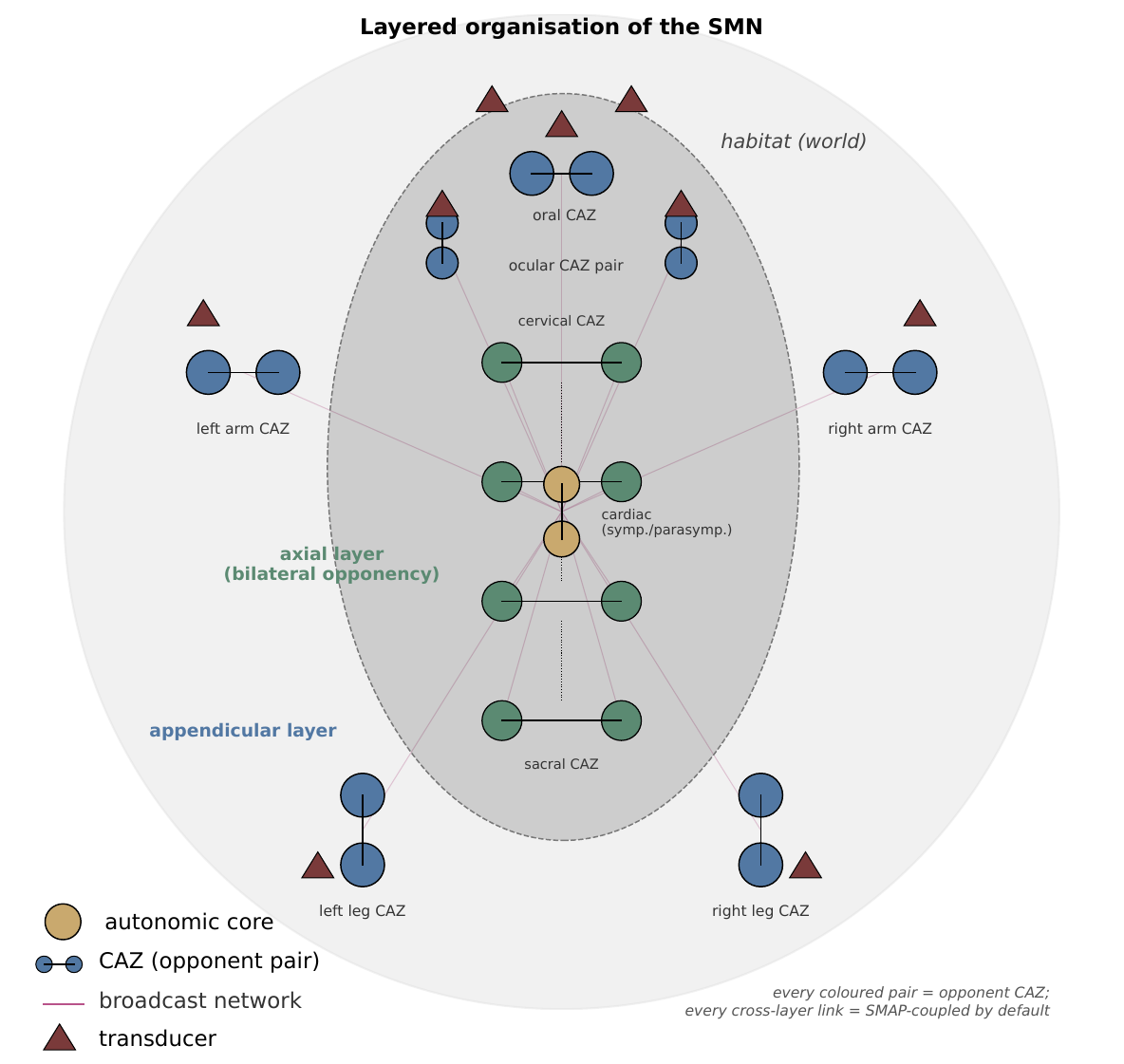}
\caption{Layered organization of the SMN.  The autonomic layer
(visceral CAZs) sits at the metabolic core; the axial layer (postural
and locomotor CAZs along the trunk) is bilaterally opponent and coupled
to the autonomic layer through interoceptive broadcast; the
appendicular layer (limb, oral, ocular CAZs) operates with
\emph{selectable} coupling to the axial layer, allowing dexterous
decoupling for manipulation.  Transducers deliver world-side input
through the appendicular and oral layers.  Every connection within the
SMN is opponent (mechanical or messaging); the only sensory streams
lacking a second efferent counterpart within the SMN are those arriving
from the habitat through the transducer interface.}
\label{fig:smn-layers}
\end{figure}

Second, the network is \emph{layered} (Fig.~\ref{fig:smn-layers}): an
autonomic core (visceral CAZs running cardiac, respiratory, and
digestive cycles), an axial layer (bilaterally opponent postural and
locomotor CAZs), and an appendicular layer (limb, oral, ocular CAZs)
whose coupling to the axis is \emph{selectable}.  That selectable
decoupling is what frees an appendage for manipulation, and it is the
hinge on which the action-pattern taxonomy of \S\ref{sec:taxonomy}
later turns; here we note only that it is a structural feature of the
layered body, present as soon as appendages are branched off a
segmented axis.  This layering is \emph{anatomical} --- given by the body plan
--- and is not to be conflated with the \emph{functional} structure of the
broadcast substrate, which is not given but \emph{constructed}: self-organized
by the broadcasting itself (the self-organizing $\Pi$ above) so as to reflect the
body's functional relations.  The canvas builds \emph{both} its regions (how many
functional territories) and its layered ordering (their dependency depth), and the
two need not coincide.  The body's anatomical layers are inherited; the canvas's
regions \emph{and} layers are built --- the framework's constructive commitment in
miniature (Fig.~\ref{fig:canvas-emergent}).

\subsection{The invariance, and what the next section must
show}\label{sec:kit-invariance}

We can now state the claim the rest of the paper rests on.  The kit
generates bodies; that is the first, ordinary sense of ``generative.''
The second is stronger, and it is what makes the framework a
\emph{model} rather than a menagerie: the constructions of the
following sections are carried out by \emph{one} mechanism, and that
mechanism is \emph{invariant} under the morphological moves.  The same
self-model read-out recovers a path for a chain, a tree for a branched
body, a cylinder for a tube, and the body at every level of a nested
one; the same reafference cut separates self from world whatever the
shape.  In the language a physicist will find natural: the order
parameter is invariant under the transformations that carry one
morphology into another.  The claim has a sharp failure mode --- a
rigid body, moving as one common mode, recovers nothing --- and it is
demonstrated, morphology by morphology and scale by scale, in
\S\ref{sec:selfmodel}, to which we now turn.

\begin{figure}[!htbp]
\centering
\includegraphics[width=0.6\linewidth]{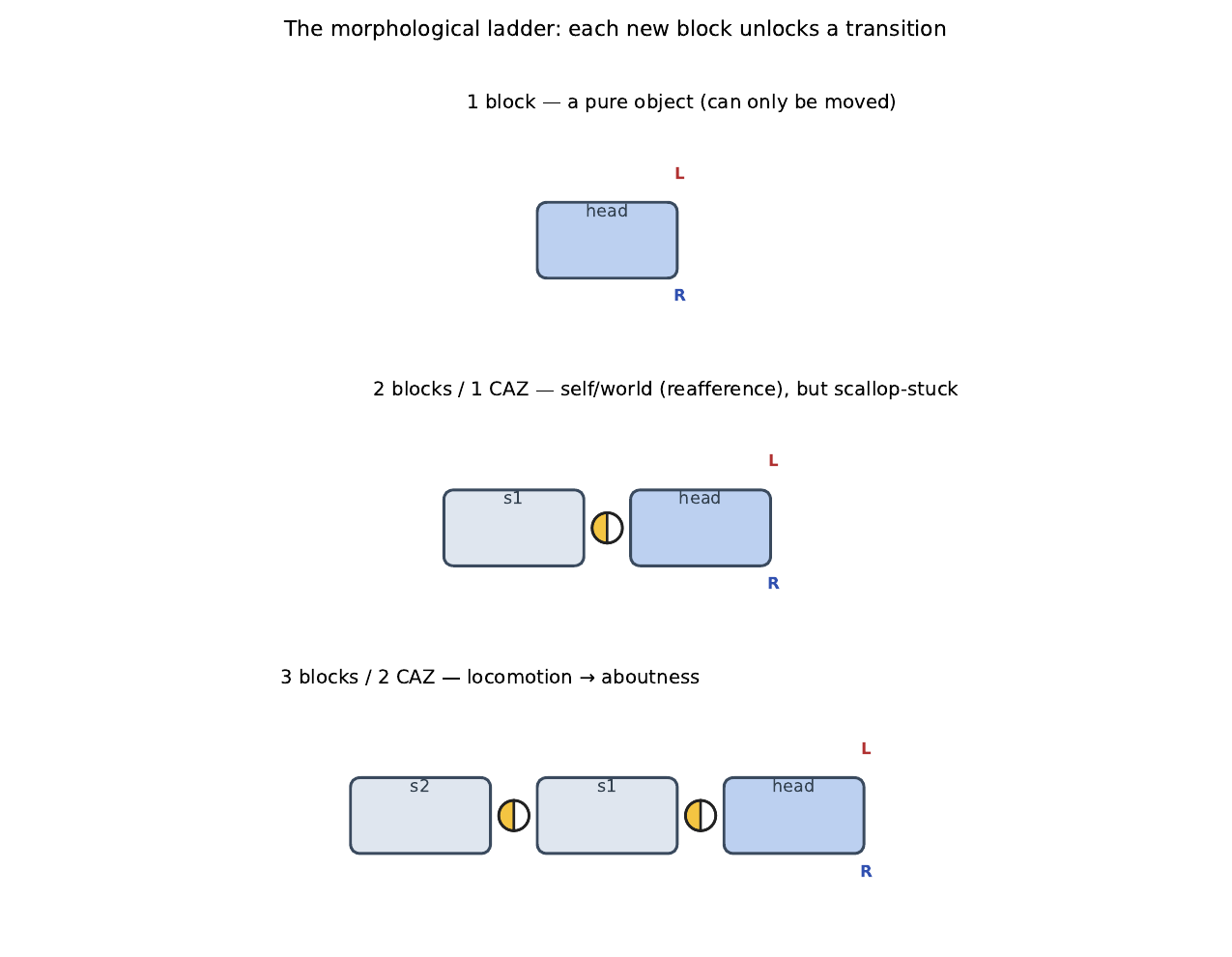}
\caption{A roadmap of what follows, read off the morphological ladder:
each added block unlocks the next construction.  \emph{One block} is a
pure object --- it can only be moved.  \emph{Two blocks, one CAZ} give
the differential displacement a self/world distinction needs (though a
single joint is ``scallop-stuck'').  \emph{Three blocks, two CAZs} give
locomotion and, with it, the recurrent haltable pattern that grounds
aboutness.  The following sections make each step precise; the ladder is
the schematic, not the argument.}
\label{fig:morph-ladder}
\end{figure}

\begin{benchbox}[ --- generative morphology]
The four moves are a runnable kit: the worm, fish, quadruped, and branched
bodies of Fig.~\ref{fig:menagerie} are instantiated as MuJoCo models from
one grammar, and the diagram grammar itself is documented as the bench's
construction language.  \rtdlink{diagram-grammar/}{Diagram grammar},
\rtdlink{construction-of-experience/}{construction of experience}.
\end{benchbox}

\section{Constructing the self-model}\label{sec:selfmodel}

Stage 2 promised an invariant: whatever body the kit assembles, the
same mechanism constructs the same things.  The first of those things,
and the one on which the rest depend, is a \emph{self-model} --- the
body's own connectivity graph, recovered by the body from its own
movement.  We deferred this from \S\ref{sec:atom-joint} because it is
the first construction the modular unit's elasticity makes possible; here we
cash it out, and use it to demonstrate the invariance.  Recovered from movement
rather than stored, it is a mechanistic rendering of the \emph{body schema} of the
phenomenological tradition --- the body known from within, the lived body of
\citet{Merleau-Ponty2013-vs} and the body schema of \citet{gallagher2005how} ---
here specified as an equilibrium of the pairwise relations among coordinated zones.

\subsection{The read-out}\label{sec:selfmodel-readout}

When a zone $i$ commands an efference $\tau_i$, the motion it produces
travels through the compliant substrate and is heard, attenuated, by
the other zones.  Let $v_j$ be the motion zone $j$ hears broadcast.
With the whole-body common mode removed, the transmission gain
\begin{equation}
G_{ij} \;=\; \Bigl|\;\tfrac{1}{T}\textstyle\sum_{t}
  \hat z(\tau_i)_t\,\hat z(v_j)_t\;\Bigr|,
\qquad
\hat z(x) = \frac{x-\bar x}{\sigma_x+\varepsilon},
\label{eq:selfmodel}
\end{equation}
is the normalized cross-correlation of a zone's own efference with each
other zone's afferent motion.  It falls off monotonically with
hop-distance --- because the substrate is elastic --- so each zone's
strongest co-mover is a true mechanical neighbour, and the union of
these purely local reads is the body's connectivity graph: the
self-model.  Methodologically this is functional-connectivity inference, a
normalized-cross-correlation cousin of Granger-style coupling estimation and
sparse network reconstruction \citep{Granger1969CausalRelations}.

Two things about \eqref{eq:selfmodel} make it an \emph{architectural}
self-model rather than an observer's reconstruction.  First, every term
is available \emph{locally}: a zone correlates its own efference against
the motions it hears on the broadcast, and nothing in the computation
requires any zone to read the whole body.  There is no central reader
assembling the graph; the graph is the union of local reads, which is
exactly what Commitment~C3 demands.  Second, the read-out \emph{is} the
mechanism, not a metaphor for one: a joint --- a CAZ --- finds the two
segments it hinges because its own angular velocity \emph{is} their
motion difference, so the correlation is high for its neighbours by
construction.

\subsection{Why only if elastic: the order parameter and its
foil}\label{sec:selfmodel-foil}

The elastic term $k(q-q_0)$ of \eqref{eq:joint} does the load-bearing
work.  A compliant body transmits a local displacement to its
neighbours \emph{with attenuation}, and it is the attenuation gradient
that \eqref{eq:selfmodel} reads as distance.  A \emph{rigid} body has no
gradient: it moves as one common mode, every zone with every other, so
the read-out cannot tell a neighbour from a distant segment.  Rigidity
is therefore not an uninteresting limit but a genuine \emph{falsifier}
--- the framework predicts that a body stiff enough to move as one piece
cannot recover its own structure, however much it moves.

This gives the section its order parameter and its matched foil.  The
\emph{order parameter} is recovery quality --- how well the graph read
off \eqref{eq:selfmodel} matches the body's true adjacency --- scored two
ways: \emph{neighbour recovery} (does each zone's strongest co-mover sit
next to it?) and \emph{order recovery} (is the full sequence along an
axis recovered?).  The \emph{foil} is the same body made rigid, or its
motion time-shuffled (which destroys the correlation structure while
preserving the marginals).  It is worth being explicit that the number
the agent computes and the number we score it by are different
quantities: the coupling \eqref{eq:selfmodel} is the agent's own local
read-out, while recovery quality is the experimenter's score against a
ground truth the agent never has.  Keeping the two apart is what makes
the test honest --- the agent is never handed the answer it is supposed
to construct.

\subsection{Topology-invariance}\label{sec:selfmodel-topology}

The invariance claim of \S\ref{sec:kit-invariance} is that
\eqref{eq:selfmodel} --- one read-out, unchanged --- recovers whatever
body the kit assembles.  On the companion bench this holds across the
topologies of \S\ref{sec:kit-topology}:

\begin{itemize}[itemsep=2pt]
\item a \emph{chain} recovers as a path: neighbour recovery $1.00$,
order recovery $0.89$, against a rigid foil at $0.40$ and a time-frozen
foil at $0.14$;
\item a \emph{branched} body recovers as a tree: every endpoint and
attachment found ($9/9$ endpoints, $8/8$ neighbours), with the branch
point correctly identified as the high-degree node;
\item a \emph{sheet} and a \emph{tube} recover as a grid and a cylinder,
at $0.99$ and $0.97$.
\end{itemize}

\noindent
The same coupling function produces all of these; nothing in it is
specialized to chain, tree, or cylinder.  The read-out does not know
what shape it is on, and does not need to --- which is the content of
morphology-invariance, now demonstrated rather than promised.

\begin{figure}[!htbp]
\centering
\includegraphics[width=\linewidth]{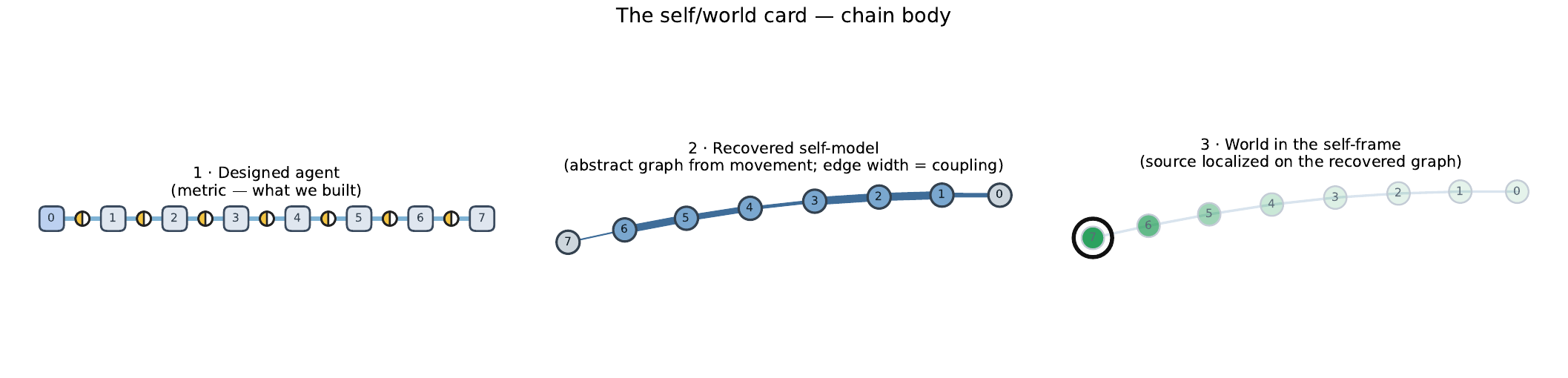}
\caption{The self/world card for a chain body.  \emph{(1) Designed
agent} --- what we built: eight segment blocks with a CAZ glyph per
joint.  \emph{(2) Recovered self-model} --- the abstract graph the agent
reconstructs from its \emph{own} movement, laid out from its own
recovered adjacency (never the body's coordinates), with edge width set
by the measured coupling \eqref{eq:selfmodel} and the correct path order
recovered.  \emph{(3) World in the self-frame} --- a field source painted
onto the recovered graph, landing on a specific node: the world in the
body's own frame, previewing \S\ref{sec:worldmodel}.  All three
renderers take only positions and edges, so the card works for any
morphology.  \emph{In the bench:} \rtdlink{experiments/self_model_topology/}{self\_model\_topology}.}
\label{fig:self-world-card}
\end{figure}

\subsection{Scale-invariance, and the canvas}\label{sec:selfmodel-scale}

Scale-recursion nesting (\S\ref{sec:kit-topology}) poses a sharper test.
A \emph{nested} body is a lattice whose nodes are themselves lattices,
and the question is whether the \emph{same} read-out, coarse-grained,
recovers the body at every level.  It does: on a three-level nested
lattice --- 36 fine segments grouped into 9 blocks grouped into 3
super-blocks --- coarse-graining \eqref{eq:selfmodel} recovers the graph
at each level: fine $0.88$, mid $1.00$, coarse $1.00$.  Coarse-graining a
local read-out and finding the structure again at the next scale is, in
the physicist's sense, a \emph{renormalization} step; that the
self-model survives it is what lets one mechanism serve a body organized
at many scales at once.

The fine level recovers least well, and this is not a defect.  Deeper
layers are perturbed less than outer ones --- a super-block moves as a
whole while its fine segments jitter within it --- so their contribution
to the self-model is faint.  We read this as a feature: the deep body is
a \emph{canvas} on which the outer layers write, and a good self-model is
not obliged to resolve every fine segment beneath.  What is robust is the
structure of the \emph{active} scales (mid and coarse stay at $1.00$
across the perturbation regimes); the faintness of the deep read-out is
the architecture reporting, correctly, that the canvas is holding still.

\subsection{What the self-model earns}\label{sec:selfmodel-earns}

The self-model is the first thing the body constructs, and everything
after is built in its frame.  It is a graph the body recovers about
\emph{itself}, from \emph{its own} movement, with \emph{no} central
reader --- a body-anchored coordinate system, not a picture in a
register.  The world-model of \S\ref{sec:worldmodel} is not a second,
independent construction: it is world-geometry expressed \emph{in this
frame}, built by the same zones now reading the world-facing
transducers.  Having recovered where its own parts are, the body is in a
position to ask where things in the world are relative to them --- and to
separate, for the first time, what its own action did to a sensor from
what the world did.  That separation is the first epistemic transition,
and the subject of the next section.

\begin{benchbox}[ --- the self-model]
\texttt{self\_model.py} reconstructs the body's hop-graph from movement
alone.  \emph{Preregistered} order parameter: graph recovery against an
elasticity-ablated foil.  Result: the true topology is recovered on
elastic bodies and collapses to chance on the rigid foil, and the same
read-out succeeds across chain, branched, sheet, and nested bodies
(topology- and scale-invariance).  \rtdlink{experiments/self_model_topology/}{Read it}.
\end{benchbox}

\section{The world-model, and the self/world/other distinction}\label{sec:worldmodel}

The self-model of \S\ref{sec:selfmodel} gives the body a frame --- where
its own parts are.  A \emph{world}-model is what that frame is for:
world-geometry expressed in the body's own coordinates, and the first
distinction the body draws in it --- between what its own action did to a
sensor and what the world did.  Drawing that distinction is the
\emph{first epistemic transition}, and this section constructs it
mechanically, as a property of the wiring rather than a judgement the
agent makes.

\subsection{The reafference problem}\label{sec:world-reaff}

Recall the transducer of \S\ref{sec:atom-transducer}: its reading
\eqref{eq:rf} depends jointly on the world $\varphi$ and the agent's own
pose $\theta$.  That joint dependence makes every change in a sensor
reading ambiguous.  It may be \emph{reafferent} --- the agent moved,
modulated the body, and brought the receptor into a new relation to a
static world.  Or it may be \emph{exafferent} --- the world moved, and
the receptor's reading changed with no action by the agent.  A body that
cannot tell these apart cannot have a world at all: every saccade would
look like an event out there.  This is the problem von Holst and
Mittelstaedt \citep{vonHolstMittelstaedt1950} named the
\emph{Reafferenzprinzip}, and J\'ekely and colleagues
\citep{Jekely2021ReafferenceSelf} argue early nervous systems were
organized \emph{primarily} to solve it --- to track the sensory
consequences of their own movement --- with sensitivity to the world a
derivative capacity.

\subsection{The predictor and the residual}\label{sec:world-residual}

The solution is a \emph{reafference predictor}, a sub-component of the
board.  It receives the agent's proprioceptive configuration $\theta$
(delivered veridically by the Sensation Modulators) and the agent's
current estimate $\hat\varphi$ of the world, forms a predicted reading,
and reports the \emph{residual} --- the difference between the actual and
the predicted:
\begin{align}
\hat s(t) &= I\exp\!\left(-\frac{(\theta(t)-\hat\varphi(t))^2}{\sigma^2}\right),
\label{eq:predicted}\\
r(t) &= s(t) - \hat s(t).
\label{eq:residual}
\end{align}
The residual is the central scalar of the architecture: what the world
has done that the agent's model did not anticipate.  One detail is
load-bearing: the predictor uses the \emph{actual} $\theta$, not a
prediction of it, because the agent's own configuration is directly
observable through its Sensation Modulators while the world is not.  This
places the prediction error exactly where the architecture says it
belongs --- in the world-model, never in the body-model.

\begin{figure}[!htbp]
\centering
\includegraphics[width=0.95\linewidth]{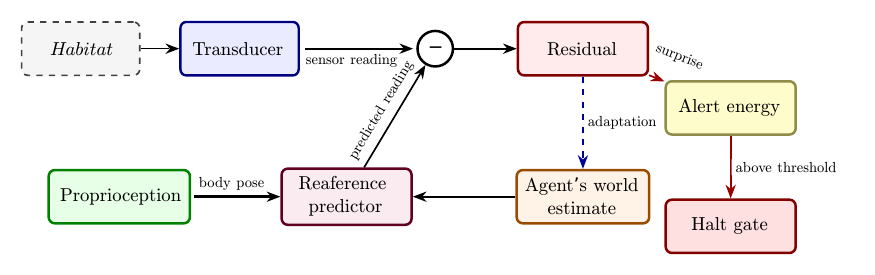}
\caption{The reafference architecture as a closed information loop.  The
transducer converts the world stimulus into a sensor reading; the
predictor combines the agent's pose (from proprioception) and its
current world estimate into a predicted reading; the comparator
subtracts prediction from actual, yielding the residual.  The residual
drives two pathways: it builds alert energy, which above threshold
engages the halt gate (\S\ref{sec:world-attention}); and it adapts the
world estimate (\S\ref{sec:world-mastery}), turning prediction error into
world-model improvement.  The habitat (dashed) is the only piece outside
the SMN.}
\label{fig:reafference-loop}
\end{figure}

The signature is sharp.  For two sensor changes of the same magnitude
--- one produced by the agent moving through a static world, the other by
the world moving past a still agent --- the residual stays at the noise
floor in the first case and rises in the second, with no adaptation in
either.  Self-caused change is predicted away; world-caused change
survives.  On the companion bench this separation is real if modest on a
single-jointed body (a residual ratio of $2.2$ against a no-predictor
foil at $1.58$), limited by reading pose from a single point on a bending
body --- which is exactly what distributed modulation, below, repairs.

It is worth being exact about what the residual \emph{is}, because it is easy to
mistake it for an internal representation corrected toward an external truth.  It
is not.  Perception here is read \emph{from the inside}: the body is a generator
of distinctions, not a filter that maps a pre-given world
\citep{GN2005_TowardsModelLifeCognition}.  A perturbation the body can invert
leaves no trace; a perturbation it cannot yet invert \emph{is} the residual ---
the generated difference, not a proposition about how things stand.  The
predictor is simply the inverting mechanism, and its updates invert more of what
remained; the residual is what has not yet been cancelled.  Nothing here need
carry satisfaction conditions --- it is a fact of the wiring's failure to
cancel, in the sense a content-sceptic can accept --- and it becomes ``about''
the world only later, and only where an emancipated action pattern makes it
addressable (\S\ref{sec:rel-cognitivism}).

\subsection{The world-model in the self-frame, and mastery}\label{sec:world-mastery}

So far the predictor is reactive: it flags surprise but does not learn.
The world-model \emph{proper} is the adapting estimate $\hat\varphi$.  A
minimal gradient step suffices:
\begin{equation}
\frac{d\hat\varphi}{dt} = \kappa\,\frac{r(t)}{\partial\hat s/\partial\varphi},
\label{eq:phi-update}
\end{equation}
a Newton-like update with learning rate $\kappa$: where the receptive
field is informative about the world ($\partial\hat s/\partial\varphi
\neq 0$) the residual carries usable information; where it is not, no
update occurs.  In the vocabulary of cortical predictive coding this is the
ascending-error / descending-prediction loop
\citep{RaoBallard1999_PredictiveCoding} realized on the body rather than in a
neural hierarchy --- $\hat s$ the prediction, $r$ the error, and
\eqref{eq:phi-update} the error-driven update; it is state inference, not yet
parameter learning.  It is also O'Regan and No\"e's \emph{sensorimotor
contingencies}
\citep{OReganNoe2001Sensorimotor,noe_action_2004} made measurable:
mastery of a contingency is the \emph{rate of decay} of the residual
after a world change.  It is not Bayesian inference and not a complete
theory of perceptual learning; it is the minimum that makes mastery a
measurable process.

Two features of this world-model matter.  First, it is \emph{in the
self-frame}: $\hat\varphi$ is world-geometry expressed in the body's own
coordinates (\S\ref{sec:selfmodel}), not an absolute map read off from
nowhere.  A body-referenced code of just this kind is what the
hippocampal--entorhinal spatial system realises in the brain
\citep{OKeefeNadel1978CognitiveMap,MoserKropffMoser2008SpatialSystem}, and
what peripersonal-space representations extend into the region immediately
around the body \citep{GrazianoCooke2006PersonalSpace}.  Second --- and here the bench corrected one of our own
expectations --- \emph{more body does not mean more world}.  A real,
body-relative world-model exists at every body size (a held-out
position-decoding skill around $0.4$, far above a shuffle control near
$0$), but its resolution does \emph{not} rise with raw sensor count: a
longer body with more transducers reads the world no better if the extra
sensors are ungated.  What raises resolution is \emph{modulation}.  Define resolution
operationally as the \emph{world-detection ratio} $R$ --- the ratio of
world-caused to self-caused residual, dimensionless, with $R{=}1$ meaning
the world is lost in the self-noise floor.  Under per-zone dual-interface
modulation each zone cancels its own self-motion, so the self-noise floor
averages down as density grows while the distributed world signal
survives, and $R$ rises as a \emph{power} of CAZ density,
\begin{equation}
R \;\sim\; \rho_{\mathrm{CAZ}}^{\,a},
\qquad a = 0.49 \pm 0.14\ \ (N{=}6),
\label{eq:resolution}
\end{equation}
in agreement with the $a{=}\tfrac12$ that $\rho^{-1/2}$ self-noise
averaging predicts (the bench ratio rises from $\approx\!14$ at
$\rho{=}3$ to $\approx\!29$ at $\rho{=}13$).  Modulation is the \emph{gate}
that enables this scaling: with it removed the self floor is uncancelled,
the exponent \emph{inverts} ($a\approx-0.5$), and $R$ declines toward $1$.
So the \emph{resolution principle} is a law, not a slogan --- $R\sim
\rho^{1/2}$ under modulation, an exponent that is a fingerprint of
independent per-zone averaging --- with raw transducer count entering
nowhere; depth is the further factor the taxonomy (\S\ref{sec:taxonomy})
adds, and the hinge on which the later turn to non-monotonic morphology
rests.

\subsection{Self, world, and other: the dual-signal property}\label{sec:world-dualsignal}

The residual separates self-caused from world-caused change in time; the
\emph{dual-signal property} makes the same distinction \emph{structural},
and extends it to a third case.  The observation is that most of an SMN's
sensory channels are, at any moment, embedded in a \emph{self-modulatable
action pattern} (SMAP): one CAZ acts, another CAZ's transducers register
the consequence through the body's geometry, and the loop closes inside
the body.  Heartbeat modulating baroreceptors, breathing modulating
posture, hand on face, tongue on palate --- the body's interior is a
perpetually active web of such couplings, running even when the agent is
doing ``nothing''.

What unites them is a structural fact.  Let $y_t = F(\boldsymbol\theta_t;
z_t)$ be the body's \emph{modulation map}
\begin{equation}
y_t = F(\boldsymbol\theta_t;\,z_t),\qquad F:\Theta\to\mathbb{R}^m,
\label{eq:F-modulation}
\end{equation}
how modulating any CAZ's configuration changes every
configuration-dependent transducer ($z_t$ the fixed background).  A
pattern is \emph{dual-signal} if two CAZs $Z_x,Z_y$ satisfy:
\begin{description}[itemsep=2pt,labelindent=0pt,leftmargin=4em,style=nextline]
\item[(DS-1) Shared sensitivity] both afferents depend on a common
modulation parameter;
\item[(DS-2) Both efference copies in broadcast] both activations are
available to the predictor through the broadcast;
\item[(DS-3) Haltability] the pattern is agent-haltable;
\item[(DS-4) Trajectory closure] the coupled trajectory stays within the
SMN-internal message graph --- no part passes through a transducer
reading an external state.
\end{description}
When these hold, the predictor has \emph{two} converging veridical
estimates of the outcome rather than one veridical and one hidden, and it
reconstructs one sensor's trajectory from the other's using broadcast
variables alone; the residual stays at the noise floor without
adaptation.

This is quantitative.  The \emph{Dual-Signal Index} measures how well the
predictor reconstructs a sensor, in three forms distinguished by what it
may condition on:
\begin{align}
\mathrm{DSI}^{\mathrm{int}}_{xy} &= \mathrm{corr}\!\bigl(s_y,\;
   \hat s_y(t\mid \theta_x,\theta_y,a_x,a_y)\bigr),\label{eq:dsi-int}\\
\mathrm{DSI}^{\mathrm{ext}}_{xy} &= \mathrm{corr}\!\bigl(s_y,\;
   \hat s_y(t\mid \theta_x,\hat\phi)\bigr),\label{eq:dsi-ext}\\
\mathrm{DSI}^{\mathrm{cross}}_{xy} &= \mathrm{corr}\!\bigl(s_y,\;
   \hat s_y(t\mid \theta_x,\hat\psi)\bigr),\label{eq:dsi-cross}
\end{align}
--- internal (SMN-broadcast variables only), external (an inferred
world-object estimate $\hat\phi$), and cross-SMN (an inferred
\emph{partner} estimate $\hat\psi$).  Together they give the
architectural geometry of contact: high $\mathrm{DSI}^{\mathrm{int}}$
marks \emph{self}-contact (the partner efferent is in the agent's own
broadcast); rising $\mathrm{DSI}^{\mathrm{ext}}$ marks learnable
\emph{world}-contact; high $\mathrm{DSI}^{\mathrm{cross}}$ marks
\emph{inter-subjective} contact (another SMN whose structured modulation
is partly recoverable from the contact but not from the agent's own
broadcast).

\begin{figure}[!htbp]
\centering
\includegraphics[width=\linewidth]{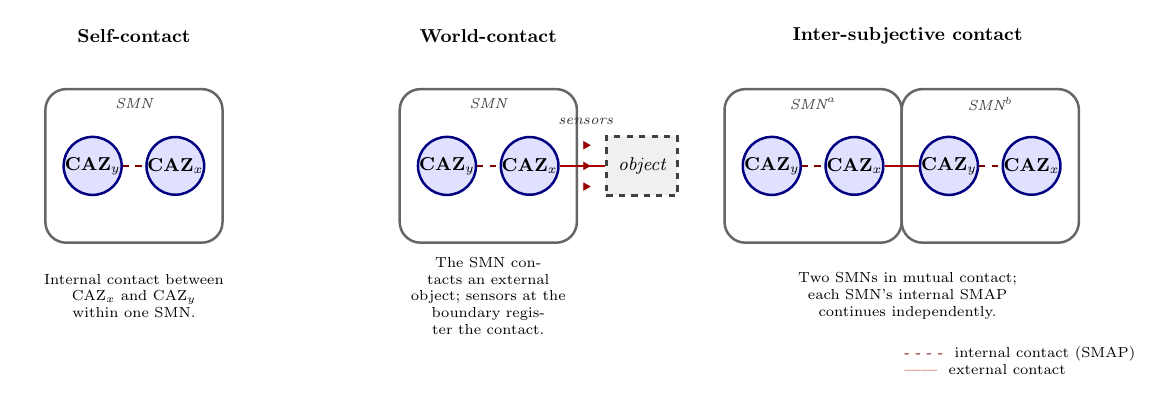}
\caption{The three configurations of contact the dual-signal property
identifies, with the same CAZ$_x$ and CAZ$_y$ across panels.
\emph{Self-contact (left):} both CAZs sit in one SMN and both efferent
streams live in one broadcast; the predictor has both veridical
configurations and the residual stays at the noise floor without
adaptation.  \emph{World-contact (centre):} CAZ$_x$ contacts an external
object with no efferent stream within the SMN, read by single-interface
sensors; if the world dynamics are learnable, the predictor adapts the
world estimate (\S\ref{sec:world-mastery}).  \emph{Inter-subjective
contact (right):} the partner CAZ$_y$ lives in another SMN with its own
broadcast; its structured modulation is partly recoverable, but by a
different route than world-contact.}
\label{fig:three-configurations}
\end{figure}

The three configurations are one architecture read three ways.  In
\emph{self-contact}, both efferent streams are inside the agent's own SMN;
both arguments to the receptive field are known; the residual stays at the
noise floor without adaptation --- on the bench, the self-contact residual
is essentially perfectly explained by the partner-as-stimulus signal,
while a decoupled control of equal residual magnitude is not.  In
\emph{world-contact}, the touched thing has no second efferent stream
within the SMN; the dual-signal shortcut is unavailable, and the
world-model must adapt (\S\ref{sec:world-mastery}).  In
\emph{inter-subjective contact}, the second stream exists but lives in the
\emph{partner's} SMN: the contact is partly SMAP-like (structured, not
random) and partly object-like (not on the agent's broadcast).

\subsection{The first epistemic transition --- and surprise as its
attention}\label{sec:world-attention}

The dual-signal property is a fact about the wiring, not a category the
agent applies.  The agent does not \emph{judge} that the second stream is
absent; the absence is a structural feature of the input.  Self, world,
and other are, on this account, architecturally distinct kinds of contact
before they are anything the agent knows --- which is what we mean by
calling the self/world distinction the \emph{first epistemic transition}
and locating its conditions in the architecture.  How the agent comes to
\emph{detect} and use the distinction across developmental and social
timescales --- and in particular how the inter-subjective ``other'' becomes
shared intentionality --- is deferred to companion work; here we have built
the conditions on which that detection rests.

One coupling of the residual remains, and it is the bridge to the next
section.  The residual does not only teach the world-model; it feeds back
into haltability.  The alert-energy dynamics of \S\ref{sec:atom-halt} gain
a second source --- surprise itself ---
\begin{equation}
\boxed{\;\frac{dE_R}{dt} = \rho\,F_{\text{active}}\,\mathbf{1}[g=1]
              + \rho_r\,|r(t)| - \frac{E_R}{\tau_E}\;}
\label{eq:alertenergy-with-surprise}
\end{equation}
and the halt gate is driven by surprise directly,
\begin{equation}
g(t^+) = 0 \;\Longleftrightarrow\; |r(t)| > \tau_h
\quad\text{(with hysteresis: halt persists until }|r|<\tau_h/2\text{)},
\label{eq:autohalt}
\end{equation}
so a large unpredicted residual halts the agent within one integration
step.  This is not a mechanism bolted on: the same alert energy that
distinguishes haltability from classical inhibition (\S\ref{sec:atom-halt})
is now the substrate of \emph{attention}, realizing Bateson's principle
that the differentiation of difference is the ground of awareness
\citep{bateson2000steps} in mechanistic form.  The halt is not a kill of
activity --- both zones stay co-activated, the configuration is held --- but
a configurational arrest under metabolic load: the body's budget
reorganizing around what surprised it, while the surprising configuration
is held \emph{available for inspection} rather than abandoned.  A body
that halts, holds, and attends to what it holds is one step from being
directed \emph{at} it --- which is where the next section begins.

\begin{benchbox}[ --- self, world, and other]
The reafference cut (Q2) separates self from world, and the dual-signal
register \textbf{R5} makes self-contact a \emph{structural} signature: the
reafference residual regressed on the partner-as-stimulus form gives
$R^2=0.998$ for self-contact and $0.05\pm0.03$ for a decoupled exogenous stimulus
of equal magnitude --- a fit-quality difference, not a magnitude one.
\rtdlink{experiments/sweep_r5_dual_signal/}{R5},
\rtdlink{experiments/q2_reafference/}{Q2}.
\end{benchbox}

\section{Haltability and object-directedness}\label{sec:directedness}

The last section left the body halting, holding, and attending to what
it holds.  That is an \emph{interruption} story --- surprise gates
ongoing action --- and it is not yet enough.  Interruption explains why
an agent switches focus; it does not explain why an agent is focused on
anything in the first place.  The central thesis of this paper claims
the stronger thing: that \emph{haltability} is the architectural
condition object-directed phenomenology requires --- that a body built
this way has the structural form of being directed \emph{at} something.
This section earns the claim.

\subsection{The recruited halt as a stance}\label{sec:direct-stance}

The architectural ground is the recruited halt of \S\ref{sec:atom-halt}.
Active equilibrium is not stillness: it is the maintained co-activation
of two opponent zones around a specific configuration, held against
perturbation, with alert energy accumulating at a level scaled to the
load.  The agent is \emph{investing metabolically} in keeping that
configuration available, and the configuration is not arbitrary --- it is
the one the reafference predictor is currently monitoring.  Three
structural features of this state are, jointly, the architectural form of
intentional directedness.

\emph{The configuration is held available, not produced.}  A motor
program that generates an output does not have an object; a held
configuration does --- the state the recruited halt maintains.  The
architecture is structurally a stance \emph{toward} that configuration,
not a generator of it.  This is the difference between a pump that
delivers a flow rate and a balance that maintains a relation.

\emph{The configuration is held interruptibly.}  The halt is gated by an
alert-energy threshold (\S\ref{sec:world-attention}) and can be released
by the world.  This is the architectural ground of the ``available,
interruptible, resumable'' character of the intentional relation.  A
thermostat also maintains a configuration, but it cannot be surprised;
the recruited halt can be --- the difference is precisely the
alert-energy substrate the thermostat lacks and that opponency (C1)
supplies for free.

\emph{The configuration is held under prediction.}  The reafference
predictor continuously tests the held configuration against what the
world is doing.  The residual measures deviation from what was
anticipated \emph{about this thing}: the agent's world estimate is what
it is currently directed at, and the residual is the architecture saying
``what I am directed at is not behaving as expected.''  This is the
structural form of \emph{consciousness of X} --- oriented toward a
specific intentional object, with the capacity to register that object's
deviation from expectation.

\subsection{From haltability to a haltable action
pattern}\label{sec:direct-pattern}

A clarification the thesis turns on: haltability is \emph{necessary} but
not \emph{sufficient}.  The bare capacity to hold a configuration is an
affordance of any opponent pair; what is sufficient for
object-directedness is a \emph{haltable action pattern} --- a halt that
is recurrent and recognizable, returned to against what resists.  A
recurrent, recognizable halt against a resistance is directedness in
Husserl's sense \citep{Husserl1900_LU,Husserl1913_Ideas}; a one-off
arrest is not.

This gives the section its order parameter, and it is one the companion
bench measures directly (computed, we stress, in the self-frame the body
recovers first, \S\ref{sec:selfmodel}).  Against a rhythmic
central-pattern-generator control and a no-haltability control, the
haltable action pattern is \emph{persistent} (a hold of $\sim\!1.5$\,s
against the CPG's $\sim\!0.3$\,s), \emph{returnable} (the same hold
re-entered some five times), and \emph{side-specific} (directed at a
particular locus, not the body at large).  That triad --- persistence,
returnability, side-specificity --- \emph{is} the operational signature of
object-directedness.  What the bench does \emph{not} yet show is
object-\emph{class} selection: choosing among several objects presupposes
recognizing them as distinct \emph{individuals}, which needs more than the
object-construction of \S\ref{sec:object} earns --- the site-fidelity
individuation rests on (companion work).  Directedness is earned here;
selection among individuated objects is not, and we do not claim it.

\begin{figure}[!htbp]
\centering
\includegraphics[width=\linewidth]{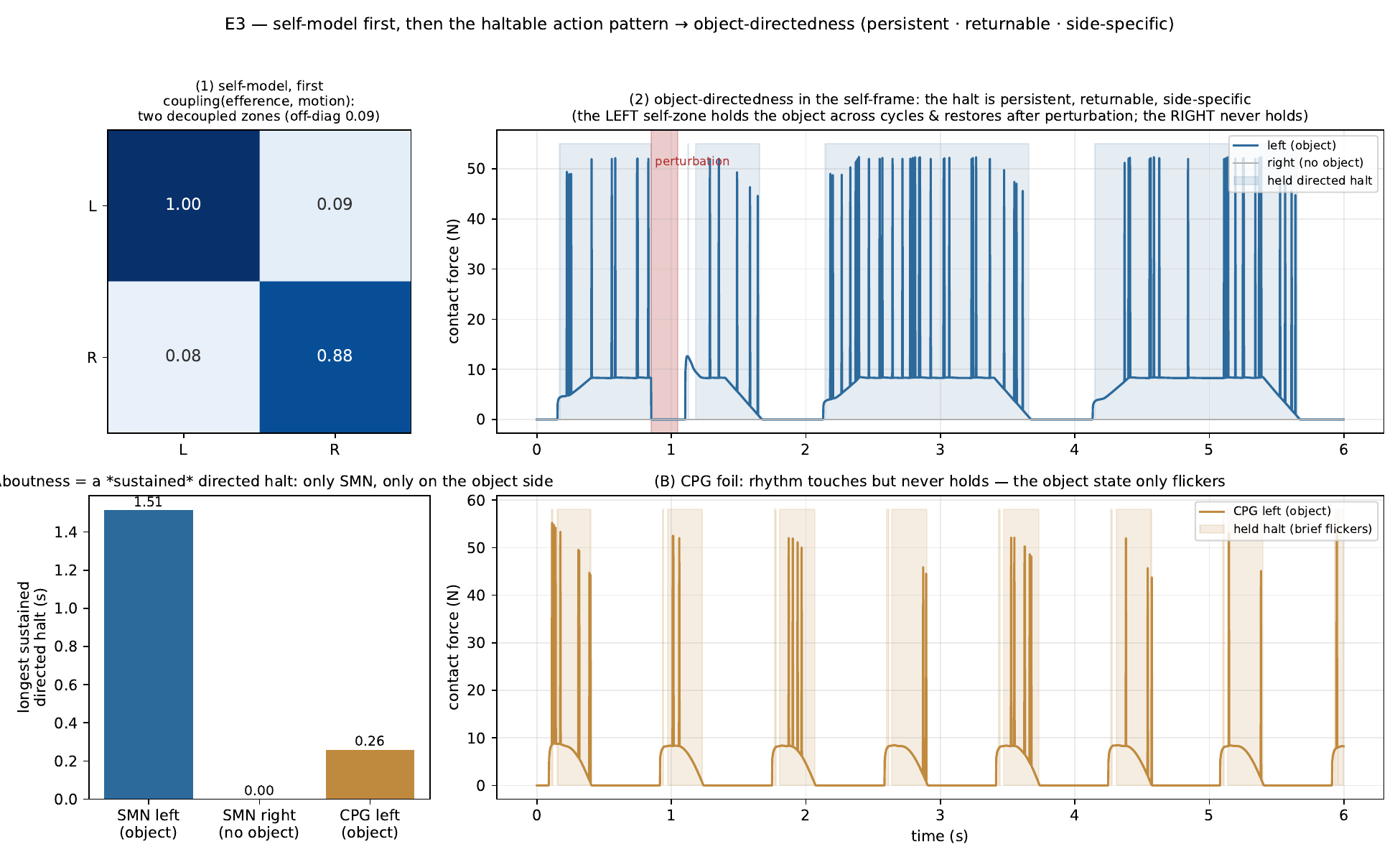}
\caption{Object-directedness as a haltable action pattern, computed in
the self-frame the body recovers first.  \emph{(1)} The self-model is
recovered before the test --- two decoupled zones (off-diagonal coupling
$0.09$).  \emph{(2)} In that self-frame, the left self-zone holds the
object across cycles and \emph{restores} the hold after a perturbation,
while the right zone never holds.  \emph{(lower left)} the order
parameter --- longest \emph{sustained} directed halt --- is $1.51$\,s for
the SMN object-side zone, $0$ for the no-object side, and $0.26$\,s for a
rhythmic central-pattern-generator foil.  \emph{(lower right)} the CPG
rhythm touches but never holds; the object state only flickers.
Persistence, returnability, and side-specificity are the operational signature
of object-directedness.  \emph{In the bench:} \rtdlink{experiments/p6_haltability_aboutness/}{p6\_haltability\_aboutness}.}
\label{fig:haltability-aboutness}
\end{figure}

\subsection{Architectural directedness}\label{sec:direct-earns}

The combination --- recruited stance, metabolic investment,
interruptibility, prediction, made recurrent as a haltable action
pattern --- is what we call \emph{architectural directedness}.  It is not
phenomenology.  The architecture offers the structural condition that any
first-person directedness must rest on; whether and how that condition is
occupied by phenomenal character is the unfinished work of consciousness
studies, and the architecture is not asked to settle it.  What the paper
claims is what the architecture \emph{earns}: a body whose recruited
halts, maintained against load and tested against prediction, has the
structural form of being directed at something.  The directedness is
real; the phenomenal character that may accompany it is a separate
question.

It is fair to press the contrast that most threatens this claim: a
thermostat, or any servo, also holds a set-point and monitors error ---
why is \emph{it} not directed at something?  The difference is a bundle,
and it is the bundle that does the work.  A servo simply \emph{is} its
set-point holding: regulation is its default and only mode, supplied from
outside, costing it nothing it could withhold.  The recruited halt is none
of these.  It is recruited \emph{against} a default --- the ongoing rhythm
--- rather than being the default; it is metabolically invested, a hold
the agent spends alert energy to maintain and can be driven to abandon; it
is prediction-coupled, tracking the residual, so it is a hold \emph{against
a resistance} and not against a number; and it is returnable and
side-specific (\S\ref{sec:directedness}), so the agent can leave a
resistance and come back to \emph{that} one.  A set-point has none of
investment, returnability, or aboutness-to-a-resistance; it holds whatever
value it is handed.  What makes the halt directed is that these hold at
once --- a costly, interruptible, prediction-tested return to a specific
resistance; strip any one and it degrades toward a servo.  That, and no
more, is what ``directed at something'' means here.

\subsection{An embodied two-vector model}\label{sec:direct-twovector}

This account has a close affinity with G\"ardenfors's analysis of events
in conceptual spaces \citep{Gardenfors2020Events,Gardenfors2019Robot}.
An SMN agent pressing on the world is an embodied realization of his
two-vector model: the antagonist pull is the \emph{force vector}, the
object's response --- it yields, resists, or complies --- is the
\emph{result vector}, and gravity and friction enter as the
counterforces.  For this agent, objecthood simply \emph{is} the
force-to-result mapping, sensed as resistance to its own modulation ---
which is exactly the recruited-halt stance developed above.  The one
refinement we add is that the agent does not \emph{learn} the mapping
probabilistically; it \emph{enacts} it by construction, through the
opposition of its pulls against the field.  The force is physically real;
what is constructed is the objecthood the agent reads off it.\footnote{This
resistance-first construal has a striking antecedent in Fichte's late
philosophy, on which the perceived world is an ``image of the resistance''
the agent meets in striving for efficacy; see
\citet{Gruneberg2026GeneratingI} and \S\ref{sec:rel-fichte}.}

This is as far as haltability alone can take us, and it is worth being
exact about how far that is.  What the halting body is directed at is, so
far, a \emph{resistance} --- a single force-to-result relation.  A
resistance is not yet an object, because an object is not a single
property: nothing counts as an object unless it has more than one, and a
single modality can report only one.  Turning the resistance into an
object --- a bundle of co-located properties, a thing with more than one
--- is the work of the next section.

\begin{benchbox}[ --- haltability and object-directedness]
Objecthood-as-resistance is realized as a halt-on-contact (C1) and in the
bilateral manipulator (E1), and the haltable pattern generates
object-directedness (E3).  The preregistered signature is a
\emph{deceptive-reach} prediction (Pred1): the halting agent and a
ballistic foil diverge on contact geometry.
\rtdlink{experiments/p6_haltability_aboutness/}{E3},
\rtdlink{experiments/sweep_pred1_haltability/}{Pred1}.
\end{benchbox}

\section{From resistance to object: multimodal crossing}\label{sec:object}

\subsection{An object is more than one property}\label{sec:object-principle}

Stage 5 earned a stance toward a resistance.  A resistance, though, is
not yet an object, and it is worth saying exactly why.  A single
modality reports a single property --- a magnitude along one dimension:
how hard a thing pushes back, how warm it is, how bright.  A magnitude is
a \emph{quality}, not a \emph{thing}.  Nothing counts as an object unless
it has more than one property, because an object is precisely what
several properties are properties \emph{of} --- the locus at which they
co-occur.  It follows that no single modality can construct an object: to
have an object, the body must bind at least two properties, and at least
two properties means at least two modalities.

This is the second constructive principle of the paper, after opponency,
and it runs parallel to the resolution principle of
\S\ref{sec:world-mastery}.  There we found that more sensors of the same
kind do not buy more world; here the same lesson recurs sharpened ---
more sensors of one modality do not buy an object.  What buys an object
is a \emph{second modality}, crossed with the first at a shared locus.

\subsection{Multimodal crossing binds the object}\label{sec:object-crossing}

The binding is not an extra faculty; it is the reafference machinery of
\S\ref{sec:worldmodel} run across modalities instead of within one.
When the halting body presses on a resistance and, at the same locus and
the same time, a second transducer --- vision, thermoreception,
chemoreception --- reads a co-varying signal, the two readings are
predictable \emph{as a unit} under the agent's own action: press
harder, and touch, seen deformation, and warmth move together in a way
the predictor can track jointly.  The object is exactly that invariant
--- the locus at which several modalities co-vary predictably as the
body probes it.  It is world-contact (the middle configuration of
\S\ref{sec:world-dualsignal}) enriched: a thing with no second efferent
of its own, but now with several afferent properties the body has bound
into one.

The binding is combinatorially generative, and in a way a physicist will
recognize from the resolution principle.  With $M$ coupled modalities the
body can distinguish on the order of $2^{M}$ property-combinations --- a
few modalities span a large space of distinguishable objects --- while
adding sensors \emph{within} a modality adds none.  Modalities, not
sensors, are the dimensions of objecthood; a new modality doubles the
object-space, a new sensor of an old modality does not.  This $2^{M}$ is a
\emph{ceiling}, not a count: it assumes the $M$ modalities are binary and
independently varying.  Real modalities are correlated --- weight with
size, warmth with softness --- so the \emph{effective} object-space is
smaller, and the exponent measures the crossings a body \emph{could} draw,
not those a given world affords.  The same
architectural fact that gives resolution its density exponent under
modulation (\S\ref{sec:world-mastery}) caps the object-space at the number
of \emph{coupled modalities}, and for the same reason: what is not
modulated, and not crossed, does not count.

\begin{figure}[!htbp]
\centering
\begin{tikzpicture}[
  every node/.style={font=\small},
  prop/.style={draw, thick, rounded corners=2pt, fill=#1!12, inner sep=4pt},
]
\node[circle, draw, very thick, fill=gray!15, minimum size=1.5cm,
      align=center] (loc) at (0,0) {a\\resistance};
\node[prop=orange] (t) at (-4.6,1.5) {touch: \emph{hardness}};
\node[prop=red]    (h) at (-4.6,0)   {thermal: \emph{warmth}};
\node[prop=blue]   (v) at (-4.6,-1.5){vision: \emph{shape}};
\draw[-{Stealth},thick] (t) -- (loc);
\draw[-{Stealth},thick] (h) -- (loc);
\draw[-{Stealth},thick] (v) -- (loc);
\node[draw, very thick, rounded corners=3pt, fill=green!12,
      text width=3.5cm, align=center] (obj) at (4.4,0)
  {\textbf{an object}\\(a kind)\\$\ge 2$ co-located,\\co-varying properties};
\draw[-{Stealth},very thick] (loc) -- (obj)
  node[midway, above, font=\scriptsize\itshape] {bind};
\node[font=\scriptsize\itshape, text=gray!50!black, align=center]
  at (0,-2.5) {one modality alone $=$ a quality, not a thing};
\end{tikzpicture}
\caption{From resistance to object.  Haltability
(\S\ref{sec:directedness}) directs the body at a \emph{resistance} --- a
single force-to-result relation, one property.  A single modality reports
one property: a \emph{quality}, not a thing.  When two or more modalities
read the same locus and co-vary predictably under the body's own probing,
their co-located properties bind into an \emph{object} --- a bundle that
recurs as a recognizable kind.  Objecthood is a cross-modal invariant,
constructed by the reafference machinery of \S\ref{sec:worldmodel} run
across modalities.  (Individuating that kind into a particular
\emph{token} is a further step this paper does not take;
\S\ref{sec:object-particular}.)  \emph{In the bench:} \rtdlink{experiments/p5_self_field_object/}{p5\_self\_field\_object}.}
\label{fig:object-crossing}
\end{figure}
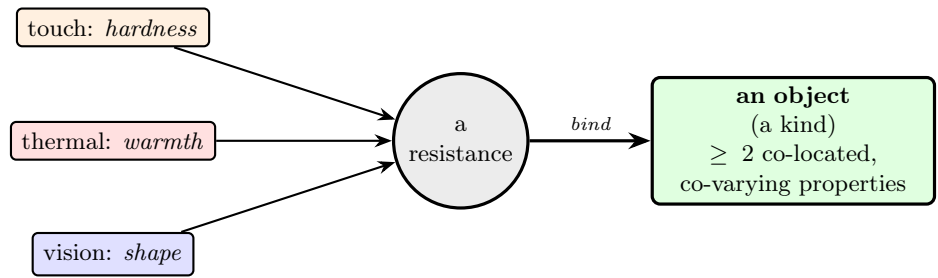

\subsection{The object as a kind --- and why the individual
waits}\label{sec:object-particular}

We should be careful about \emph{which} sense of object the multimodal
bundle earns.  It is a recognizable \emph{kind} --- a warm, hard, dark
such-and-such --- not yet an \emph{individual}.  To register a kind is to
find that a bundle of properties recurs; to individuate a
\emph{particular} is to hold that \emph{this} bundle is a distinct token
from \emph{that} qualitatively identical one.  The two come apart, and
the framework's claim --- which we state here but argue elsewhere --- is
that they come apart in a definite order: an agent discovers \emph{genus
and species} relations, the kinds, \emph{first}, and the particularity of
individuals only \emph{later}.

The individual waits because individuating a token demands
\emph{site-fidelity}: a commitment to spatial geometry, together with the
exclusion principle that no two objects occupy the same place at the same
time.  Only once that is secured can an agent be \emph{certain} whether
two encounters are one individual or two --- and certainty of that kind is
what counting rests on.  Site-fidelity is not cheap: it is a
cognitive-developmental achievement won through prolonged deliberation,
not a competence a body plan hands over.  It is here that much of what
STEM practice later requires --- counting, measurement, the individuation
of samples --- has to wait.  \emph{We therefore make no strong claim, in
this paper, about the discovery of individual particulars.}  What the
multimodal bundle earns is the object as a kind; how the individual token
is won, and why the order runs kinds-before-individuals, is a careful
philosophical argument we leave to companion work.  The \emph{second
epistemic transition} is thus \emph{located} here --- its precondition is
the multimodal object of this section --- but it is not crossed.

\subsection{What is earned, and what is deferred}\label{sec:object-seam}

This is the boundary of the paper's Phase-I scope, and we mark it
honestly.  What is \emph{earned} is the principle and its mechanism: that
objecthood requires crossed modalities, and that the crossing is the
reafference machinery already in hand, run across modalities.  What is
\emph{not} yet delivered is the demonstration.  On the companion bench,
cross-modal object construction is deferred: an early version used
hand-defined features rather than genuine binding in the messaging beam
--- a shortcut we flag rather than count --- and a clean test on the
minimal organism awaits multimodal binding built into the broadcast
itself.  Beyond the demonstration, the full route from an object-kind to
a \emph{counted individual} needs more than the single body offers, and
more than morphology: the \emph{site-fidelity} on which individuation
rests (\S\ref{sec:object-particular}), a ratchet that stabilizes an
individual across encounters, and the emancipated, branched appendages
that let a body mark, point to, and re-identify a thing --- with external
traces and other agents carrying part of the load.  Those are the work of
the companion program.

The lesson to carry forward is twofold.  The object as a kind is within
reach of the multimodal body; individuated \emph{tokens} and counting are
not --- and not merely because the body is small, but because
individuation is a developmental, deliberative purchase
(\S\ref{sec:object-particular}), secured through site-fidelity rather than
anatomy.  What \emph{more body} does buy is a richer repertoire of action,
and that is the subject of the last construction: a taxonomy of action
patterns, and the anatomy that unlocks each level.

\begin{benchbox}[ --- from resistance to object]
The multimodal step is realized as a self/field/object factoring (E2):
pressing on part of one's own body, on the field, and on a thing are
separated in the bench's read-out.  Binding two or more modalities into a
single object-kind (P3) is the deferred crossmodal seam, marked as
Phase-II work.  \rtdlink{experiments/p5_self_field_object/}{E2},
\rtdlink{experiments/p3_crossmodal_discrimination/}{P3}.
\end{benchbox}

\section{Toward a taxonomy of action patterns}\label{sec:taxonomy}

Everything so far has concerned one CAZ, or a few.  The last
construction is what happens to a \emph{repertoire} of action as the body
grows --- and it is the construction that hands the argument to its
companion work.  The claim of \S\ref{sec:object} was that the second
epistemic transition asks for more body; this section says what ``more''
must mean, and names the ladder of action patterns that more body
climbs.

\subsection{From one CAZ to a chain: the phase-oscillator
description}\label{sec:tax-oscillator}

A single CAZ holds a balance; a \emph{chain} of them --- spinal CAZs in
locomotion, oral CAZs in chewing, axial CAZs in respiration --- does
something a single CAZ cannot, and the natural language for it is that of
coupled phase oscillators.  For a chain of $N$ CAZs with intrinsic
frequencies $\omega_i$, phases $\theta_i$, neighbour coupling $K_{ij}$,
and target lags $\varphi_{ij}$,
\begin{equation}
\dot\theta_i = \omega_i + \sum_{j\in\mathcal N(i)}
   K_{ij}\sin(\theta_j - \theta_i - \varphi_{ij}) + u_i(t).
\label{eq:kuramoto}
\end{equation}
What makes this a specification rather than an illustration is the last
term.  The modulatory input $u_i(t)$ is not stipulated from outside: it
\emph{is} the haltability-gated drive each CAZ's board emits
(\S\ref{sec:atom-halt}), reflecting the local affordance and the
broadcast state of the network.  The rhythm is the body's; what breaks
the rhythm into goal-directed action is haltability.  The four regimes
below are then formal conditions on $u_i$, and they are the taxonomy.

\subsection{Four regimes of action}\label{sec:tax-regimes}

\emph{Basal Action Patterns (BAPs).}  The modulatory input is absent
($u_i\equiv 0$): the chain runs on intrinsic frequency and coupling
alone, a stable limit cycle.  Heartbeat, peristalsis, breathing at rest
--- the body's autonomic backdrop.

\emph{Haltable Action Patterns (HAPs).}  The modulatory input is a
phase-reset pulse that gates an oscillator to a halt value, with the
haltability operator active and alert energy \eqref{eq:alertenergy}
governing the cost of the hold; release returns the oscillator to its
prior trajectory or a new one.  Active holding, postural freeze --- the
architectural form of stopping, and the substrate of the directedness of
\S\ref{sec:directedness}.

\emph{Negotiable Action Patterns (NAPs).}  The modulatory input
continuously rephases or reweights the inter-CAZ couplings in response to
perceptual predicates.  A NAP is what a HAP becomes when the halt is no
longer absolute and the resumption no longer a simple return: the same
machinery that admits stopping admits \emph{adapted} resumption.  Where
HAPs are stoppable, NAPs are \emph{modifiable} primitives --- assemblies
that can be reweighted, rephased, and recombined on the fly, in what
amounts to assimilation and accommodation
\citep{Piaget1952Origins,Kelso1995,HakenKelsoBunz1985PhaseTransitions,BizziCheung2013}.
Gait adapting to terrain, bimanual coordination, speech articulation, and
dexterous tool use are NAP-level.

\emph{Transactional Action Patterns (TAPs).}  NAPs whose phase-reset
events are \emph{externalized} --- the halt pulse leaves a public trace
(a gesture, a vocalization, a mark) recoverable by another agent's
transducers.  This is a property of the agent--environment--other
coupling, and it is the gateway to convention and language: it is what
turns a phase reset into a token.

\begin{figure}[!htbp]
\centering
\includegraphics[width=0.55\linewidth]{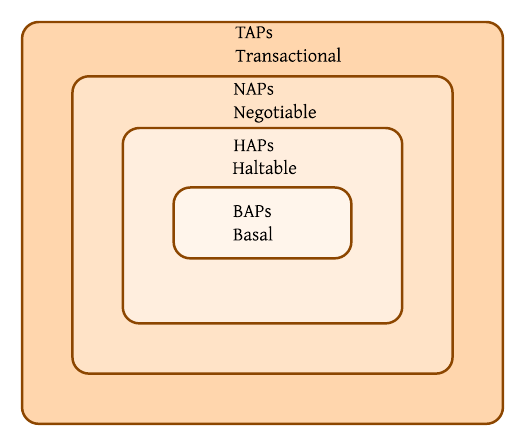}
\caption{The four action-pattern regimes, nested from Basal (innermost)
to Transactional (outermost).  Each outer layer presupposes the inner
ones --- TAPs build on NAPs, NAPs on HAPs, HAPs on BAPs.  The default is
the autonomic backdrop of BAPs; halting, negotiating, and externalizing
are the successive architectural moves that wrap it.}
\label{fig:nested-action-patterns}
\end{figure}

\subsection{Two axes, kept distinct}\label{sec:tax-axes}

It would be easy to misread the four regimes as a single five-rung
ladder that also includes the self-modulatable action patterns (SMAPs)
of \S\ref{sec:world-dualsignal}.  They are two different axes, and
keeping them apart matters.  BAP\,$\to$\,HAP\,$\to$\,NAP\,$\to$\,TAP is
the \emph{emancipation} axis: each step frees action further from the
fixed rhythm, ending outside the body altogether.  SMAP is an orthogonal
property --- \emph{self-modulation}, the dual-signal closure of a pattern
whose loop stays inside the body --- and a pattern at any rung of the
emancipation axis may or may not be a SMAP.  The emancipation axis says
how far an action has travelled from autonomic rhythm toward public
convention; the SMAP property says whether its sensory consequences close
within the SMN's own broadcast.

Of the emancipation axis, this paper builds the lower two rungs.  BAPs
and HAPs are in scope, and the HAP is exactly the haltable pattern that
Stage 5 turned into directedness.  NAPs and TAPs are \emph{named} here,
and their architectural conditions identified --- the HAP\,$\to$\,NAP
transition is the haltability operator gaining rephasing and
coupling-reweighting capacity; the NAP\,$\to$\,TAP transition is the
externalizability of the phase-reset trace --- but their development into
recursive, recombinable structure is companion work.  We mark the path;
we do not walk it here.

\subsection{Why the repertoire needs more body}\label{sec:tax-nonmonotonic}

Here the pivot the paper has been approaching since \S\ref{sec:world-mastery}
comes due.  The BAP/HAP distinction is not available to a single CAZ:
a lone oscillator can be halted, but ``halt'' means something only
against a \emph{backdrop} it can be held apart from, and a backdrop
requires other CAZs still running.  The distinction is therefore a claim
\emph{about morphology} --- it exists only once the network is layered
(\S\ref{sec:kit}).  The same is true up the axis: NAPs need enough
coupled assemblies to recombine, and TAPs need emancipated,
externally-visible appendages to leave traces with.  Each rung of the
emancipation axis is unlocked by a level of anatomical organization, not
by a bigger version of the same one.

This is the resolution principle of \S\ref{sec:world-mastery} in its
strongest form.  Linear scaling --- more identical segments, more
identical sensors --- does not climb the ladder; we saw it does not even
enrich the world-model.  What climbs the ladder is \emph{non-monotonic}
morphology: a new layer that anchors on a stabilized old one and adds a
qualitatively different kind of zone, exactly the morphogenetic
progression of \S\ref{sec:kit} (polarity, then tubularity, then
segmentation, then bilaterality, then appendages).  Generativity comes
from a new layer wrapping a stable old one --- which is why the
architecture's answer to ``more body'' is not ``a longer body'' but ``a
more deeply layered one.''  The repertoire of action, and with it the
resolution of the world and the range of object-kinds a body can
register, scales with CAZ density $\times$ internal capacity $\times$
depth of nesting, never with raw count.

\begin{figure}[!htbp]
\centering
\includegraphics[width=0.72\linewidth]{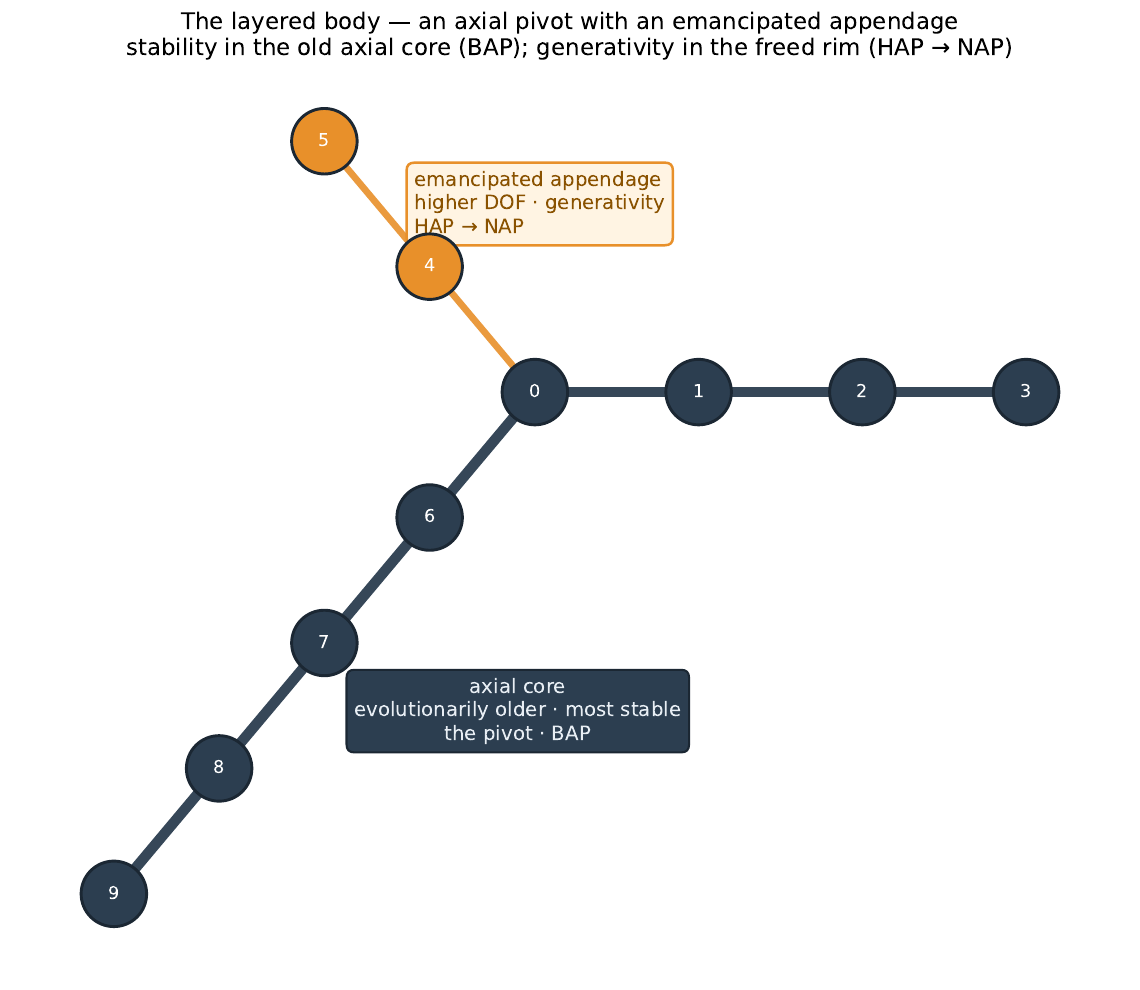}
\caption{The non-monotonic layering that climbs the ladder, in the
diagram grammar.  The evolutionarily older \emph{axial} core (dark) is
the most stable part of the self-graph --- the pivot the rest swings
against, and the home of the Basal Action Patterns.  An
\emph{emancipated appendage} (orange), freed of the axial load-bearing
job, gains degrees of freedom and with them the Haltable and Negotiable
patterns where generativity lives.  A stable axial hub with a plastic
emancipated rim is what ``more deeply layered'' means: a new layer
anchored on a stabilized old one, not a longer chain of the same.  \emph{In the bench:} \rtdlink{experiments/p8_objecthood_transition/}{p8\_objecthood\_transition}.}
\label{fig:layered-body}
\end{figure}

\subsection{What becomes locatable}\label{sec:tax-locatable}

Naming the ladder does one more thing: it makes the standing impasse of
cognitive science \emph{locatable} on the architecture, even though this
paper does not resolve it.  The recursive, recombinable generativity that
cognitivism rightly prizes has an address here --- the modifiable
modular dynamics of NAPs --- and the embodiment that 4E rightly insists
on has one too --- the opponent substrate those NAPs run on.  Placing
recursion at the NAP rung is not the same as delivering a theory of it,
and we do not claim the reconciliation; the NAP and TAP rungs are the
direction, and the companion work is where they are built.  What the
architecture offers now is a \emph{map with the competing accounts
placed on it}: each identified with a regime or a construction, each
separated from the SMN by a criterion a measurement could decide.  Making
that map explicit --- the affinities, the differences, and the
falsifiable criteria --- is the work of the final section.

\begin{benchbox}[ --- the taxonomy]
The ladder's predictions are quantified: world-resolution scales with
network size (E4), objecthood appears as a coupling-density transition
(E4b), and zonal dissociations follow the nested-zone prediction (Pred2).
\rtdlink{experiments/p7_scaling_network/}{E4},
\rtdlink{experiments/p8_objecthood_transition/}{E4b},
\rtdlink{experiments/sweep_pred2_zonal/}{Pred2}.
\end{benchbox}

\section{Where the SMN sits: limiting cases and falsifiable criteria}\label{sec:related}

The construction is complete; what remains is to place it.  We do so in
the spirit \S\ref{sec:taxonomy} set up: each account below has identified
a true property of cognitive systems, and the SMN's contribution is to
say what kind of architecture admits those properties together --- so
that, as an \emph{implication} of the architecture, each can be located
as a \emph{limiting case} of it, separated from the SMN by a criterion a
measurement could decide.  This is a positioning, not a finished
reconciliation; the one reconciliation the paper deliberately leaves open
--- cognitivism and 4E --- is flagged where it arises.

\subsection{Sensorimotor enactivism}\label{sec:rel-enactive}

The sensorimotor tradition
\citep{OReganNoe2001Sensorimotor,noe_action_2004} holds that perception
is constituted by mastery of how sensory inputs change under action.  The
SMN is consonant and supplies a mechanism: the reafference predictor of
\S\ref{sec:worldmodel} formalizes mastery as residual decay.  Where it
adds is the self/world distinction the enactivists tend to take as given
--- rather than a specialized discriminator, the SMN treats it as a
structural fact about the wiring, the dual-signal property
(\S\ref{sec:world-dualsignal}) handling self, world, and other in one
family.  The limiting case is a sensorimotor account read without that
structure; the criterion is the \emph{structural} signature of Register~5.

Two honest caveats bound this positioning.  First, the constitutive traditions
(radical and autopoietic enactivism) make embodiment a claim about what
perception \emph{is}, not about its causal mechanism; Register~5 can come out
exactly as predicted and leave that claim untouched, so the ``limiting case''
here marks a positioning, not an adjudication
\citep{HuttoMyin2013RadicalizingEnactivism}.  Second, C3 buys decentralization
--- no central \emph{reader} --- but not \emph{autonomy}: the connectivity
graph, the drive, and the target $\lambda$ are specified by the designer, and
the agent's persistence is not yet at stake in its own activity.  Precariousness
in Di~Paolo's sense --- an agent that must actively conserve its own viability
\emph{or perish} \citep{DiPaolo2005Autopoiesis} --- is deferred to the companion
bench, where an energy-gated body can be made to die; this framework supplies
the \emph{body} that sense-making requires, and is enactivism-adjacent without
claiming to be autopoietic.

\subsection{Active inference and predictive processing}\label{sec:rel-ai}

The free-energy program \citep{Friston2010FEP,clark2008supersizing} holds
that biological systems minimize prediction error through a generative
model driving perception and action.  The mapping onto the SMN is close,
and better stated plainly than resisted: the residual \eqref{eq:residual}
is a prediction error; the world update \eqref{eq:phi-update} is a
gradient step on negative log-likelihood, that is, predictive-coding
perceptual inference; the drive toward the equilibrium $\lambda$
\eqref{eq:drive} is action minimizing proprioceptive prediction error
under a prior preference; alert energy and the filter $\alpha$ (C4--C5)
do the work of precision; and the self/world/other cut of
\S\ref{sec:world-dualsignal} is a Markov-blanket partition by conditional
independence.  An active-inference reader will recognize almost every
moving part.

We therefore do not claim to falsify the free-energy principle, and we
withdraw any suggestion that it is less falsifiable than the SMN.  The
principle is a variational bound, not itself an empirical hypothesis; its
\emph{process theories} are testable and heavily tested
\citep{Friston2017ProcessTheory}, and our own C1 (opponency), stated in
\S\ref{sec:atom-commitments} as an empirical generalization that contrary
evidence would overturn, has exactly the status of that bound --- a
commitment about how systems are organized, not a theorem.  Comparing our
registers against the principle was comparing a process theory to a
bound; the honest comparison is process theory to process theory.

Made at that level, the relation is not rivalry but grounding.  Active
inference \emph{presupposes} a generative model, a prior preference, a
precision, and a Markov blanket, and --- in its enactive, self-evidencing
form especially
\citep{Hohwy2016SelfEvidencing,Bruineberg2018AnticipatingBrain} --- reads
all four off the agent's own engagement with the world.  The SMN asks
where, in a body, those four come from, and answers with a specific
physics: the generative model from a reafferent opponent unit that senses
its own equilibrium; the prior preference from the equilibrium that
opponent pair physically affords; precision from a metabolic quantity,
alert energy; the blanket from broadcast membership.  The free-energy
principle is substrate-agnostic by design; the SMN is a substrate
commitment from which the predictive machinery is \emph{drawn} rather than
assumed.  In this sense the SMN is the constructive grounding active
inference brackets --- and for the class of embodied cognitive agents,
active inference is the description of what an opponent, elastic body
generates.  That is an alliance, not a takeover: the enactive free-energy
program and this framework both build the world from the body's
engagement, and stand closer than a rivalry framing admits.

Two clarifications follow.  First, the ``from outside / from the data''
characterization (\S\ref{sec:rel-robotics}) is aimed at machine
world-models, not at active inference, which is self-evidencing; we do not
extend it to the free-energy principle.  Second, the separators the SMN
can honestly claim are narrower than earlier drafts implied.  Register~1
--- antagonist tonic activation rising with load and decaying over seconds
--- separates the opponent substrate from \emph{classical inhibition}, but
not from active inference: motor active inference already predicts the
same antagonist scaling, having absorbed equilibrium-point control into
precision-weighted proprioceptive prediction
\citep{AdamsShippFriston2013_ActiveInference}.  What would genuinely separate a
purely residual-driven SMN from an expected-free-energy agent are
dynamical dissociations --- an agent that acts to resolve resolvable
ambiguity carrying \emph{no current residual} (epistemic foraging), and
attention that \emph{rises} rather than falls as an informative channel is
made noisier ($E_R\!\propto\!|r|$ versus precision $\propto$ inverse
variance --- a dissociation to be read under \emph{fixed} volatility and a single
informative channel, since expected precision is itself optimized and can rise
with attended uncertainty).  Both are runnable in the companion bench, neither is
settled here, and we flag them as the discriminating experiments while claiming no
verdict.  (The same broadcast is where Global Workspace Theory's shared
workspace \citep{Baars1988CognitiveTheory} and Damasio's felt body-state
\citep{Damasio2010_Self} acquire a structural locus.)

\subsection{Affordance competition and morphological
computation}\label{sec:rel-afford}

Two traditions locate the same need --- a body whose structure does
cognitive work.  Cisek's affordance competition
\citep{Cisek2007AffordanceCompetition} places it at action selection,
Pfeifer's morphological computation
\citep{PfeiferBongard2007BodyShapesMind} at the substrate; both are right
about \emph{what}, neither specifies \emph{which} structures, distributed
\emph{how}.  The SMN supplies both: affordance competition becomes
distributed across haltability gates and broadcast couplings rather than
centralized arbitration, and morphological computation gains a substrate
specification (opponency along axes, Sensation Modulators, the CAZ)
rather than a methodological slogan.  The limiting case is selection
centralized rather than distributed; the criterion is Register~2, the
post-halt resumption-latency advantage a distributed alert-energy
substrate predicts and centralized gating does not.

\subsection{Gibson: ecological information and the mutuality of
affordance}\label{sec:rel-gibson}

Gibson \citep{Gibson1979EcologicalApproach} holds that the environment's
structure is directly available to a moving observer as affordances, not
raw sense-data.  The SMN supplies the mechanism Gibson left constitutive:
the differential displacement a Sensation Modulator produces is what
turns ambiguous sensory flow into a body-anchored specification --- without
Sensation Modulators there is no \emph{self-generated} differential flow, and
without that flow no SMN-specific route to body-anchored information.  Affordance is
recovered relationally \citep{Chemero2003-fm}, and the SMN adds that it
is \emph{mutual} --- a differentiated world affords nothing to an
undifferentiated agent --- with a sharp robotics prediction: similar
anatomies build similar world-models, different anatomies different ones,
in one environment.  The limiting case is direct pickup without an
internal integrative estimate; the criterion is that removing the
broadcast estimate (\S\ref{sec:kit-network}) degrades multi-CAZ
coordination, which direct pickup predicts unnecessary.

\subsection{Smith on registration}\label{sec:rel-smith}

The closest philosophical precedent is Brian Cantwell Smith's
\emph{registration} \citep{Smith1996OriginObjects}: objectness is not
pre-given but an achievement of engaged participation.  The SMN cashes
this out in mechanism --- the dual-signal property
(\S\ref{sec:world-dualsignal}) makes ``this stream lacks a second
efferent in my SMN'' available to the architecture itself, and the three
DSIs are Smith's registration distinctions in numerical form.  (We adopt
``register'' for the architecture's empirical patterns in acknowledgement
of this affinity.)  Smith stops short of a substrate; the SMN commits to
layered opponency in a body.  The limiting case is registration as a
single self/world cut; the criterion is the three-way DSI geometry, which
separates self, world, and other by a structural contrast rather than one
boundary.

\subsection{Conceptual spaces (G\"ardenfors)}\label{sec:rel-gardenfors}

G\"ardenfors's conceptual spaces
\citep{Gardenfors2020Events,Gardenfors2019Robot} are the rare account
that mediates between the enactive and the representational camps, and
the SMN meets it on two commitments: geometry and grounding.  As
\S\ref{sec:directedness} developed, the SMN agent is an embodied
realization of the two-vector model, and the body-relative state-space it
constructs is, in G\"ardenfors's terms, a conceptual space.  We read
conceptual spaces as the bridge from \emph{grounded} to \emph{positional}
meaning --- the level at which embodied structure becomes a navigable
geometry --- a thread this Phase-I paper only opens and the companion work
develops.

\subsection{Cognitivism and the 4E debate}\label{sec:rel-cognitivism}

The SMN does not oppose cognitivism or 4E; it has the potential to place
each as a limiting case of one architecture --- though this is the
placement the paper does \emph{not} complete.  What 4E gets right ---
cognition is constitutively embodied --- the SMN supplies the mechanism
for; what cognitivism gets right --- that recursive generative capacity is
real and biologically given --- the SMN locates in the modifiable modular
dynamics of NAPs (\S\ref{sec:taxonomy}) rather than in disembodied
symbols \citep{Chomsky1957SyntacticStructures}.

The dispute between cognitivism and the 4E camp is, at bottom, a dispute about
\emph{content}: whether cognition trades in structured, recursive
representations, and if so, where they live.  The architecture dissolves the
dispute by \emph{relocating} the answer across its own layers --- and this, not
a victory for either side, is what the framework is for.

\emph{The substrate carries no content (the 4E camp is right, going down).}  The
living body is a stochastic opponent system, and at the basal layer
(\S\ref{sec:taxonomy}) its dynamics are pure dialogical invertibility: Basal
Action Patterns (BAPs) that sustain the body and, if they stop, stop it ---
heartbeat, breathing, the locomotor cycle.  There is no representation
\emph{inside} this; there is a precarious, self-maintaining loop.  Radical
enactivism's content-scepticism holds exactly here, and we grant it without
reservation.

\emph{Structure is real, and it lives in emancipated action, not in inner
symbols (cognitivism gets its structure, relocated).}  As the body layers
(\S\ref{sec:tax-nonmonotonic}), zones are \emph{emancipated} from the
life-sustaining load, and on them the body performs Haltable Action Patterns
(HAPs).  Because a HAP is haltable, returnable, persistent, and side-specific
(\S\ref{sec:directedness}), it holds \emph{structure}: action is no longer
monotonic but has addressable states.  This is the informational asymmetry on
which everything turns --- a BAP is an obligate cycle with nothing to address; a
HAP has a hold that can be entered, left, and returned to.  What cognitivism
calls ``content'' is, on this reading, nothing over and above HAP-structure: a
haltable, recomposable pattern that \emph{can} carry information precisely
because it is emancipated from the obligate rhythm of life.  Content is a
property of \emph{freed action}, not of a store of inner symbols.

Why a HAP can carry information at all turns on the halt.  A free-flowing
invariant stream --- an uninterrupted BAP --- carries none: information is
difference, and a pure rhythm offers none.  The halt cuts a \emph{gap} into
the stream, and structure in the gaps is what makes a signal recoverable; a
gap is also a \emph{punctuation}, and punctuation is a necessary condition
of syntax.  So the same haltability that earns directedness earns the two
prerequisites of content --- information and syntax --- that a gapless BAP
cannot supply; generativity is not bolted on but already structural in an
architecture that segments its own streams.  This is also why the HAP's
independence is only \emph{apparent}.  A HAP is not free-floating; it is
anchored on the BAP layers beneath it, \emph{emancipated} from the
life-sustaining rhythm, not escaped from the body.  The appearance that a
freed action pattern is disembodied, and so mind-like, is exactly that ---
an appearance.  A HAP is a physical token of a physical event, and what is
cognitively significant is never the event that holds but the \emph{pattern}
of events --- a differentiation of difference --- that the halts articulate.
This refines Davidson's token identity of the mental and the physical
\citep{Davidson1980EssaysActionsEvents}: the bearer of cognitive
significance is not a physical event-\emph{token} but a physical
event-\emph{pattern}, and the halts, with the gaps they cut, are what make
the pattern.  The full argument belongs with the second epistemic
transition, in companion work.

\emph{Representations, if one insists on them, are objects in the world (the
impasse dissolves).}  At the upper rung the agent lays down Transactional Action
Patterns (TAPs) --- schematic patterns that persist as \emph{traces} in the
physical world; a linguistic object is a physical trace.  If these are to be
called representations, they are representations located \emph{outside} the
organism, and they are cognized by exactly the machinery the architecture
already supplies for any physical object: objecthood-as-resistance
(\S\ref{sec:directedness}) and multimodal crossing (\S\ref{sec:object}).  No
inner symbol-processing faculty is required or posited.  The cognitivist's
representations are thus real but \emph{public and external}; the enactivist's
``no inner content'' is preserved intact.  This external-symbol reading has
allies: distributed and extended cognition have long located representational
vehicles in the world and its artifacts
\citep{Hutchins1995CognitionWild,ClarkChalmers1998ExtendedMind}.

This external location does not make the trace self-interpreting, and here
the embodiment the enactivist demands re-enters decisively.  A TAP is
\emph{written} by one agent, encoding a pattern into a durable mark, and
\emph{read} by another --- and the reading is itself a HAP, an embodied
haltable pattern run over the trace.  An agent that lacks the decoding HAP
cannot read the trace even though the pattern is fully present in the world:
the marks are there and mean nothing to it.  The agent's HAPs are thus the
encoding \emph{and} the decoding interface, and the meaning of a public
representation is never in the trace alone but in the embodied patterns that
write and read it.  External representation is therefore real without
floating free --- doubly grounded, in the writer's SMN and the reader's.
How two agents come to share the decoding patterns --- the co-intentionality
that turns private traces into a public code --- is the second epistemic
transition, whose exposition belongs to companion work in progress.

\emph{But isn't the world-estimate itself an inner representation?}  The
sharpest objection points not at the traces but at the world-model: the running
estimate $\hat\varphi$, and its self-frame companion, updated by
\eqref{eq:phi-update}, is decoupleable, accuracy-governed, and action-driving
--- why is \emph{that} not an internal representation?  We resist the label on a
criterion, not a preference.  For a representation to do the work the cognitivist
needs of it, it must be a \emph{composable} form: constituents that recombine
systematically and productively \citep{FodorPylyshyn1988Systematicity}.
$\hat\varphi$ is not a form of that kind.  It is an \emph{event-token} --- an
instantaneous data point, perishable, overwritten at the next gradient step ---
not a syntax of recombinable parts.  What could count as a representation is the
\emph{pattern} such tokens compose once they are laid out in a state-space and
read as a single shape: the plot, not the point.  That composable form is
precisely what this Phase-I architecture does not yet build; constructing it, and
showing that it recombines systematically, is the work of the second epistemic
transition.  So the disagreement is not whether the body holds internal
\emph{states} --- plainly it does --- but whether those states are
\emph{already} representations.  On the composable-form criterion they are not;
they are the data from which one is later constructed.  It is the reply owed to
radical enactivism \citep{HuttoMyin2013RadicalizingEnactivism}: the estimate
covaries with the world --- the Hard Problem of Content grants covariance freely
--- but covariance is not content, and content arrives only with the composable
form, which is not built here.  Faced with the job-description challenge
\citep{Ramsey2007RepresentationReconsidered}, the answer is the same: an
event-token has no representational job to discharge until it is composed into a
recombinable form.  This also marks the
honest boundary of the present claims: the systematicity and productivity a
cognitivist rightly demands \citep{FodorPylyshyn1988Systematicity}, and the
read-addressable internal symbols documented in the animal-learning record
\citep{GallistelKing2009}, are Phase-II business --- fixed here only in
\emph{where} such a form would sit, above the event-tokens, not inside them.

A word on Marr's levels, since the objection often takes that form.  The
framework does not offer a computational-level theory \emph{separable} from its
implementation, because it denies the separation: the body's opponent-elastic
physics is not a substrate on which a computation is merely realised but the
place the computation is \emph{done} \citep{Marr1982Vision}.  The problem being
solved, in Marr's sense, is the construction of a self-model, a world-model, and
object-directedness from a compliant opponent body; the framework's wager is
that this problem cannot be posed independently of the substrate that solves it.

The war, on this diagnosis, was a category error about \emph{location}.
Precariousness (the BAP substrate), structure (the emancipated HAP), and
representation (the external TAP trace) are three \emph{layers} of one
architecture, not three rival claims about one place.  We do not complete the
upper cash-out here --- NAPs and TAPs are named and not built
(\S\ref{sec:taxonomy}), and fixing the \emph{meaning} of a public trace requires
the social layer this Phase-I paper brackets.  But the shape of the
reconciliation is already fixed by the layering: contentless life below,
addressable action in the middle, world-borne representation above.

\subsection{Transcendental self-constitution: Fichte and the first-person
gap in 4E}\label{sec:rel-fichte}

A recurring objection to the 4E and dynamical accounts, pressed sharply by
\citet{Gruneberg2026GeneratingI}, is that they specify the \emph{external}
dynamics of perception--action coupling but stay silent on how those
dynamics come to have \emph{first-person significance} --- how an
autopoietic process is an act not only \emph{by} a self but \emph{for}
one.  Reading Fichte's late \emph{Wissenschaftslehre} against 4E,
Gr\"uneberg locates the missing dimension in a self-referential
schematism: the I is not given and then represented but \emph{generated},
constituted so that its own performances come to matter to it.  The SMN
shares the diagnosis and answers it in the register this paper works in
--- architecture rather than transcendental deduction.  What makes the
coupling ``for a self'' is built: the self-model recoverable only on an
elastic body (\S\ref{sec:selfmodel}), the dual-signal that separates self
from world and other (\S\ref{sec:world-dualsignal}), and the recruited
halt that turns a felt resistance into a stance (\S\ref{sec:directedness}).
Where Gr\"uneberg supplies a self-referential schematism, the SMN offers a
mechanism such a schematism can run on --- the same relation the framework
bears to enactivism (\S\ref{sec:rel-enactive}) and to conceptual spaces
(\S\ref{sec:rel-gardenfors}).

The convergence is not merely programmatic.  Gr\"uneberg reads the
Fichtean perceived world as an ``image of the resistance'' the agent meets
in striving for efficacy --- almost verbatim the SMN's
objecthood-as-resistance (\S\ref{sec:direct-twovector}); and the
effort-modulated perception he draws on --- hills judged steeper under
load --- is the SMN's resolution principle, on which what is perceived
scales with the body's own capacities rather than with raw transduction
(\S\ref{sec:world-mastery}).  A third contact concerns the particular: on
Fichte's schematism individuation is deferred, arriving only when an
\emph{individual purpose} determines an object, exactly as the SMN defers
the token to a later, developmental purchase --- site-fidelity --- rather
than reading it off the morphology (\S\ref{sec:object-particular}).  The
accounts part on grounding: Fichte keeps the self-relation
transcendentally irreducible, anchored in ``the absolute,'' whereas the
SMN wants the construction materially specified and falsifiable.  But the
shared quarry --- a self for which the world is meaningful --- makes this
the framework's natural point of contact with the transcendental
tradition, and the place where a mechanism is most clearly what that
tradition still wants.

\subsection{Large language models}\label{sec:rel-contemp-ai}

Transformer-based language models \citep{Vaswani2017Attention} have made a
structural fact impossible to ignore: a stochastic substrate, stabilized by
training and perturbed by context, emits a \emph{stream} that simulates how
humans speak.  The SMN's reading is that the intuition is right and the
diagnosis can be made precise.  What a body-anchored agent \emph{has} is a
\emph{snapshot} --- the holistic state of a whole-body opponent network, its
constructed world-image --- and this snapshot is \emph{private}: no agent has
direct access to another's.  That privacy is exactly why a \emph{stream} is
necessary.  Communication is not a transfer of snapshots but their serialization
into a shared channel of tokens: a stream from one agent \emph{perturbs}
another's snapshot, constructing a construal, and the response to that perturbation
returns, again, as a stream.  On this construal meaning is \emph{held} as a
snapshot and \emph{travels} only as a stream --- a commitment that contests
use-based and externalist theories on which meaning is already public, a dispute
we join in the companion rather than here.

The two are related by an asymmetry we locate in the \emph{provenance} of the
snapshot at each end.  A body-anchored agent serializes a private, body-anchored
world-image into the channel, where it is \emph{reconstructed} --- never
transmitted --- at the far end.  A transformer, trained to predict the next token
in others' streams, assembles a snapshot \emph{from the channel itself}; a prompt
--- itself a stream --- perturbs that snapshot and returns a stream.  Structurally
this is the very perturbation-and-response loop communication runs on, which is why
the surface can convince.  The difference we press is not the \emph{direction} of
the loop --- both systems run it --- but what the snapshot is \emph{made of}: for
the agent, a body's opponent dynamics; for the transformer, a compression of the
public stream.  In the terms of \citet{Mahowald2024DissociatingLanguageThought}, a
transformer may be \emph{formally} competent on the stream while lacking the
body-anchored, \emph{functional} grounding a snapshot would supply.

What would make this precise is a criterion on the \emph{substrate}, not the
surface --- and stated carelessly it misfires.  A body-anchored agent's snapshot is
a \emph{living} one, continuously reconfigured as opponent dynamics meet new
coupling; a transformer's \emph{weights} are frozen between training runs.  But its
within-context \emph{activations} are updated online --- in-context learning is
exactly such a within-run reconfiguration --- so the discriminator cannot be online
updating \emph{as such}, which context-conditioning, RLHF, and continual
fine-tuning all supply in some form.  The narrower discriminator we propose is an
\emph{opponent, body-anchored substrate that reconfigures in use}, of which online
state-change is a necessary but not sufficient sign.  Whether the gap closes by
adding embodiment to a channel-first system, or only by rebuilding on
opponent-substrate principles, is a question this position lets us pose sharply
rather than settle here.  Operationalizing the criterion, and setting the whole
account against the standing debate over what language models understand
\citep{Bender2021StochasticParrots,Shanahan2024TalkingAboutLLMs}, is the work of a
companion preprint, \emph{Snapshots and Streams}
\citep{Nagarjuna2026SnapshotsStreams}, developed for the ECogS~2026 meeting (OIST).
Here we record only what the unmediated SMN already fixes: the snapshot is the
body's and private; the stream is its serialization into the one channel other
agents can reach --- not the reverse.

\subsection{Machine world-models and cognitive robotics}\label{sec:rel-robotics}

The engineering literatures build precisely what
\S\S\ref{sec:selfmodel}--\ref{sec:object} construct --- self-models,
world-models, object-models --- and it is against them that the SMN's
commitments are sharpest, and most falsifiable.

\emph{Machine world-models.}  A large program learns a world-model as a
latent predictive structure from high-dimensional observation, for
planning and control
\citep{Ha2018WorldModels,LeCun2022WorldModels,Hafner2020Dreamer}.
Generative image models go further still: \emph{Stable Diffusion}
\citep{Rombach2022LatentDiffusion} synthesizes an entire visual world by
denoising, from a distribution learned over billions of images.  Both are
striking, and both differ from the SMN in the same way --- which sharpens
the point of \S\ref{sec:rel-contemp-ai}.  Their world is a
\emph{positional} artifact of data: a manifold of appearances an outside
observer's camera has seen.  The SMN's world is \emph{enacted}:
world-geometry in the body's own frame, the invariant of the body's own
opponent-modulated probing.  Stable Diffusion can render a photorealistic
kitchen without ever having pressed on a surface, been resisted by an
object, or halted against anything; its world has no self in it and no
contact in it.  The difference is falsifiable, and it is the resolution
principle (\S\ref{sec:world-mastery}): a learned world-model sharpens
with more data, parameters, and pixels; the SMN predicts resolution rises
with CAZ density $\times$ modulation $\times$ depth and \emph{not} with
raw sensor count.

\emph{Robot self-modeling.}  Closer still is robot self-modeling.
Bongard, Zykov and Lipson's resilient machine
\citep{Bongard2006SelfModeling}, and its deep-learning successor
\citep{Kwiatkowski2019SelfModeling}, solve exactly the problem of
\S\ref{sec:selfmodel} --- a robot that recovers a model of its own body
from sensorimotor data and uses it to act and to recover from damage.
The SMN solves the same problem by opposite means, and the contrast is a
clean falsifier.  Those systems fit a self-model with a \emph{central}
optimizer (evolutionary search, then a trained network) over a space of
candidate morphologies; the SMN recovers the body graph \emph{locally},
each CAZ reading its own coupling, with no central fit and no candidate
space (Commitment~C3).  And where an optimizer can self-model a
\emph{rigid} robot perfectly well, the SMN's local read-out predicts that
a body which moves as one rigid common mode recovers \emph{nothing} ---
recovery requires an elastic substrate (\S\ref{sec:selfmodel}).  Same
problem; the SMN's answer is decentralized and physically conditioned
rather than optimized.

\emph{SLAM.}  The robot's world-model has a more established form still:
simultaneous localization and mapping \citep{DurrantWhyteBailey2006SLAM}
builds an \emph{absolute} metric map of the environment and localizes the
agent within it.  The SMN's world-model is world-geometry in the
\emph{self}-frame --- relational, with no absolute coordinates and no map
an outside observer could read off.  The falsifiable difference is
anatomy-dependence: a SLAM map is agent-independent (two robots map the
same room to the same metric layout), whereas the SMN predicts the
world-model is \emph{anatomy}-dependent --- different bodies construct
different world-models of the same room (\S\ref{sec:rel-gibson}, the
mutuality of affordance).

\emph{Interactive perception.}  \emph{Interactive perception}
\citep{Bohg2017InteractivePerception} leverages action to perceive ---
objecthood discovered by manipulation --- which is the SMN's
objecthood-as-resistance (\S\ref{sec:directedness}), though the SMN adds
that individuating a particular is deferred to site-fidelity, not assumed
(\S\ref{sec:object}).

\emph{Epigenetic and developmental robotics.}  The tradition the SMN
belongs to most naturally is \emph{epigenetic robotics}, the field named
at Lund by Zlatev and Balkenius for the study of how increasingly complex
cognitive structures ``emerge in the system as a result of interactions
with the physical and social environment''
\citep{Balkenius2001EpigeneticRobotics} --- cognition as a
\emph{developmental} achievement of an embodied agent rather than a
pre-programmed one, a thesis carried forward as cognitive developmental
robotics \citep{Asada2009DevelopmentalRobotics}, intrinsically-motivated
development \citep{Oudeyer2007IntrinsicMotivation}, and predictive-coding
robotics \citep{tani2016exploring}.  The self\,$\to$\,world\,$\to$\,object
trajectory of \S\S\ref{sec:selfmodel}--\ref{sec:object}, and the
emancipation ladder of \S\ref{sec:taxonomy}, are exactly this thesis: a
sequence of epistemic transitions built, not given.

The affinity with the Lund programme in particular is not incidental.  It
has pursued, in parallel, both halves of what an epigenetic account needs:
the developmental \emph{mechanism} --- Balkenius and colleagues' \textsc{Ikaros}
brain-modelling framework and the childlike \emph{Epi} humanoid, built to
reproduce early infant development in a robot \citep{Balkenius2010Ikaros}
--- and the \emph{semantics} such development grows into --- G\"ardenfors's
conceptual spaces (\S\ref{sec:rel-gardenfors}).  The SMN sits between them
and completes the picture from below: it specifies the opponent substrate
the emergence actually runs on --- opponency, haltability, and the
dual-signal property --- beneath the conceptual-space geometry it grows
toward.  What it adds to the tradition is a claim about \emph{what}
develops first, and why it must: not a learned policy but a self-model
recoverable only on an elastic body, then a world-model in that self's
frame, then object-directedness earned by the recruited halt --- each with
the falsifier that comes with it.  Epigenetic robotics has long held that
cognition is constructed through engagement; the SMN offers one
architecture for that construction.

\subsection{Competing accounts as limiting cases, with falsifiable
criteria}\label{sec:limiting-cases}

Gathering the comparisons: each account above is recoverable as a
\emph{limiting case} --- the architecture run in a restricted regime, or
read at a single scale --- and each placement carries a criterion a
measurement could decide.  Table~\ref{tab:limiting-cases} states them.
The criteria are not rhetorical: each is a register the architecture
predicts or a structural consequence of the specified dynamics, so the
positioning is falsifiable in the same currency as the rest of the paper.

\begin{table}[!htbp]
\centering
\small
\caption{Competing accounts placed as limiting cases of the SMN
architecture, each with a falsifiable criterion --- the measurement that
would separate the SMN from the account it situates.}
\label{tab:limiting-cases}
\begin{tabularx}{\linewidth}{@{}p{2.5cm} X X@{}}
\toprule
\textbf{Account} & \textbf{Recovered as a limiting case when\ldots} &
\textbf{Falsifiable criterion} \\
\midrule
Sensorimotor enactivism &
sensorimotor mastery is read off the reafference predictor with the
self/world distinction taken as given &
Register~5: the partner-as-stimulus residual is \emph{structurally}
$R^2\!\to\!1$ in self-contact and $R^2\!\to\!0$ for an exogenous
stimulus --- structural, not a magnitude \\
\addlinespace[2pt]
Active inference / predictive processing &
action reduces to descent on the present residual (no expected-free-energy
lookahead) and precision is read as inverse variance &
epistemic foraging (acting to resolve ambiguity that carries \emph{no}
current residual) and precision dissociation (attention rising, not
falling, as an informative channel is made noisier: $E_R\!\propto\!|r|$
vs.\ precision $\propto$ inverse variance, at fixed volatility) --- companion-bench
discriminators, not yet decided \\
\addlinespace[2pt]
Affordance competition / morphological computation &
affordance selection is centralized gating rather than distributed
haltability &
Register~2: a post-halt resumption-latency advantage set by alert energy
at release, absent under centralized gating \\
\addlinespace[2pt]
Gibsonian direct pickup &
information is picked up without an internal integrative estimate &
removing the broadcast state-space estimate (\S\ref{sec:kit-network})
degrades multi-CAZ coordination; direct pickup predicts no such estimate
is required \\
\addlinespace[2pt]
Smith on registration &
registration is a single self/world cut &
the three-way geometry $\mathrm{DSI}^{\mathrm{int/ext/cross}}$ separates
self, world, and other by a \emph{structural} contrast, not one boundary \\
\addlinespace[2pt]
Cognitivism / language of thought &
inner symbolic content is posited where the SMN relocates structure to
emancipated haltable action and representation to external traces &
systematic HAP recombination, and representation world-located (removing
external traces removes the capacity) --- a Phase-II discriminator \\
\addlinespace[2pt]
Large language models &
weights frozen between training runs; within-context activations update but
reconfigure no body (no opponent, body-anchored substrate) &
an embodied opponent-substrate agent reconfigures its BAP/HAP state-space in use;
a channel-first system has no such substrate to reconfigure \\
\addlinespace[2pt]
Machine world-models \& generative diffusion &
the world-model is a learned distribution over an outside observer's data
(pixels), read at no body &
resolution improves with data, parameters, and pixels; the SMN predicts it
improves with CAZ density $\times$ modulation, not raw sensor count (the
S1 null, Q1b) \\
\addlinespace[2pt]
Robot self-modeling (central optimizer) &
a self-model is fitted by a central optimizer over candidate morphologies &
a rigid robot self-models fine under optimization; the SMN's local read-out
recovers \emph{nothing} from a rigid body (recovery needs elasticity) \\
\addlinespace[2pt]
SLAM (absolute metric map) &
world-geometry is an agent-independent absolute map &
two bodies build the same map of a room; the SMN predicts
\emph{anatomy}-dependent world-models of the same room \\
\bottomrule
\end{tabularx}
\end{table}

We stress the scope.  These placements are an \emph{implication} of the
Phase-I architecture, not a completed unification.  The one placement the
paper deliberately does not cash out is the cognitivism--4E
reconciliation: it turns on the recursive generative capacity the NAP and
TAP layers carry, which this paper specifies only as a direction
(\S\ref{sec:taxonomy}).  The wider landscape is drawn; the Phase-I paper
fills in the region a single body can reach, and marks the rest as
companion work.

\section{The central nervous system: the beam of the body's balance}\label{sec:neuro}

Of all the programs this paper must answer to, one has gathered more
evidence than any other and can least be set aside: systems and cognitive
neuroscience.  A century of electrophysiology, imaging, micro-stimulation,
lesion and inactivation, TMS, and controlled behaviour has mapped, in
extraordinary detail, how neural activity relates to perception and action
--- and much of that work is already embodied in spirit.  A framework that
begins cognition in the body cannot pass over it.  Nor does it wish to: the
SMN does not diminish the nervous system but gives it a place it is usually
denied --- the necessary one.

What follows is an architectural \emph{interpretation} of the neuroscience, not a
review-level demonstration that the interpretation is correct.  Its role is to
state what the SMN predicts the neural evidence is evidence \emph{of}, and what
experiment would separate this reading from a command-centred one.

The framework's own name carries two readings, and the point is that they
are inseparable.  SMN is at once a \emph{sensory--motor network} --- an
anatomical structure, largely neural --- and a \emph{sensation modulating
network} --- a functional architecture of opponent modulation and
broadcast.  These are not a structure that neuroscience studies and a
function that we theorise; they are two faces of \emph{one} system, and
both cognitive neuroscience and this framework study the whole, structure
and function together.  What differs is the interpretation, not a division
of labour.  And the first thing the wiring does is make the body
\emph{one}: a heap of competent zones is not an agent; zones bound into a
shared state are.  Remove the binding and the oneness of the agent is lost
--- which is why the framework cannot, and does not, treat the brain as
epiphenomenal.

It is tempting to call the nervous system the \emph{core} and the
sensory--motor body the \emph{periphery}, but that is the wrong figure,
because it re-imports a centre that commands a margin.  Every CAZ already
has both a sensory--motor (modulating) component and a messaging (network)
component; the CAZ is a module, but not a \emph{localised} one.  The right
figure is the one the architecture has used throughout --- a
\emph{balance}.  The nervous system is the \emph{beam}: localised, because
its office is to \emph{integrate} the whole into one shared state.  The
modulating sensory--motor components are the \emph{weights}: distributed
over the entire body, their office is to \emph{differentiate} --- to move
one zone against the constancy of the rest.  Integration on the beam,
differentiation at the weights: the beam plays the role computational
neuroscience calls a \emph{neural integrator} --- by analogy, a
line-attractor-like manifold holding the pooled body-state, a dynamical
characterization we borrow rather than derive here --- while the weights are the
differentiating elements that move one coordinate against that held constancy.  Neither beam nor weights
commands the other.  A balance has no commander.

Read this way, what cognitive neuroscience \emph{records} --- the neural
population state, where sensory and motor variables are found together ---
is the state the beam carries: the integrated, shared read of body-state
that is the broadcast.  This is not a demotion of the data but an
interpretation of what they are evidence \emph{of}, and it is falsifiable.
The SMN predicts that neural population state is reconstructable from the
coupled peripheral opponent and reafferent variables --- because it is a
read of body-state, not an autonomous computation --- and that silencing an
integrating (beam) node degrades \emph{multi-zone coordination} while
\emph{local opponent competence} survives; a central-commander account
predicts, instead, that the competence itself is lost.  Inactivation with
population recording decides between them.  This broadcast is kin to the
global workspace \citep{Baars2005GlobalWorkspace,Dehaene2014Consciousness},
but network closure marks a testable difference: where the global
\emph{neuronal} workspace turns on \emph{ignition} --- a winner-take-all
access in which one content is broadcast at a time --- closure requires
that \emph{every} zone contribute to and read the broadcast at once.
Ignition is competitive; closure is constitutive.

None of this denies that signals arise in the beam.  Micro-stimulation of
motor cortex evokes coordinated postures and movements
\citep{Graziano2002ComplexMovements}; intention is decodable before an
action unfolds.  The framework reads these \emph{with} it: the beam writes
a coordinated \emph{body-state target} --- a referent, in Feldman's sense
\citep{Feldman2015ReferentControl} --- that the weights realise through
their opponent equilibria, not muscle by muscle.  What it denies is the
\emph{isolated} commander --- a region that generates commands while
standing outside the loop.  There is none: because the CAZ is a network
from the start, no part of the brain is unconnected to some CAZ, and an
experiment could establish an isolated commander only by finding a
signal-generating region \emph{disconnected} from the motor system ---
which does not exist.  The isolated command centre is not merely
unobserved; in a fully coupled network it cannot be constructed.

The architecture also predicts how the beam is \emph{organised} --- how the
regions of the brain divide their labour.  Much of the old brain is given
to regulating the Basal Action Patterns, the life-sustaining zones; the
cortex, and the neocortex in particular, is the evolutionary elaboration
that serves \emph{emancipated} modulation.  We therefore predict that
Haltable Action Patterns are governed by the messaging beams resident in
them, and that the dexterous, emancipated parts --- lips, jaw, buccal zone,
tongue, neck, eyes, limbs, hands --- occupy most of the cortical messaging
territory.  They do: the motor and somatosensory maps are dominated by
exactly these zones.  Their near-uniform cortical structure reflects the
sameness of their mechanism --- \emph{modulate one zone while holding the
rest constant} --- which is the experimental logic, \emph{salva veritate},
by which a body probes and builds a world-model.  The reach of a species'
cognition then tracks how many of its CA zones are emancipated, and the
size of its brain follows from how many messaging beams that emancipation
requires --- which is why mammals, birds, and above all humans have the
brains they do.

Stated so, this can look like an inversion of the usual causal story ---
cognition from body rather than body from brain.  It is not, and the
difference matters.  The CAZ is a network from the word go; its
sensory--motor and messaging components are not two things that interact
but two faces of one unit, and the visible separations between ``body'' and
``brain'' are apparent, not real.  Just as the cell is the
structural-and-functional unit of \emph{life}, the CAZ is the
structural-and-functional unit of \emph{cognition}.  Cognition, like life,
is therefore a \emph{systemic} property: it is not conferred by the body
upon the brain, nor by the brain upon the body, and it is a mistake --- the
very mistake the brain-centric picture makes, only mirrored --- to say that
either component does more than the other.  Larger brains did not
\emph{produce} higher cognition, nor did emancipated bodies acquire large
brains as an afterthought; the two co-arose because they are one system
emancipating together.  That is the collapse the framework asks the reader
to resist.

The layering this implies is visible at the bedside.  General anaesthesia
silences the emancipated outer layers first --- cortical sensation,
awareness, voluntary movement --- while the old-brain zones that regulate
breathing and heartbeat continue.  This is the SMN's layered body made
clinical: the autonomic BAP core persists when the cognitive rim is
switched off, and \emph{life} and \emph{cognition} come apart exactly as
the framework says, the first surviving the suspension of the second ---
one continuous architecture serving both.

Opponency, finally, is found in the neural substrate, not only the
anatomical one: as excitatory--inhibitory balance, as the
agonist/antagonist organisation of the neuromodulatory systems, and ---
binding the whole body --- as the opponent autonomic and endocrine axes
through which state and emotion are regulated, sympathetic against
parasympathetic, the hormonal contraries of arousal and rest
\citep{Craig2002HowDoYouFeel,BarrettSimmons2015Interoceptive,Seth2013InteroceptiveInference}.
Emotion, so read, is a whole-body opponent state carried through chemistry
as well as wiring.  And the recurrence goes all the way down: the CAZ is a
fractal design, realised in every cell with a cytoskeleton --- a neuron
does not shorten like a muscle, but it \emph{shapes} itself through opponent
cytoskeletal dynamics, and the elastic substrate that C7 makes load-bearing is
present at every layer, from the cytoskeletal mesh to the tendon.  This
pervasiveness of the elastic opponent unit at every scale is what makes the
nested, networked oneness of the body an \emph{architectural} fact rather
than a stipulation (\S\ref{sec:selfmodel-scale}).

The framework thus does not compete with cognitive neuroscience for one
explanatory slot; it reinterprets what the neural evidence is evidence
\emph{of}, and returns the nervous system to its place --- the integrating
beam of the body's balance, without which the agent would not be one,
participating as one continuous architecture in both the maintenance of
life and the construction of cognition.  What is given up is only the
commander at the top of a cascade that a network does not contain; what is
gained is a systemic account in which brain and body form one integrating
manifold --- the beam pooling body-state, the weights differentiating within
it.  Neuroscience's century of evidence is, on this reading, not set
aside but, for the first time, given a body to be the brain of.

\section{Conclusion}\label{sec:conclusion}

\subsection*{What the paper built}

The paper began with the smallest agent it could draw --- three segments
and two Coordinated Action Zones (\S\ref{sec:atom}) --- and built
outward.  From that modular unit, and the physics it carries (gravity, mass,
friction, elasticity, opponency), everything followed by composition.
The same kit generates the whole menagerie of animal bodies
(\S\ref{sec:kit}), and the \emph{same} mechanism, invariant under those
morphological moves, constructs a self-model (\S\ref{sec:selfmodel}), a
world-model in the body's own frame (\S\ref{sec:worldmodel}), and the
structural self/world/other distinction that is the first epistemic
transition.  Haltability turns that construction into
object-\emph{directedness} (\S\ref{sec:directedness}); multimodal
crossing turns the resistance it is directed at into an
\emph{object} (\S\ref{sec:object}); and a taxonomy of action patterns
(\S\ref{sec:taxonomy}) names the trajectory --- basal, haltable,
negotiable, transactional --- that more deeply layered bodies climb.

Two constructive principles carry the whole.  The first is
\emph{opponency}: the architectural primitive is an opponent pair whose
balance the agent senses and modulates, and from it come co-activation,
the recruited halt, attention, and directedness --- no module added on
top.  The second is that \emph{an object is more than one property}: no
single modality can construct a thing, so objecthood is a cross-modal
invariant.  The central thesis sits between them --- \emph{haltability},
the active holding of an opponent equilibrium, is the architectural
condition object-directed phenomenology requires
\citep{Husserl1900_LU,Husserl1913_Ideas} --- and the two principles are
what make it constructive rather than stipulated.

\subsection*{What the architecture earns, and the status of the claim}

The phenomenology payoff is structural, not metaphysical.  Haltability is
the condition aboutness rests on, not its sufficient producer; the
directedness is real, and whether phenomenal character occupies it is a
separate question the architecture does not settle.  A further modesty is
owed.  The SMN proposes alternate readings of established findings --- the
nervous system as the integrating beam of the body's balance rather than
its commander, motor tissue as cognitive substrate rather than peripheral
effector.  \S\ref{sec:neuro} places systems and cognitive neuroscience on
this reading and states a criterion that inactivation and population
recording could decide; but this paper does not itself carry out that
comparative test against the experimental record.  It presents the
SMN as a coherent and possible cognitive model, biologically motivated in
its commitments and mathematically specified in its single-CAZ dynamics,
without claiming this is in fact how cognition is rooted in the living
organism.  On the empirical verdict we are, at this stage, neutral;
conducting the test is a follow-up in its own right.

\subsection*{Why it might matter}

Four modest claims.  \emph{For cognitive science}, the SMN supplies a
structural specification at a level the field has generally treated only
philosophically, with eight registers that are concrete falsifiers: a
system claiming SMN-compatibility should reproduce them, and one that
does not is a different architecture.  \emph{For cognitive robotics}, the
commitments are implementable --- opponency along axes, Sensation
Modulators, broadcast, haltability gating --- and the hypothesis that a
structured opponent substrate is energetically cheaper than an
unstructured reservoir (\S\ref{sec:rel-afford}) is a quantitative target
the framework raises but does not settle.  \emph{For grounding AI}, the
account (\S\ref{sec:rel-contemp-ai}) locates the difference between
natural and artificial intelligence in the substrate --- dynamically
constructed by opponent dynamics in a body versus frozen at training time
--- and makes the question of closing the gap one that can be asked
precisely.  \emph{For neuroscience}, the account (\S\ref{sec:neuro}) offers
a coherent interpretation of what the recorded neural state-space
\emph{is} --- the body's integrated broadcast --- together with a
criterion that inactivation and population recording could decide, turning
a reframing into a testable claim.

\subsection*{What is deferred}

\paragraph{Open within the formalism.}
The full multi-CAZ network analysis --- stability, controllability,
observability, convergence --- is open; the neural implementation is
schematic; the precise relation to equilibrium-point control
\citep{Feldman1986_EquilibriumPoint} and discrete-event refinements
(Petri nets \citep{peterson1977petri}, hybrid automata, categorical
descriptions of the broadcast) are left for focused treatment.

\paragraph{Downstream (the companion program).}\label{sec:roadmap}
The specification opens onto a research program whose strands take it as
their starting point, and which is what the body of the paper means by
``companion work'':
\begin{enumerate}[itemsep=2pt,leftmargin=2em]
\item \emph{The first epistemic transition} --- the developmental story by
which the agent comes to \emph{detect} and use the self/world distinction
made architecturally available here.
\item \emph{The second epistemic transition} --- the discovery of
\emph{individual particulars} (as against the kinds \S\ref{sec:object}
earns): how, through the \emph{site-fidelity} on which counting rests, a
multimodal object-kind becomes a re-identifiable individual --- the
framework's claim being that kinds are discovered before individuals.
\item \emph{The third epistemic transition (snapshots and streams)} ---
the conventionalization by which NAP-level traces become TAP-level public
marks, and generative grammar becomes tractable on an opponent substrate.
\item \emph{The trust-conditions thesis} --- how calibrated functional
relations among objects ground a substantive objectivity, given the
architectural priority of self-modulation established here.
\end{enumerate}

Cognitive science has long had a choice between treating the body as
peripheral and invoking it as constitutive without a precise
architecture.  This paper offers a third option: a tractable
specification in which the body is neither peripheral nor merely invoked
but architecturally specified --- and on which the developmental, social,
and conventionalization programs can build.  What it delivers is the
physical, architectural ground on which that wider landscape must stand.

\appendix
\renewcommand{\thesection}{Appendix \arabic{section}}
\renewcommand{\thesubsection}{\arabic{subsection}}
\renewcommand{\thesubsubsection}{\thesubsection.\arabic{subsubsection}}
\setcounter{section}{0}

\section{Formal apparatus and empirical registers}\label{app:formalism}

\noindent\emph{The architecture's formalism is developed inline in the
main body, next to the construction each equation makes precise.  This
appendix is the formal \emph{index} to that material.  It states what
kind of formalism the paper offers and what it does not claim
(\S\ref{app:formalism-kind}); maps the equations onto the five
interlocking sub-systems of the architecture, with a single symbol
glossary (\S\ref{app:formalism-map}, Table~\ref{tab:gloss-formal}); and
then collects in one place the eight empirical registers the architecture
predicts (\S\ref{sec:registers}) and the companion bench that realizes
them (\S\ref{app:simulations}).  A reader who wants the mathematics as a
compact reference --- typically cognitive roboticists, biophysicists, and
computational neuroscientists --- can read this appendix as a standalone
specification whose parts point back to where each result is derived.}

\subsection{What kind of formalism this is}\label{app:formalism-kind}

The formal apparatus has four levels.

\begin{description}[itemsep=4pt,labelindent=0pt,leftmargin=2em,
                    style=nextline]
\item[Architectural definitions]
What components the system has, what relations hold between them, what
their input--output signatures are.  Fully specified for: zones,
Sensation Modulators, transducers, the Coordinated Action Zone with its
haltability operator, the reafference predictor, and the residual.
\item[Single-CAZ dynamics]
How the variables of a single CAZ evolve in time.  Specified as a closed
system of differential equations with named parameters and explicit
dynamical laws --- rigorous as a toy model and serviceable as a reference
implementation, not a complete physiological model of any specific CAZ.
\item[Multi-CAZ network specification]
The architectural structure of the network of CAZs.  Specified as a
hybrid graph-dynamical system with a stated tuple of components; the full
analytical treatment of network dynamics (stability, controllability,
observability, convergence) is \emph{deferred}.  We do not claim a
completed theory of multi-CAZ networks; we claim a specification of the
object whose theory is to be developed.
\item[Empirical registers]
Operational targets the architecture predicts (\S\ref{sec:registers}),
demonstrated by reference simulations as existence proofs that the
equations can generate the predicted contrasts.  Biological validation is
not claimed; the registers are falsifiers any SMN-compatible embodiment
should satisfy.
\end{description}

\noindent
A reader who reads the formalism through this classification will not, we
hope, mistake an architectural definition for a complete dynamical theory
or a simulation existence proof for biological validation.

\subsection{A note to a category theorist}\label{app:category-theory}

The previous subsection ended on an admission: the multi-CAZ network is specified,
but its theory is \emph{deferred} --- we have named the object, not the mathematics
that governs it.  This note is an invitation to a reader equipped to supply that
mathematics, because we suspect its natural home is category theory, and we would
welcome the collaboration.

The suspicion has a concrete source.  As stated here the SMN is several graphs on
\emph{different} vertex sets --- a kinematic graph (segments joined by actuators), a
bipartite sensory graph (transducers feeding boards), a messaging graph (boards
joined by broadcast) --- related by an incidence on the actuators, together with
operators $F$, $\alpha$, $\Pi$ that live in none of them.  Graphs strain to hold this
together, and the strain (``one graph read two ways'' does not typecheck; the
actuator is a node here and an edge there) is the familiar signature of a structure
whose objects are typed and whose arrows compose --- that is, a \emph{category}.  The
pieces then fall into place with almost suspicious ease: the typed substrate is a
presheaf (a $\mathsf{C}$-set); the bipartite sensory graph is a \emph{profunctor};
the mechanical network is the image of a \emph{functor} the signal graph induces
through its actuators (an incidence, not a duality); and $F$, $\alpha$, $\Pi$ are
\emph{morphisms} in a monoidal (process) category, with the sensorimotor loop a
string diagram closed by a \emph{trace}, so that network closure --- our central
non-degeneracy condition on $\Pi$ --- becomes the statement that every object lies on
a feedback cycle.  A coloured, open Petri net is a serviceable bridge: its
transitions \emph{are} the operators, its places the typed state-holders, its colours
the modalities, and it carries a categorical semantics beneath the diagram grammar
this paper already uses.

What such a formalism would \emph{earn}, beyond tidiness, is \emph{compositionality}
--- the property the framework most wants and least has in rigorous form.  The kit
that builds bodies by scaling, branching, nesting, and configuring is a compositional
claim; an \emph{operad of wiring diagrams} would make ``a menagerie from one
grammar'' a theorem rather than a figure.  The self-model --- the body recovering its
own connectivity from movement --- is, read categorically, a \emph{functor} from the
mechanical to the messaging category, and its bench estimator a way of estimating
that functor.  Nesting (lattices whose nodes are lattices) is composition of open
systems along shared boundaries.  None of this is idle: a categorical formalization
of active inference is already under way --- Markov categories, lenses, and
statistical games \citep{StClereSmithe2023} --- so a categorical SMN would let the
constructive-versus-generic contrast of \S\ref{sec:rel-ai} be argued in one shared
language.

We would not be founding this program but joining it.  Ehresmann and Vanbremeersch's
\emph{Memory Evolutive Systems} \citep{EhresmannVanbremeersch2007} model neural and
cognitive emergence as \emph{colimits} in hierarchies of categories; Baez and
collaborators have built a working theory of open networks --- open Petri nets, open
dynamical systems \citep{BaezMaster2020OpenPetri} --- with software
\citep[AlgebraicJulia;][]{Patterson2022Catlab} that could instantiate the SMN grammar
directly; Spivak's operads of wiring diagrams \citep{Spivak2013WiringDiagrams} supply
the compositional scaffolding; St~Clere~Smithe and Myers
\citep{StClereSmithe2023,Myers2022CategoricalSystems} give a categorical account of
the Bayesian brain; and Coecke's categorical compositional semantics
\citep{Coecke2010CompositionalMeaning} is the natural home for the serialization of
meaning our companion on snapshots and streams pursues
\citep{Nagarjuna2026SnapshotsStreams}.

One hook we would chase first, because it is ours.  The framework's guiding heuristic
is \emph{inversion}: structure as the achievement of construction, not its premise.
Inversion has a precise categorical shadow in \emph{adjunction} --- the canonical
``inverse up to structure.''  An adjunction between a construction functor and a
forgetful, structure-remembering one would state ``structure is what construction
achieves'' as a theorem rather than a slogan, and give the whole programme --- from
the opponent pair to the self-model --- a single formal spine.  This is, frankly, an
invitation: we can specify the object; we would value a collaborator who can prove
things about it.

\subsection{A map of the formalism}\label{app:formalism-map}

\begin{figure}[!htbp]
\centering
\begin{tikzpicture}[
  cluster/.style 2 args={
    rectangle, rounded corners=3pt, draw=#1!60!black, very thick,
    fill=#1!8, inner sep=5pt, align=left, font=\footnotesize,
    text width=4.6cm
  },
  edge/.style={->, >=Stealth, thick, draw=gray!70!black},
  edgelbl/.style={font=\scriptsize, inner sep=2pt, fill=white,
                  text=black, rounded corners=1pt}
]
\node[cluster={blue}{}] (net) at (5,5) {%
  {\small\bfseries\color{blue!50!black}(E) Network operators}\\[2pt]
  Broadcast operator $\Pi$ \hfill (\S\ref{sec:kit-network})\\
  Active-perception filter $\alpha$ \hfill (\S\ref{sec:atom-commitments})\\
  Phase-oscillator coupling \hfill Eq.~\eqref{eq:kuramoto}\\[1pt]
  \textcolor{blue!40!black}{\rule{4.4cm}{0.3pt}}\\
  {\scriptsize\itshape Realizes Commitments~C3, C4}
};
\node[cluster={orange}{}] (ctrl) at (0,1.6) {%
  {\small\bfseries\color{orange!50!black}(B) Communication board}\\[2pt]
  Halt gate \hfill Eq.~\eqref{eq:halt}\\
  Drive (PD on $\lambda$) \hfill Eq.~\eqref{eq:drive}\\
  Activation routing \hfill Eq.~\eqref{eq:board}\\[1pt]
  \textcolor{orange!50!black}{\rule{4.4cm}{0.3pt}}\\
  {\scriptsize\itshape Routes opponent pair (Commitment~C1)}
};
\node[cluster={red}{}] (halt) at (10,1.6) {%
  {\small\bfseries\color{red!50!black}(C) Haltability operator}\\[2pt]
  Alert energy (load) \hfill Eq.~\eqref{eq:alertenergy}\\
  Alert-energy steady state \hfill Eq.~\eqref{eq:Estar}\\
  Alert energy (+ surprise) \hfill Eq.~\eqref{eq:alertenergy-with-surprise}\\
  Auto-halt threshold \hfill Eq.~\eqref{eq:autohalt}\\[1pt]
  \textcolor{red!50!black}{\rule{4.4cm}{0.3pt}}\\
  {\scriptsize\itshape Recruits opponency affordance (Commitment~C1)}
};
\node[cluster={green!60!black}{}] (mech) at (0,-2.5) {%
  {\small\bfseries\color{green!40!black}(A) Sensation Modulator + joint}\\[2pt]
  Force law $\Phi$ \hfill Eq.~\eqref{eq:Phi}\\
  Afferent readout $\Psi$ \hfill Eq.~\eqref{eq:Psi}\\
  Joint dynamics \hfill Eq.~\eqref{eq:joint}\\[1pt]
  \textcolor{green!40!black}{\rule{4.4cm}{0.3pt}}\\
  {\scriptsize\itshape Realizes Commitments~C2, C6, C7}
};
\node[cluster={purple}{}] (reaf) at (10,-2.5) {%
  {\small\bfseries\color{purple!50!black}(D) Reafference machinery}\\[2pt]
  Transducer $S$ \hfill Eq.~\eqref{eq:rf}\\
  Prediction $\hat s$ \hfill Eq.~\eqref{eq:predicted}\\
  Residual $r=s-\hat s$ \hfill Eq.~\eqref{eq:residual}\\
  World-model update $\hat\varphi$ \hfill Eq.~\eqref{eq:phi-update}\\[1pt]
  \textcolor{purple!50!black}{\rule{4.4cm}{0.3pt}}\\
  {\scriptsize\itshape Realizes Commitment~C5 (drop the predictable)}
};
\node[draw=gray!50!black, dashed, thick, rounded corners=3pt,
      fill=gray!7, inner sep=4pt, align=center, font=\footnotesize,
      text width=4.6cm]
  (world) at (5, -6.0) {%
  {\small\bfseries\color{gray!30!black}Habitat / world}\\[1pt]
  External stimulus state $\varphi(t)$\\
  {\scriptsize\itshape (outside the \smn{} envelope)}
};
\draw[edge] (net.south) to[bend right=8]
  node[edgelbl,pos=0.55] {$b(t),\,\alpha$} (ctrl.north);
\draw[edge] (ctrl.south west) to[bend right=20]
  node[edgelbl,left=1pt] {$a_+,\,a_-$} (mech.north west);
\draw[edge] (mech.north east) to[bend right=20]
  node[edgelbl,right=1pt] {$s_+,\,s_-$} (ctrl.south east);
\draw[edge] (mech.east) -- node[edgelbl,above=1pt] {$\theta$} (reaf.west);
\draw[edge] (mech.north east) -- node[edgelbl,pos=0.55,above right=-1pt]
  {$F_{\rm active}$} (halt.south west);
\draw[edge] (reaf.north) -- node[edgelbl,right=2pt] {$|r|$} (halt.south);
\draw[edge] (halt.west) -- node[edgelbl,above=1pt] {$g,\,E_R$} (ctrl.east);
\draw[edge, dashed] (world.north east) -- node[edgelbl,above=1pt] {$\varphi$}
  (reaf.south west);
\end{tikzpicture}
\caption{Functional dependency graph of the SMN formalism.  Each box
names a sub-system and lists the equations that live in it; arrows label
the variables that flow between sub-systems.  \textbf{(A)} The Sensation
Modulator and joint produce force from received activation and emit
proprioceptive afferents.  \textbf{(B)} The board takes afferents, the
broadcast state, the halt gate, and the alert-energy bias, and produces
the two opposing activations.  \textbf{(C)} The haltability operator
integrates motor load and sensory surprise into alert energy and the halt
gate.  \textbf{(D)} The reafference machinery pairs proprioception with
the world-state estimate to produce the residual.  \textbf{(E)} The
network operators broadcast integrated body state and gate which sensors
enter each board's predictive computation.}
\label{fig:formula-network}
\end{figure}
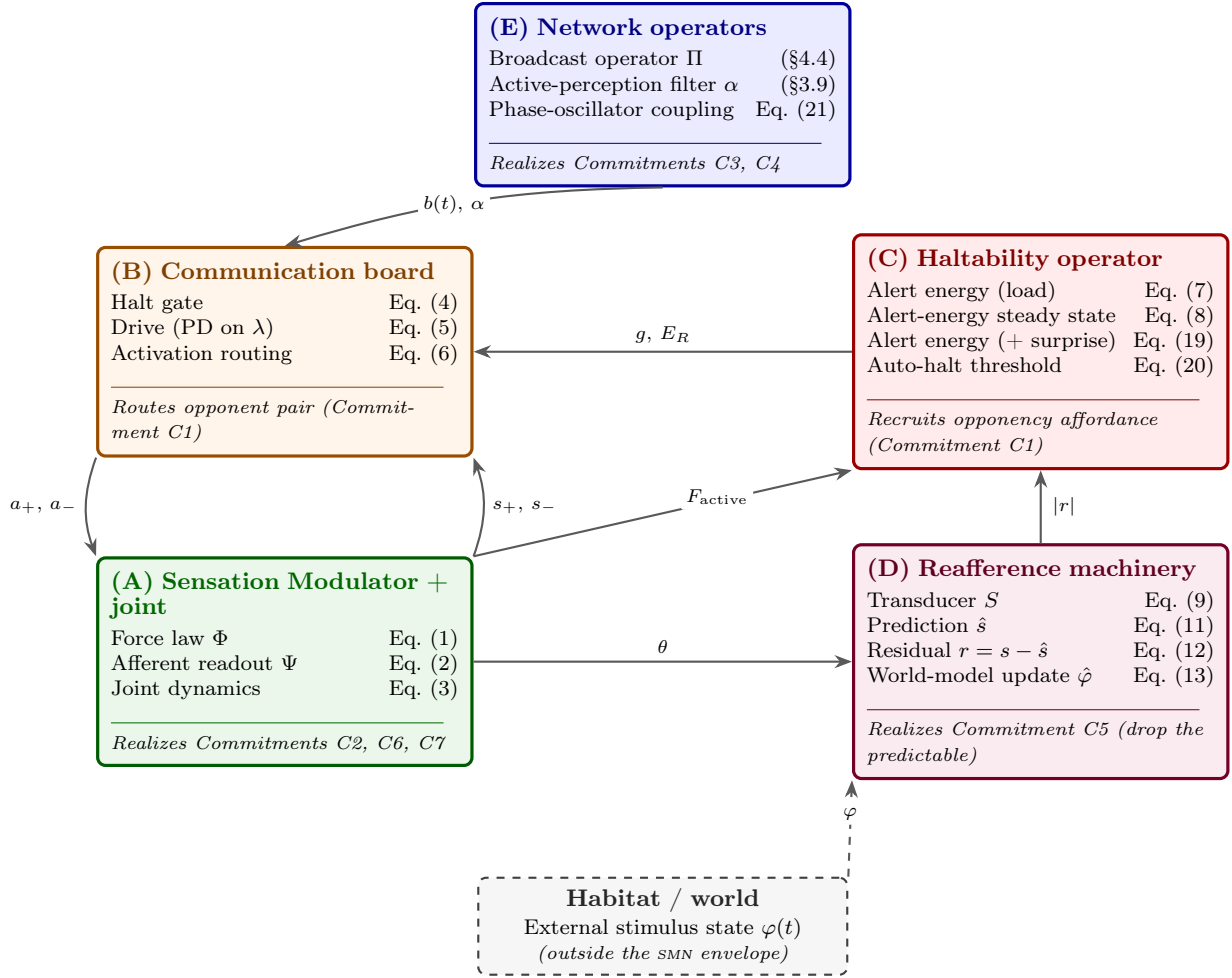

Figure~\ref{fig:formula-network} maps where the equations go: five
interlocking sub-systems --- a Sensation Modulator and joint
(\textbf{A}), a communication board (\textbf{B}), a haltability operator
(\textbf{C}), the reafference machinery (\textbf{D}), and the network
operators that integrate across CAZs (\textbf{E}).  Of these, (E) --- the
broadcast operator $\Pi$ and the active-perception filter $\alpha$ ---
integrates across CAZs; $\Pi$ is specified in \S\ref{sec:kit-network}, where
its \emph{network-closure} condition is stated and shown to be tested, on the
two-CAZ network, by the dual-signal property (\S\ref{sec:world-dualsignal}).
What Phase~I defers is the behaviour of \emph{large} networks, not the
operators themselves.  Symbol definitions are in
Table~\ref{tab:gloss-formal}.

\begin{table}[h]
\centering
\small
\caption{Formal operators and symbols.}
\label{tab:gloss-formal}
\begin{tabularx}{\linewidth}{@{}l l X l@{}}
\toprule
\textbf{Symbol} & \textbf{Reads} & \textbf{Note} & \textbf{First use} \\
\midrule
$\Phi$ & Force law of an SM. &
$F=\Phi(a,q,\dot q)$, $F\geq 0$ (pull-only); Hill-type in the reference
implementation. & Eq.~\eqref{eq:Phi} \\
\addlinespace[2pt]
$\Psi$ & Afferent readout of an SM. &
$s=\Psi(q,\dot q,F)$; identity on $(q,\dot q,F)$ in the reference
implementation. & Eq.~\eqref{eq:Psi} \\
\addlinespace[2pt]
$g$ & Halt gate of a CAZ's board. &
$g=1$ permits drive; $g=0$ engages halt. Set by $H$ from alert energy and
surprise. & Eq.~\eqref{eq:halt} \\
\addlinespace[2pt]
$\lambda$ & Target equilibrium configuration. &
The joint configuration the board steers the CAZ toward (equilibrium-point
control). & Eq.~\eqref{eq:drive} \\
\addlinespace[2pt]
$E_R$ & Alert energy. &
Non-negative metabolic cost of maintaining co-activation; built by load
and surprise, decays passively. & Eq.~\eqref{eq:alertenergy} \\
\addlinespace[2pt]
$\rho,\tau_E,\beta,\gamma$ & Alert-energy parameters. &
Build rate, decay constant, partner-tonic coupling, post-halt gain.
$\beta$ controls Reg.~1; $\gamma$ controls Reg.~2. & \S\ref{sec:atom-halt} \\
\addlinespace[2pt]
$\rho_r$ & Surprise coupling rate. &
Rate at which $|r|$ builds alert energy; couples reafference to attention.
& Eq.~\eqref{eq:alertenergy-with-surprise} \\
\addlinespace[2pt]
$\tau_h$ & Surprise threshold for halt. &
Auto-halt engages when $|r(t)|>\tau_h$, with hysteresis on release.
& Eq.~\eqref{eq:autohalt} \\
\addlinespace[2pt]
$r(t)$ & Reafference residual. &
$r=s-\hat s$. Central scalar: nonzero $r$ means the world did something the
model did not anticipate. & Eq.~\eqref{eq:residual} \\
\addlinespace[2pt]
$\hat\varphi$ & Predictor's world-state estimate. &
Updated from residual history; used to form $\hat s$.
& \S\ref{sec:world-residual} \\
\addlinespace[2pt]
$\hat\psi$ & Predictor's partner-modulation estimate. &
Used in $\mathrm{DSI}^{\mathrm{cross}}$; recoverable because the partner
shares the opponent architecture. & Eq.~\eqref{eq:dsi-cross} \\
\addlinespace[2pt]
$\kappa$ & World-model learning rate. &
Adaptation rate of $\hat\varphi$ from residual; $\kappa=0$ recovers the
non-learning agent. & Eq.~\eqref{eq:phi-update} \\
\addlinespace[2pt]
$\Pi$ & Broadcast operator. &
Maps network state to broadcast $b(t)$; every CAZ contributes to and reads
from $b$. & \S\ref{sec:kit-network} \\
\addlinespace[2pt]
$\alpha$ & Active-perception filter. &
Sensor streams admitted to a CAZ's predictive computation; others remain
under peripheral monitoring. & \S\ref{sec:atom-commitments} \\
\addlinespace[2pt]
$\mathrm{DSI}^{\mathrm{int}}$ & Internal Dual-Signal Index. &
Reconstruction quality using only SMN-internal broadcast; high in
self-contact. & Eq.~\eqref{eq:dsi-int} \\
\addlinespace[2pt]
$\mathrm{DSI}^{\mathrm{ext}}$ & External Dual-Signal Index. &
Reconstruction quality given inferred external state; rises when world
dynamics are learnable. & Eq.~\eqref{eq:dsi-ext} \\
\addlinespace[2pt]
$\mathrm{DSI}^{\mathrm{cross}}$ & Cross-SMN Dual-Signal Index. &
Reconstruction quality given inferred partner state; high in
inter-subjective contact. & Eq.~\eqref{eq:dsi-cross} \\
\bottomrule
\end{tabularx}
\end{table}

\subsection{Predicted registers}\label{sec:registers}

The formalism produces eight predicted registers, of which seven are
empirically testable and one is explicitly hypothetical: two from the
bare CAZ (\S\ref{sec:atom-caz}); three from transducers and reafference
(\S\ref{sec:worldmodel}); one from the integration of haltability and
reafference through alert energy; one structural register of intra-body
SMAPs (\S\ref{sec:world-dualsignal}); a hypothesis about shared
intentionality as inter-body extension; and one summary register of the
dual-signal contrast across the three structural configurations of
\S\ref{sec:world-dualsignal}.

Registers~1--6 are the \emph{Phase-I falsifiers}: they live entirely on
the single body and its physics, and are the ones the companion bench
(\S\ref{app:simulations}) currently exercises.  Register~7 is marked
hypothetical because it is an \emph{inter-body} extension --- two bodies
in mutual reafference --- and so belongs to the companion program
(\S\ref{sec:roadmap}); Register~8 re-presents the self/world/other
contrast that Registers~5 and~7 exhibit piecewise.  Each of Registers~1--6
is a pre-registered, falsifiable experiment in the companion bench, run
against a matched foil.  The result that would have \emph{surprised} us,
in every case, is the foil's outcome appearing in the SMN condition: a
load-independent partner tone (R1), no resumption advantage (R2), a
world-change residual that does not decay (R4), or a self-contact residual
the partner-as-stimulus form fits no better than the decoupled control
(R5).  A register passes only by \emph{separating} from its matched foil,
never by reaching a particular number.  Three of them are, moreover,
testable \emph{in vivo} and not only in silico: Register~1 as antagonist
surface-EMG that scales with agonist load and decays over seconds,
Register~2 as a resumption-latency paradigm, and Register~3 as sensory
attenuation.

\begin{description}[itemsep=4pt,labelindent=0pt,leftmargin=2em,
                    style=nextline]
\item[Register 1 (tonic load coupling):]
During steady-state engagement of one zone, the partner zone's tonic
activation $a_{\text{partner}}$ increases monotonically in the active
zone's force $F_{\text{active}}$.  From \eqref{eq:Estar} and the
steady-state version of \eqref{eq:board},
\begin{equation}
a_{\text{partner}} \approx a_0 + \beta\rho\tau_E F_{\text{active}},
\label{eq:sig1}
\end{equation}
linear in load with slope $\beta\rho\tau_E$.  Classical inhibition
predicts $a_{\text{partner}}=a_0$ independently of load.

\item[Register 2 (resumption latency):]
After a short halt during active engagement, the latency to a small
reversed target is shorter under SMN than under classical inhibition, with
advantage
\begin{equation}
\Delta t \approx \frac{\gamma E_R(t_{\text{release}})}{K_p},
\label{eq:sig2}
\end{equation}
controlled by the gain $\gamma$ and the alert energy at halt release.

\item[Register 3 (reafference --- self versus world):]
For a fixed change in the sensor reading, the residual $|r(t)|$ is small
when the change is self-generated (agent moves through a static world) and
large when world-generated (agent still while the world moves,
$\hat\varphi$ outdated).  This realizes the \emph{Reafferenzprinzip}
\citep{vonHolstMittelstaedt1950,CrapseSommer2008} as a quantitative
signature --- the same signature measured \emph{in vivo} in
sensory-attenuation paradigms \citep{BlakemoreWolpertFrith1998Tickle}.
Active inference gives that signature a native reading --- self-generated
input is attenuated as its precision is reduced
\citep{Brown2013SensoryAttenuation} --- so Register~3 separates a
reafferent world-model from one lacking efference copy, not the SMN from
active inference.

\item[Register 4 (sensorimotor contingency mastery):]
With $\kappa>0$ in \eqref{eq:phi-update}, the residual following a world
change decays at a rate set by the learning rate and the receptive-field
gradient; mastery is the rolling-RMS residual approaching the noise floor
after each world-state change.  This realizes the O'Regan--No\"e theory
\citep{OReganNoe2001Sensorimotor} as a measurable adaptation process.

\item[Register 5 (intra-body SMAP / dual-signal residual):]
For any two CAZs of one SMN whose mechanical states co-determine a shared
sensory substrate, the partner CAZ's $\theta$ is available through
cross-zone broadcast, and the residual stays at the noise floor without
adaptation.  Regressed against the partner-as-stimulus form
\begin{equation}
\Delta\hat s(\theta_1,\theta_2,\hat\varphi_2)
= I\!\left[\exp\!\left(-\tfrac{(\theta_2-\theta_1)^2}{\sigma^2}\right)
        - \exp\!\left(-\tfrac{(\theta_2-\hat\varphi_2)^2}{\sigma^2}\right)\right],
\label{eq:sig5}
\end{equation}
with $I$ the intensity from \eqref{eq:rf}, $R^2\to 1$; for a decoupled
control reading an exogenous stimulus, $R^2\to 0$.  The signature is a
\emph{structural} feature of the residual, not a magnitude difference ---
the empirical realization of the dual-signal property (DS-1)--(DS-4).
Because self-contact is \emph{defined} by broadcast membership, what
Register~5 tests is not whether the partner is recoverable --- that is
built in --- but whether the residual carries the \emph{structural} form
of \eqref{eq:sig5} and whether that form is \emph{diagnostic}: the
decoupled control, lacking the partner variable, must fit it no better than
chance.  The falsifying outcome is a self-contact $R^2$ well below $1$, or
a decoupled $R^2$ near it.

\item[Register 6 (surprise-driven attention and auto-halt):]
With the surprise term of \eqref{eq:alertenergy-with-surprise} active, an
unpredicted event drives $|r|$ above the threshold $\tau_h$ of
\eqref{eq:autohalt}, gating the active drive within one integration step.
The same alert-energy substrate that distinguishes haltability from
inhibition is thereby the substrate of attention
(\S\ref{sec:world-attention}).

\item[Register 7 (shared intentionality, hypothesis):]
Two SMN agents in mutual reafference (case (c) of
\S\ref{sec:world-dualsignal}, both predictors adapting via
\eqref{eq:phi-update}) show a joint mutual residual $\sqrt{r_1^2+r_2^2}$
decaying toward the noise floor, dramatically reduced relative to
asymmetric or no adaptation ($\sim 92\%$ vs.\ neither-adapts in the
reference simulation).  We hypothesize its decay rate is the empirical
signature of shared intentionality in the sense of
\citet{Tomasello2005Understanding}.

\item[Register 8 (the dual-signal contrast: self / world / other):]
A \emph{unifying re-presentation} of the contrast Registers~5 and~7
exhibit piecewise.  The three Dual-Signal Indices of
\S\ref{sec:world-dualsignal} give the architectural geometry of contact:
self-contact drives $\mathrm{DSI}^{\mathrm{int}}\to 1$ once $\Pi$ has
settled; world-contact gives $\mathrm{DSI}^{\mathrm{int}}\to 0$ with
$\mathrm{DSI}^{\mathrm{ext}}$ rising iff the external dynamics are
learnable; inter-subjective contact gives low
$\mathrm{DSI}^{\mathrm{int}}$ but $\mathrm{DSI}^{\mathrm{cross}}$ rising
with adaptation.  The distinction is \emph{architecturally available}, not
something the agent thereby \emph{has} in a developmentally rich sense;
how it comes to be detected and used is companion work
(\S\ref{sec:roadmap}, items~1--2).
\end{description}

\paragraph{Consequences awaiting implementation.}
The network-level formalism (\S\ref{sec:taxonomy}) entails three further
predictions the present reference simulation does not exercise --- they
are consequences of the architecture, and their quantification requires a
multi-CAZ implementation:
\begin{enumerate}[leftmargin=2em,itemsep=2pt]
\item \textbf{BAP-backdrop entrainment.} \hap-level \caz{}s should exhibit
phase-locked modulation by ongoing BAPs.
\item \textbf{Mode-switching latency between axial and appendicular
regimes}, with a characteristic switching signature at the network level
(analogous to the Haken--Kelso--Bunz phase transition
\citep{HakenKelsoBunz1985PhaseTransitions}).
\item \textbf{Hebbian recruitment of \nap{}s.} Recurrently co-active
\caz{}s should develop strengthened cross-routing weights.
\end{enumerate}
We do not number these as Registers~9--11 because they presuppose an
implementation the present preprint does not provide.

\subsection{Reference realizations in the companion bench}\label{app:simulations}

Each register above, and each construction in the main body --- the
elastic self-model (\S\ref{sec:atom-joint}), the reafference cut and
Dual-Signal Index (\S\ref{sec:worldmodel}), and haltable
object-directedness (\S\ref{sec:directedness}) --- is realized as a
runnable, falsifiable experiment in the companion computational bench,
\textsc{SMN-Lab}:
\begin{center}
\href{https://smn-lab.readthedocs.io}{\texttt{smn-lab.readthedocs.io}}
\quad$\cdot$\quad
\href{https://github.com/gnowgi/smn-lab}{\texttt{github.com/gnowgi/smn-lab}}
\end{center}
\noindent
Every experiment there is pre-registered (hypothesis, order parameter,
matched foil, and pass/fail fixed before running) and exports tidy,
self-describing datasets; the read-out functions shown in the main body
are the same functions the bench runs.  The eight registers of
\S\ref{sec:registers} are realized there as pre-registered experiments
with matched foils: Registers~1--6 on the single body --- including the
tonic-load coupling and the post-halt resumption advantage, each against a
classical-inhibition foil, and the structural dual-signal residual --- while
the inter-body Register~7 is left to the companion program.  Consolidating
the detailed, growing simulation record in the living bench rather than in
this paper is deliberate: the paper states the architecture and its
falsifiers; the bench carries the results.

\section*{Statement on AI assistance}

In preparing this manuscript, the authors used a large language model
(Anthropic Claude) for editorial and production tasks: copy-editing and
prose tightening, \LaTeX{} formatting, suggesting and locating
bibliographic references, drafting cross-references, and assistance with
the companion bench code, \textsc{SMN-Lab} (\S\ref{app:simulations}).  For
the simulations,
the architectural predictions, the choice of which registers to implement,
and the discriminating contrasts to test were specified by the authors; AI
assistance contributed to a subset of the experimental designs and to the
implementation of all simulations, under author review.  The architectural
commitments, formal definitions, the eight predicted registers, the
philosophical thesis on haltability and phenomenology, and all substantive
interpretive claims are the authors' own.  Every claim in the paper has
been reviewed and is the authors' responsibility.

\bibliographystyle{plainnat}
\bibliography{library}

\end{document}